%% file: main.tex
\documentclass{aa}
\usepackage{txfonts}
\usepackage{amsmath} 
\usepackage{amssymb} 
\usepackage{graphicx, graphics}
\usepackage{xcolor}
\usepackage{ragged2e}
\usepackage{soul}

\usepackage{hyperref}
\hypersetup{colorlinks=true, 
    linkcolor=blue, 
    filecolor=magenta, 
    urlcolor=blue, 
    citecolor=blue, 
    pdfauthor={Kartick Sarkar}, 
    pdftitle={Fermi Bubbles Review} 
}

\renewcommand{\ion}[2]{#1 \textsc{#2}}

\newcommand{\msun}{M$_{\odot}$ }
\newcommand{\kmps}{km s$^{-1}$ }
\newcommand{\pcmsq}{cm$^{-2}$ }
\newcommand{\pcc}{cm$^{-3}$ }
\newcommand{\mpcc}{$m_{\rm p}$ \pcc}
\newcommand{\psec}{s$^{-1}$ }
\newcommand{\ergps}{erg \psec}
\newcommand{\mpy}{\msun yr$^{-1}$ }
\newcommand{\ovii}{\ion{O}{vii} }
\newcommand{\oviii}{\ion{O}{viii} }
\renewcommand{\deg}{^\circ}

\newcommand{\gray}{$\gamma$-ray }
\newcommand{\grays}{$\gamma$-rays }
\newcommand{\fermilat}{\textit{Fermi-LAT} }
\newcommand{\rosat}{\textit{ROSAT} }
\newcommand{\erosita}{\textit{eROSITA} }
\newcommand{\planck}{\textit{Planck} }
\newcommand{\sgra}{Sgr A$^*$ }

\begin{document}

\title{The Fermi/eROSITA Bubbles:}
\subtitle{A look into the nuclear outflow from the Milky Way}
\author{Kartick C. Sarkar}
\authorrunning{Kartick Sarkar}

\institute{School of Physics and Astronomy, Tel Aviv University, Israel \\
Dept of Space, Planetary \& Astronomical Sciences and Engineering, Indian Institute of Technology Kanpur, India\\
\email{kcsarkar@iitk.ac.in, kartick.c.sarkar100@gmail.com}\\\\
$^*$ \textit{For image copyrights, please follow the copyright information from the original sources.}
}

\abstract{
Galactic outflows are ubiquitous in galaxies containing active star formation or supermassive black hole activity. The presence of a large-scale outflow from the center of our own Galaxy was confirmed after the discovery of two large ($\sim 8-10$ kpc) \gray bubbles using the \textit{Fermi-LAT} telescope. These bubbles, known as the Fermi Bubbles, are highly symmetric about the Galactic disk as well as around the Galactic rotation axis and appear to emanate from the center of our Galaxy. The sharp edges of these bubbles suggest that they are related to the Galactic outflow. These bubbles are surrounded by two even bigger ($\sim 12-14$ kpc) X-ray structures, known as the eROSITA bubbles. Together, they represent the characteristics of an outflow from the Galaxy into the circumgalactic medium. Multi-wavelength observations such as in radio, microwave, and UV toward the Fermi Bubbles have provided us with much information in the last decade. However, the origin and the nature of these bubbles remain elusive. In this review, I summarize the observations related to the Fermi/eROSITA Bubbles at different scales and wavelengths, and give a brief overview of our current understanding of them. 
}

\maketitle
\nolinenumbers

\tableofcontents

\input{chap_1_introduction}

\input{chap_2_multiwavelength_view}

\input{chap_4_GC_activity}
\input{chap_5_proposed_origin}
\input{chap_6_summary}

\input{chap_7_acknoledgement}
%
\bibliographystyle{aa}
\bibliography{review}
%
%
%
%
%
\end{document}

%% file: chap_1_introduction.tex
\section{Introduction}
\label{chap1:Introduction}

The evolution of galaxies is undeniably connected to the feedback from the galaxy, be it in the interstellar medium (ISM) or in the circumgalactic medium (CGM). Feedback sources include supernovae (SNe) or active galactic nuclei (AGN). Feedback controls the mass and energy transfer between the galaxy and its environment which further affects the star formation in the galaxy. It is believed that below the critical mass ($M \sim 10^{12}$ \msun), the SNe feedback dominates the feedback whereas, the black hole feedback takes over at higher masses \citep{Silk1977, Dekel1986, Silk2012}. Coincidentally, our Galaxy is at the critical mass where one expects both the SNe and black hole feedback to be active. Therefore, outflows from the center of our galaxy provide an excellent opportunity to study the nature of outflow and its effect on the Galaxy. 

Studying the observational aspects of the feedback in different environments is often clouded by the lack of proper observational tracers. Therefore, observations that probe the feedback mechanism rely on partial information such as optical/infrared emission or absorption (for galaxies), radio emission (for galaxy clusters), or X-ray emission. It is rarely the case that we get to observe a multi-wavelength picture of a system that is undergoing feedback. In the past two decades, it has become increasingly clear that the Milky Way is such a system with a large-scale ($\gtrsim$ kpc) outflow where we can possibly have a multi-wavelength observation of the feedback process. 
The discovery of the X-ray hourglass by ROSAT \citep{Snowden1997, Bland-Hawthorn2003}, the microwave haze \citep{Dobler2008}, and the two giant gamma-ray bubbles, called Fermi Bubbles (FBs; \citealt{Dobler2010, Su2010}), toward the Galactic Center (GC) brought a revolution in this field and boosted the motivation to study the outflow in our Galaxy in much more detail. 
With the enhanced engagement in the past decade, the scientific community has been able to obtain a plethora of observational data of the regions surrounding the Fermi Bubbles in multiple wavelengths, ranging from radio, optical/UV, and X-ray to gamma-rays. The origin of the bubbles, however, still remains debated. 

In this review, I will provide an overview of the progress made to understand the origin of these bubbles and the related physical processes. The review will start with a brief introduction to the outflow process in galaxies in order to familiarize the readers with certain technical terms. The review will then summarize i) the existing observations towards the FBs and the Galactic center, ii) conclusions from the independent observations (often contradictory), and iii) theoretical models trying to make sense of the observations and searching for the origin of these bubbles. 

\subsection{Description of an outflow}
\label{chap1:subsec:outflow-general}
Galactic outflows (or winds) have generally been defined as the outflow of cold/warm ($\sim 10^4$ K) gas from the galaxy. The first example of galactic outflow was discovered in M82 using optical and radio observations \citep{Lynds1963, Burbidge1964}. The observations bore characteristics similar to the Crab Nebula, a SN remnant. Therefore, it was proposed that the origin of such an outflow in M82 is due to the SNe explosions at the center of the galaxy. It was later confirmed that the central $\sim 200$pc of M82 indeed shows an enhanced star-formation activity at a rate of $\approx 10$ \mpy over the last $\sim 15$ Myr \citep{Barker2008, Konstantopoulos2009}. Galactic outflows are regularly discovered in external galaxies either via emission from the outflowing material or via absorption caused by the material against a background light source. \cite{Heckman1990} noticed double-peaked emission lines at a distance of $\sim$ kpc from Far Infrared Galaxies that are thought to host significant star formation. The velocity of the gas was estimated to be $\sim 200-600$ \kmps, comparable to the velocities of the optical filaments in M82 \citep{Shopbell1998}. X-ray emission from hot ($\sim 0.5$ keV) gas has also been observed to be coinciding with the optical emission in the M82 wind, indicating a multiphase nature of the galactic outflow. The energetics of the optical filaments and X-ray gas suggests that the wind is being driven by SN explosions at the central star-forming region of M82 \citep{Watson1984, Kronberg1985, Strickland1997, Strickland2004, Lopez2020}.

Another example of a well-studied outflow is in the nearby galaxy NGC 3079. The galaxy hosts a $\sim$ kpc large bubble that is observable in H$\alpha$, [\ion{N}{ii}], and radio emission \citep{Veilleux1994, Cecil2001} and appears to be inside an even larger ($\sim 60$ kpc) X-ray/UV bubble \citep{Hodges-Kluck2020}. The $\sim 60$ kpc bubble is thought to have originated due to the central star-formation activity as suggested by the O/Fe abundance ratio in the bubble. The total star formation rate (SFR) in NGC 3079 is estimated to be $\approx 10$ \mpy \citep{Yamagishi2010}. However, the origin of the $\sim$kpc bubbles is debated since NGC 3079 also hosts a $\sim 15$ pc long radio jet at the galactic center. Moreover, the star formation falls short of producing the required energy. Since the observed AGN jet and the bubble are misaligned by $\approx 65^\circ$ in this galaxy, it has been suggested that a precessing jet could have produced the $\sim$ kpc scale bubble or at least contributed to some of the energy along with the SNe \citep{Irwin1988, Veilleux1994}.

\begin{figure*}
    \centering
    \includegraphics[width=\textwidth, clip=true, trim={0cm 4.5cm 0cm 2cm}]{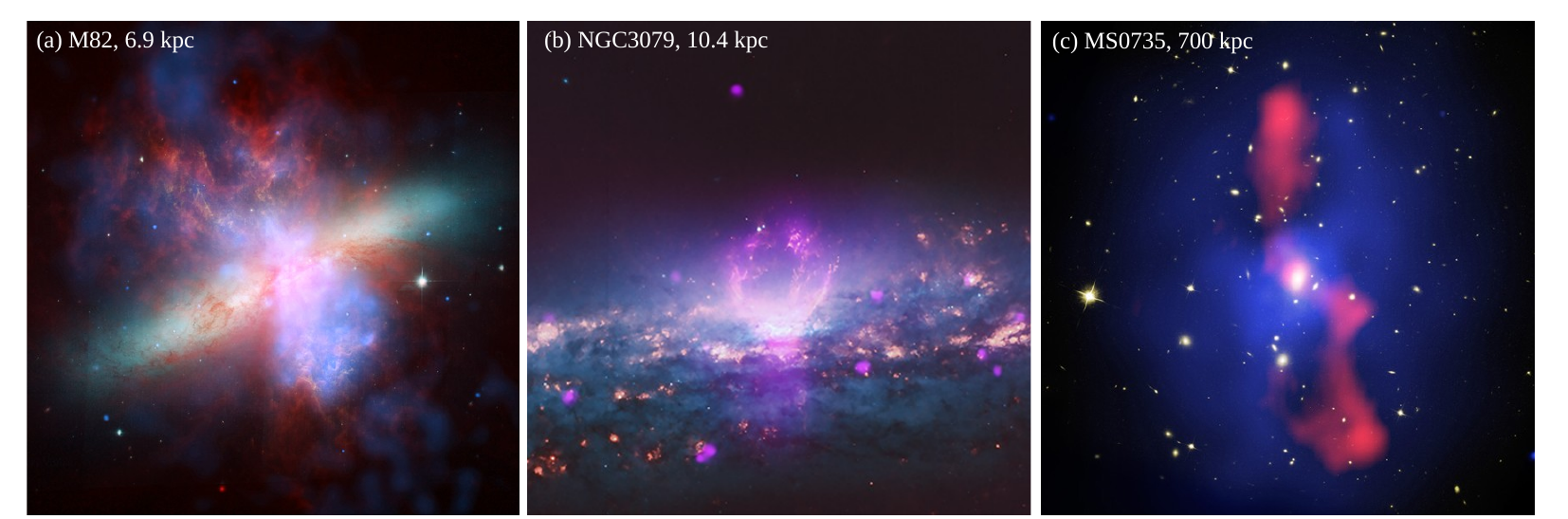}
    \caption{Bubbles and winds in different systems. (a) galaxy M82, mass $\sim 5\times 10^{10}$ \msun, SFR $\approx 10$ \mpy. Credit: \href{https://chandra.harvard.edu/photo/2006/m82/}{X-ray: NASA/CXC/JHU/D.Strickland; Optical: NASA/ESA/STScI/AURA/The Hubble Heritage Team; IR: NASA/JPL-Caltech/Univ. of AZ/C. Engelbracht}. (b) galaxy NGC 3079, mass $\sim 10^{12}$ \msun, SFR $\approx 10$ \mpy, AGN power $\sim 7\times 10^{42}$ \ergps. \href{https://chandra.harvard.edu/photo/2019/ngc3079/}{Credit NASA/CXC/SAO}, and (c) galaxy cluster MS 0735, mass $\sim 10^{15}$ \msun, AGN power $\sim 2\times 10^{46}$\ergps. Credit: \href{https://chandra.harvard.edu/photo/2006/ms0735/}{X-ray: NASA/CXC/Univ. Waterloo/B.McNamara; Optical: NASA/ESA/STScI/Univ. Waterloo/B.McNamara; Radio: NRAO/Ohio Univ./L.Birzan et al}. The physical extent of each image is noted at the top of the image. 
    }
    \label{fig:3-bubbles}
\end{figure*}

The importance of AGN jets in driving powerful outflows increases significantly as we go to bigger objects, such as galaxy clusters. The bright central galaxies of the clusters often host supermassive black holes (SMBH) of mass $\sim 10^9-10^{10}$ \msun. Such massive black holes can produce powerful AGN jets/bubbles that are observable in radio emission \citep{Burns1990}. The jets are often correlated with massive X-ray cavities of size $\sim 10-100$ kpc in the clusters \citep{Bohringer1993, McNamara2000, McNamara2001, Mittal2009, Gitti2012, Biava2021}. It is believed that the shocks and cavities driven by the jets regulate the cooling and heating of the intracluster medium and thus regulate the growth of the SMBH itself \citep{Jones2002, Fabian2003, Voit2005}. 

Figure \ref{fig:3-bubbles} shows a visual comparison of the winds and bubbles in different systems. From left to right, the systems are more and more massive, and the central energy source changes from starburst to AGN, as expected \citep{Dekel1986, Silk2012}. While M82 has typical circular velocity of $v_{\rm rot} \sim 80$ \kmps \citep{Goetz1990, Greco2012}, NGC 3079 and MS 0735 have typical velocities of $\sim 200$ \kmps and $\sim 2000$ \kmps \citep{Koda2002,McNamara2005,Cavagnolo2009}. Although it is clear that the MS 0735 bubbles are most probably driven by a supermassive black hole, the importance of the SMBH in driving the wind/bubble in NGC 3079 is similar to the SNe \citep{Sebastian2020}. Out of these three cases, the Milky Way (MW) is most probably similar to NGC 3079 which also has a central black (mass $\sim 2\times 10^6$ \msun; \citealt{Trotter1998, Kondratko2005}), very similar to the MW black hole, $M_{\rm bh} \approx 4\times 10^6$ \msun \citep{Genzel1996, Schodel2002, Ghez2003, EventHorizon2022_IV}. The only major difference is that the star formation in the inner region of MW is much smaller than NGC 3079. This improves the chance that MW AGN is, in principle, more active and probably contributes to the formation of the FBs. This is one of the main reasons behind the decade-old debate - whether the Fermi Bubbles originated from SNe or from an AGN in the GC. Given that both sources are equally possible in the MW, let us briefly look into the energies produced by SNe and AGN. For a detailed review of AGN and stellar feedback-driven bubbles/winds, please refer to \citealt{Veilleux2005, DZhang2018, Silk2012}.

\subsubsection{SNe-driven outflows}
\label{chap1:subsubsec:SNe-outflow}
The stellar feedback is mostly dominated by SNe explosions (typically core-collapse type) after the initial $\sim 3-4$ Myr of the stellar evolution \citep{Leitherer1995,Leitherer1999}. Core-collapse SNe occur when massive stars ($M_\star \gtrsim 8$ \msun) fail to produce enough radiation by nuclear fusion to support their own weight against gravity. This generally happens after the stellar core, where fusion takes place, exhausts its Hydrogen fuel and successively starts fusing heavier elements such as Helium, Carbon, Nitrogen, and Oxygen together. 
The heavier elements burning phase is much shorter compared to the Hydrogen burning phase and hence the end of the Hydrogen burning phase in a star marks the collapse of the core. 
For a massive star ($\gtrsim 8$ \msun), fusion continues until Iron since further fusion reactions are endothermic. At this stage, due to the absence of energy production, the core collapses further to create a proto-neutron star and generates about $\sim 10^{53}$ erg of energy, most of which is carried away by neutrinos. Only a small fraction (about a \%) of this neutrino energy is transferred to the collapsing material in the form of mechanical energy which then causes the entire star to explode as a SN. The explosion leaves behind a compact core, that may turn into a black hole if the progenitor mass is $\gtrsim 25$ \msun \citep{Woosley2005, Hirschi2007}. The number of such core-collapse SNe in a general stellar population can be calculated by finding the number of stars with mass, $M_\star \gtrsim 8$ \msun in a young stellar population. Assuming a typical initial mass function for the young stars, this number is $\sim 1/(100$ \msun), i.e. about one SN explosion for every $100$ \msun that goes into star formation. Given that each SN produces $\approx 10^{51}$ erg of mechanical energy, the rate of SNe generated energy is $\approx 10^{49}$ per \msun of young stars formed. Along with the energy, a SN also deposits a significant fraction of the stellar mass into the surrounding medium. Considering typical stellar synthesis models \citep[such as][]{Leitherer1999}, we can write down the total energy and mass output from a star-forming region to be 
$\dot{E}_\star \approx 7\times 10^{41} \mbox{ \ergps}$ (SFR/\mpy) and $\dot{M}_\star \approx 0.3$ SFR, respectively.
However, not all this energy is available for driving an outflow. A significant fraction of this energy is lost via radiative cooling during the SNe evolution in the ISM. Although the retained energy ($\alpha$) for individual SNe is $\sim 5\%$ \citep{Cox1972, CGKim2015, Sarkar2021b}, clustered SNe can easily retain $\sim 20-40\%$ of the initial energy \citep{Strickland2009, Gentry2017, Vasiliev2017, Fielding2018}. Mass in the outflow, on the other hand, can increase as the outflow entrains/sweeps more mass as it propagates. The mass entrainment is represented by the mass loading factor, $\beta$. Together, one can write down the mass and energy available for the wind to be
\begin{eqnarray}
\label{chap1:eq:Edot-sfr}
\dot{E}_{\rm wind} &\approx&  7\times 10^{41}\,\,\alpha\, \mbox{ \ergps} \frac{\mbox{SFR}}{\mbox{\mpy}} \nonumber \\
\dot{M}_{\rm wind} &\approx& \beta\,\, \mbox{SFR} \,.
\end{eqnarray}
For future calculations, I will assume $\alpha \approx 0.3$. The mass loading factor, $\beta$, have values in the range of $0.1-10$ \citep{Strickland2009, Arribas2014, Heckman2015}. In an energy-driven outflow, as is the case for an outflow expanding into a low-density medium, it is only the energy that decides the large-scale structure and evolution of the outflow. Mass loading factor may have an effect on $\alpha$ but it does not directly affect the impact of the outflow with the CGM. 

\subsubsection{AGN-driven outflow}
\label{chap1:subsubsec:agn-outflow}
Unlike the SNe explosions where most of the energy comes from nuclear fusion, the main source of the energy in an AGN is gravitational energy. Considering a black hole of mass $M_{\rm bh}$ that formed over time $t_{\rm bh}$, we can estimate the average energy released by the BH accretion over its lifetime to be $\sim \epsilon_{\rm acc}\: G\: M_{\rm bh}^2/(r_s t_{\rm bh})$, where $r_s = 2 G M_{\rm bh}/c^2$ is the Schwarzschild radius of the BH and $\epsilon_{\rm acc}$ is the efficiency of the energy production process. For black holes, this produces an average luminosity of $\sim \epsilon_{\rm acc}\: 3\times 10^{44}$ ($M_{\rm bh}/10^8$\msun) ($t_{\rm bh}$/10 Gyr)$^{-1}$ \ergps which is $\sim 3\epsilon_{\rm acc}\%$ of the Eddington luminosity, $L_{\rm edd}$. Clearly, AGNs can provide sufficient energy to drive galactic outflows. Although there are gaps in the understanding of the exact energy release process i.e. whether it is jet driven or disk wind-driven, or radiation-driven, one can write down the BH mechanical luminosity in the radiatively inefficient regime to be
\begin{equation}
\label{chap1:eq:Edot-bh}
    L_{\rm bh} \approx \epsilon_{\rm acc}\: \dot{M}_{\rm bh} c^2 \sim 0.1\:\dot{M}_{\rm acc} c^2\,,
\end{equation}
where $\dot{M}_{\rm acc} \equiv \dot{M}_{\rm bh}$ is the accretion rate of the BH. It is generally accepted that the energy is released in the form of mechanical energy if $\dot{M}_{\rm acc}/\dot{M}_{\rm edd} \lesssim 10^{-2}$, else, in the form of radiation energy \citep{Churazov2005, Guistini2019}. Here, $\dot{M}_{\rm edd} $ is the accretion rate corresponding to the Eddington luminosity, $L_{\rm edd} \equiv \dot{M}_{\rm edd} c^2 = 1.3 \times 10^{38}\, (M_{\rm bh}/M_\odot)$ \ergps. Since the current accretion rate in our galaxy is $\sim 10^{-8}$ \mpy \citep{Yuan2004}, $\dot{M}_{\rm acc}/\dot{M}_{\rm edd} \sim 10^{-7}$. Therefore, any possible AGN energy release in our Galaxy is most probably in the form of a mechanical jet/wind. There are, however, claims that suggest that the SMBH at the MW center had a much more violent history in the past \citep{Kayoma1996, Baganoff2003, Totani2006, Bland-Hawthorn2019}. See section \ref{chap4:subsec:MWBH-acc-rate} for a more detailed discussion on this topic.

\subsection{Confinement by the circumgalactic medium}
\label{chap1:subsec:cgm}
The outflow dynamics is not only dictated by the gravity of the galaxy but also by the presence of a circumgalactic medium (CGM). The CGM is a low density ($n_{\rm cgm}\sim 10^{-2\: {\rm to}\: -5}$ \pcc) and hot ($T_{\rm cgm} \sim 10^6$ K) gaseous medium surrounding a galaxy out to a radius of $\sim 100-300$ kpc. The CGM forms around galaxies due to the continuous accretion of baryonic matter onto the galaxy and the formation of an accretion shock close to the virial radius. 
Theoretically, it is expected that the accretion shocks in massive galaxies (total mass $\gtrsim 10^{12}$ \msun) are stable against radiative cooling and hence these galaxies are able to sustain a hot CGM around it \citep{White1978, Birnboim2003, Maller2004, Dekel2006}. Given that our Galaxy (virial mass, $M_{\rm vir} \approx 10^{12.1}$ \msun; \citealt{McMillan2011, McMillan2017}) lies at the boundary of the above mass limit, it was uncertain whether our Galaxy contains such a CGM around it. 
However, observations in the past two decades have made significant progress not only to prove that the MW has a hot CGM but also to constrain some of its properties. 
Detection of \ion{O}{vi}, \ion{O}{vii}, and \ion{O}{viii} absorption lines along several Galactic and extragalactic sources made it clear that MW has a reserve of hot ($\sim 10^6$ K) gas surrounding it \citep{Nicastro2003, Sembach2003, Wang2005}. More direct evidence includes the head-tail structure of \ion{H}{i} clouds at a distance of $\sim 10$ kpc in the MW which indicated that these clouds are moving in a low-density medium, i.e. the CGM \citep{Putman2011}. 
More quantitative description of the CGM came from the \ion{O}{vii} and \ion{O}{viii} absorption lines against background AGNs observed using \textit{Chandra} and \textit{XMM-Newton} telescopes \citep{Nicastro2003, Bregman2007, Gupta2012, Henley2010, Miller2015, Faerman2017, Faerman2020}. The observations indicate that i) the Galactic CGM extends till $\sim 200$ kpc and its density profile can be represented roughly as $n_{\rm cgm} \approx 1.35\times 10^{-2}$ \pcc $(r/{\rm kpc})^{-3/2}$ for $r\gtrsim 1$ kpc \citep{Miller2015}, and ii) the temperature of the CGM is quite uniform on the sky and has a value of $T_{\rm cgm} \approx 2.2 \times 10^6$ K \citep{Henley2013}, close to the virial temperature of the Galaxy, $T_{\rm vir} \approx 5\times 10^5$ K  \footnote{It is quite possible that the estimated temperature is dominated by the gas within $\lesssim 50$ kpc where the gas is heated by the SNe feedback from the Galaxy \citep{Faerman2020}. }. 
More recently, there have been reports of an even hotter ($\sim 10^7$ K) component of the CGM \citep{SDas2019a, SDas2019b, Gupta2021}. It is, however, unclear if this is just another component of the CGM or if it represents a log-normal temperature distribution in the CGM \citep{Vijayan2022}. The \ovii absorption lines in the MW also revealed that the CGM is not stationary but is rotating with a speed of $\approx 180\pm40$ \kmps, very close to co-rotation with the stellar disk \citep{Hodges-Kluck2016}. 

The importance of CGM in the propagation of outflow is simple to understand. As an example, for a galaxy with a star-formation rate, SFR, the outflow velocity in the absence of the CGM would be given as $v_{\rm fw} = \sqrt{2\dot{E}_{\rm wind}/\dot{M}_{\rm wind}} \approx 1480 \sqrt{\alpha/\beta}$ \kmps. However, the same outflow would be modified in the presence of CGM and create a shock, much like the stellar wind in the interstellar medium. The shock radius and velocity in the presence of the CGM can be written as \citep{Castor1975, Weaver1977}
\begin{eqnarray}
\label{chap1:eq:rs-sfr}
    r_s &\approx& \left( \frac{\dot{E}_{\rm wind}}{\rho_{\rm cgm} }\right)^{1/5} \: t^{3/5} \nonumber \\
    &\approx& 2.4 \mbox{  kpc  } \left(\frac{\alpha}{n_{-3}}\right)^{-1/5} \left(\frac{\mbox{SFR}}{\mbox{\mpy}} \right)^{1/5}\: t_6^{3/5}  \nonumber \\ 
    v_s &\approx& \frac{3}{5}\left( \frac{\dot{E}_{\rm wind}}{\rho_{\rm cgm} }\right)^{1/5}\:  t^{-2/5} \nonumber \\
    &\approx& 1400 \mbox{  \kmps  } \left(\frac{\alpha}{n_{-3}}\right)^{-1/5} \left(\frac{\mbox{SFR}}{\mbox{\mpy}} \right)^{1/5}\: t_6^{-2/5}
\end{eqnarray}
where, $n_{\rm cgm} = 10^{-3}\: n_{-3}$ is the particle number density of the CGM, and $t_6 = t/$Myr. Clearly, the outflow becomes slower quickly in the presence of a CGM over a time scale of $\sim$ Myr. Additionally, for a galaxy that is forming stars at a rate of SFR $= 1$ \mpy for $\sim 30$ Myr, the wind propagates only to $\sim 24$ kpc, far shorter than the extent of the CGM. This means the wind is still confined within the CGM and that too with a velocity of $v_s \sim 450$ \kmps, i.e. with Mach $\sim 2$ (assuming a sound speed of $\approx 210$ \kmps for a CGM temperature of $2\times 10^6$ K). A very similar estimate for a typical AGN outburst at the Galactic center produces a shock radius and velocity 
\begin{eqnarray}
\label{chap1:eq:rs-agn}
    r_s &\approx& \left(\frac{E_{\rm agn}}{\rho_{\rm cgm}} \right)^{1/5}\: t^{2/5} \approx 3.2 \mbox{  kpc  } \left(\frac{E_{56}}{n_{-3}} \right)^{1/5}\: t_6^{2/5} \nonumber \\
    v_s & \approx & \frac{2}{5} \left(\frac{E_{\rm agn}}{\rho_{\rm cgm}} \right)^{1/5}\: t^{-3/5} \approx 1270 \mbox{  \kmps  } \left(\frac{E_{56}}{n_{-3}} \right)^{1/5}\: t_6^{-3/5}
\end{eqnarray}
where, $E_{\rm agn} = 10^{56} E_{56}$ erg is the energy for the AGN outburst. In this case, I assumed that all the energy was released in a burst lasting for a short duration compared to the dynamical time such that the blast-wave solution \citep{Sedov1946, Taylor1950} applies. Following the previous example, we find that an outburst of $E_{56} = 6$ (for an equivalent energy of SFR $=1$ \mpy over 30 Myr) would only travel to $\sim 18$ kpc with a velocity of $236$ \kmps after a time of $\sim 30$ Myr. The numbers are very similar to the numbers from the star-formation-driven wind. 

Therefore, it is quite natural to expect an outflow extending to $\sim 10$'s of kpc in galaxies that contain sufficient star formation or AGN activity. The shock driven by such outflows may even shine in X-rays depending on the presence of a CGM (Jana et al., \textit{in prep.}). 
In the next sections, we discuss the presence of such an outflow in our own Galaxy, its discovery, and its implications. 

\begin{figure*}
    \centering
    \includegraphics[width=0.95\textwidth, clip=true, trim={1cm  1cm 1cm 0cm}]{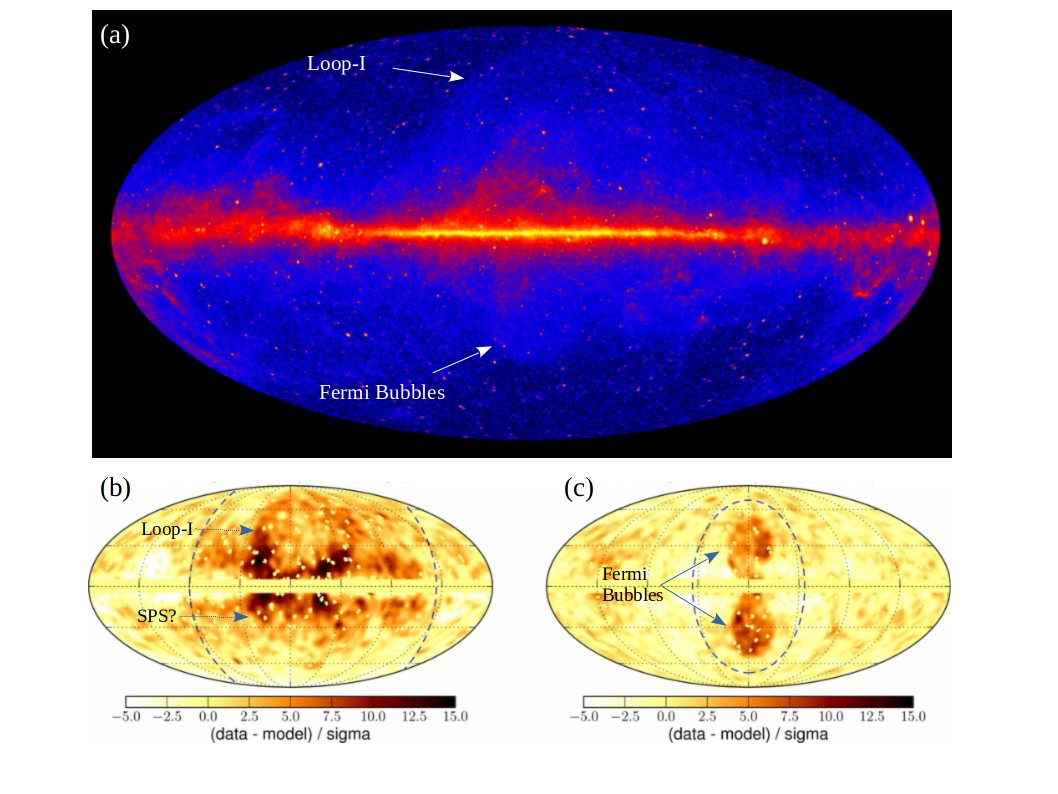}
    \caption{\fermilat all-sky maps. \textit{Panel (a)}: 12-yr view of the \gray sky ($E >1$ GeV), shown in Hammer projection. The bright central plane represents the Galactic disk where most of the \gray emission is due to the CR protons produced by supernovae shocks. Bright point sources outside the disk are external galaxies powered by AGNs. The most important structures for this review are marked as the Loop-I and the Fermi Bubbles. Credit:\href{https://svs.gsfc.nasa.gov/14090}{NASA/DOE/Fermi LAT Collaboration}. \textit{Panel (b)} and \textit{(c)}: Significance map of the residual after modeling known \gray structures in a 50-month sky map. High 'sigma' in the residual map indicates un-modeled emission and shows the presence of new structures that are not correlated with the gaseous foreground map. The left and right panels show further breakdown into components that have soft (photon flux, $dN/dE \propto E^{-2.4}$) and hard (photon flux, $dN/dE \propto E^{-1.9}$) spectral indices, respectively. The image is reprinted from \cite{Ackerman2014}, with authors' permission (copyright by AAS). For a reconstructed \gray sky map, see \cite{Scheel-Platz2023}.}
    \label{chap1:fig:Fermi-skymap}
\end{figure*}

\subsection{The Fermi Bubbles}
\label{chap1:subsec:FBs}

\subsubsection{The data and template analysis}
\label{chap1:subsubsec:FB-instrument}
Observations of diffuse \gray emission improved significantly with the Large Area Telescope (LAT) detector on board the Fermi Gamma-Ray Space Telescope, together known as the \fermilat Telescope \citep[][and \fermilat-website \footnote{\url{https://fermi.gsfc.nasa.gov/science/overview.html}}]{Atwood2009, GLAST1999}. The detector is based on the physics of pair conversion in which \gray photons interact with a tungsten plate to produce e$^-$-e$^+$ pairs. These pairs are then spatially tracked using silicon-based strips attached to the tungsten plates and finally absorbed in a cesium iodide calorimeter for measuring the total energy. The telescope detects $\gamma$-rays in the $20$ MeV-$300$ GeV energy range and has a field of view of $\approx 2$ sr (almost one-fifth of the entire sky), a spatial resolution of about a degree, and a high peak effective area ($>8000$ cm$^2$). The detector also hosts an anti-coincident detector made of specially made plastic tiles that only interact with charged particles (but not \gray photons) and help the detector to reject cosmic ray events. 

Figure \ref{chap1:fig:Fermi-skymap} (panel \textit{a}) shows the \gray sky map based on the data collected by \fermilat at $E>1$ GeV energies over 144 months. The long and bright central patch represents the \gray emission from the Galactic disk. The point-like sources inside the disk are pulsars, whereas, the point-like sources outside the disk are distant AGNs. The image also shows faint but sharp arc-like structures in the northern as well as in the southern hemispheres. The northern structure coincides with a previously known large-scale radio structure, Loop-I (see section \ref{chap2:multiwavelength-view} for more details), and the southern structure is part of the recently discovered Fermi Bubbles. A corresponding northern part of the FBs is also noticeable in this map, although much less sharp due to the bright Loop-I emission surrounding it.  

The origin of the diffuse \gray emission in our Galaxy can be attributed to three main processes - \textit{Hadronic}, \textit{Bremsstrahlung}, and \textit{Inverse Compton} (IC) \citep{Fichtel1978, Su2010, Ackerman2014}.
The hadronic process is one of the main sources of \grays in the Galactic disk. In this process, CR protons (and heavier nuclei) interact with interstellar gas nuclei to produce mostly pion particles, \textit{viz} $\pi^0$, $\pi^+$, and $\pi^-$. The neutral pion decays into two \gray photons in a very short time scale ($\sim 10^{-16}$ s). Given that, $\pi^0$ mass is $\approx 135$ MeV, this decay channel of producing \grays is expected to be inefficient at energies below $\sim 70$ MeV.  
Bremsstrahlung emission from CR electrons and positrons interacting with charged gas particles become important at $E \lesssim 10$ GeV. These electrons and positrons also interact with the local radiation field, such as the interstellar radiation field (ISRF) and cosmic microwave background (CMB) to produce energetic \grays via the IC process. Note that the charged pions, in the hadronic process, decay into muons and $\mu$-neutrinos. Detection of these neutrinos is a tell-tale sign of a hadronic origin for the \grays. 

As can be seen in figure \ref{chap1:fig:Fermi-skymap} (panel a), the \gray emission from the Galactic disk is so bright that it is often hard to locate the presence of other fainter structures. It is, therefore, necessary to remove these `foreground' structures to obtain features that are fainter. There are several methods that are used to remove the effects of such foregrounds, each employing slightly different methods for removing the foreground \citep{Dobler2010, Su2010, Su2012, Ackerman2014} using templates or sharpening the map using Bayesian inference methods \citep{Selig2015a, Selig2015b} to detect the fainter structures.  
The most commonly used is the global template analysis method where spatial templates for different components are fit together to obtain the intensity of each of the components. The most obvious template is the \gray emission correlated with the disk gas since the \gray emissivity in both the Hadronic and the Bremsstrahlung process is directly proportional to the gas density and the CR density. For the spatial distribution of the gas, a sky map of the molecular, atomic, and ionized gas column densities in our Galaxy is created based on different tracers such as dust, CO (a tracer for H$_2$), HI, and H$\alpha$ emission maps, and pulsar dispersion measure. A CR transport code, \textsc{galprop} \citep{Strong1998}, is then run to create a template for the \gray emission in the Galaxy. \textsc{galprop} is a code that numerically solves the CR diffusion equation for a given distribution of gas and CR sources in the ISM, and produces \gray intensity following the Hadronic and Bremsstrahlung processes in addition to calculating the IC emission from the daughter (secondary) particles. Estimation of the free parameters such as the CR injection rate or the diffusivity requires that the produced \gray intensity map (the template) fit with the observed sky map. This template is called the gas-correlated template for its dependence on the gas distribution in the Galaxy.
An IC emission (due to primary CR electrons) template is also created from \textsc{galprop} by assuming a certain ISRF. Templates for other known features, such as the Loop-I are assumed based on prior radio emission maps \citep[for example,][]{Haslam1982}. Once these templates are fit at every sky pixel outside a central ellipse ($|l| \lesssim 20^\circ$ and $|b|\lesssim 50^\circ$, to avoid over-fitting any unknown sources in this region) and at every energy bin, the fit is extended to the central ellipse. The residual map shows an excess emission (at $E \gtrsim 5$ GeV) in the form of two bubbles. These bubbles are called the Fermi Bubbles (FBs) following the discovery paper by \cite{Su2010} \footnote{An excess emission within the mentioned central ellipse was noticed in an earlier paper by \cite{Dobler2010} and was named as \textit{Fermi Haze} since it was not clear if the `Haze' has a sharp edge or not.}.  A template for these bubbles can then be finally included in the fitting procedure to map out the intensity and spectra for each of these templates. 

In an alternative method, called the local template analysis \citep{Ackerman2014},
one assumes a linear combination of the gas column density and a two-dimensional spatial polynomial in smaller patches of the sky to fit the \gray emission in each energy bin. The residual map is then further fit using two bivariate Gaussian profiles (2D Gaussian profiles with two Gaussian components in each direction), representing the IC emission from the disk and from the halo. The halo component is further divided into two assumed spectral components, soft (photon flux, $dN/dE\propto E^{-2.4}$, corresponding to the Loop-I) and hard (photon flux, $dN/dE \propto E^{-1.9}$), within a `suspected' bubble region (obtained from residual maps at $E > 10$ GeV) to obtain spatial maps of the bubble and the Loop-I \citep[see][for more details]{Ackerman2014}. The sky maps for these two spectral components are shown in Figure \ref{chap1:fig:Fermi-skymap} (Panel b and c). The hard component shows the well-known Fermi Bubbles. 

\subsubsection{Morphology}
\label{chap1:subsubsec:FB-morphology}
As we see in Figure \ref{chap1:fig:Fermi-skymap}, the FBs consist of two elliptical \gray emitting regions with major axes perpendicular to the Galactic plane and connected at the GC. The orientation of the FBs and their superposition with the GC strongly suggest that the FBs originated from the GC, i.e. these bubbles are $\sim 8$ kpc away from the Sun. The FBs extend to $\approx \pm 20^\circ$ in Galactic longitude, $l$, and to $\approx \pm 50^\circ$ in Galactic latitude, $b$, and cover almost $5\%$ of the whole sky. This means that each of these bubbles is $\sim 7$ kpc wide and $\sim 10$ kpc long, among the largest structures in our Galaxy. 
Although they are quite symmetric about the Galactic Pole and the Galactic plane, slight asymmetry is noticed in both bubbles. For example, the northern bubble (positive latitude) extends only $\approx 18^\circ$ to the East (positive longitude) but $\approx 23^\circ$ to the West (negative longitude). This East-West asymmetry is larger in the southern bubble. The southern bubble is also $\approx 4^\circ$ bigger than the northern one \citep{Su2010, Sarkar2019}. I will discuss the possible origin of such asymmetries in a later section. 

The intensity of both bubbles is largely uniform over the entire bubble surface except in small patches that are consistent with statistical fluctuations. However, there is an indication of consistently higher intensity ($\gtrsim 6$-sigma significance) along a line $\approx 15^\circ$ away from the Galactic pole. It has been suggested that this excess emission can represent the cocoons of an old jet activity at the GC \citep{Dobler2010, Ackerman2014}. This elongated cocoon-like emission is highly dominated by the excess emission in the South-East edge of the FBs and may alternatively be part of an un-modeled `donut-like' foreground that extends well beyond the edge of the FBs \citep{Su2010}. More recently, it has been suggested that the cocoon-like feature inside the southern FB is due to the \gray emission from the Sagittarius dwarf spheroidal galaxy and its stream and is not related to the FBs\citep{Crocker2022}. 

Apart from the FBs, figure \ref{chap1:fig:Fermi-skymap} also reveals an even larger structure extending to $|l| \sim 50^\circ$ and $|b| \sim 80^\circ$ in the northern hemisphere. The structure has the form of a shell, the inner edge of which is roughly coincident with the outer edge of the FBs. We will see later that this structure is spatially coincident with the radio-emitting Haslam 408 MHz map \citep{Haslam1982} or the X-ray-emitting North Polar Spur \citep{Snowden1997, Bland-Hawthorn2003, Predehl2020}. Although a similar \gray shell is not visible in the southern hemisphere, a horn-shaped emission just outside the base of FBs is noticed in the south. The current data do not indicate if the `horns' extend to high latitudes like the northern counterpart. Therefore, it is still debated whether the \gray shell in the northern hemisphere is in fact related to the FBs or it is completely unrelated. A more detailed discussion of this topic is given in section \ref{chap2:multiwavelength-view}. The northern \gray shell also has an East-West asymmetry, very similar to the East-West of the FBs thus indicating a possible correlation between the two. 

\begin{figure}
    \centering
    \includegraphics[width=0.45\textwidth]{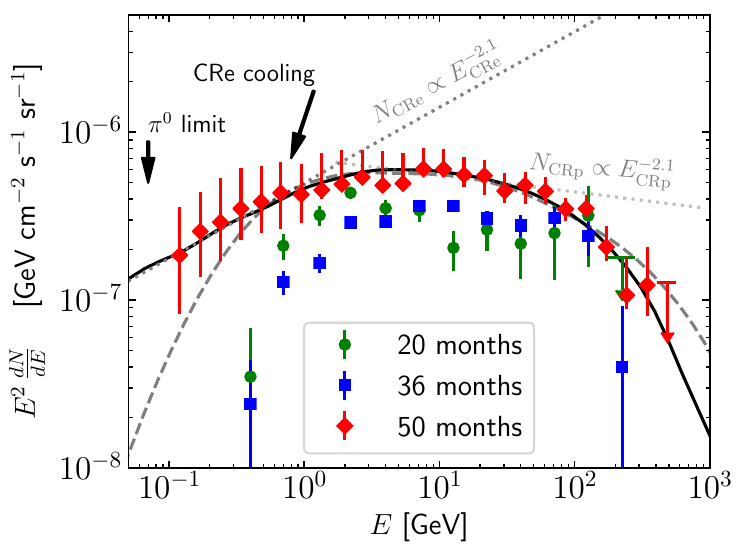}
    \caption{Spectrum of the Fermi Bubbles for different exposure times. The data from the discovery paper \citep{Su2010} is shown in green. The blue and red points show the spectrum after longer exposure times and are taken from \cite{Su2012} and \cite{Ackerman2014}, respectively. The apparent discrepancy between different papers is mostly due to uncertainties in modeling Galactic foregrounds and analysis strategy. The gray dotted lines represent the typical spectral slopes for IC (from CR electrons) and Hadronic (p-p) emission mechanisms for a given CR population. The black solid and dashed lines represent more accurate spectra for these processes (see section \ref{chap5:origin-FBs}).}
    \label{chap1:fig:FB-spectra}
\end{figure}

\subsubsection{Spectrum}
\label{chap1:subsubsec:FB-spectrum}
The spectra from both the global and local template analyses are largely consistent with each other and are shown in figure \ref{chap1:fig:FB-spectra}. The figure shows 3 data sets (points) corresponding to different data accumulation periods (exposure times) and analysis techniques \citep{Su2010, Su2012, Ackerman2014}. The error bars in the $50$-month data represent the systematic uncertainties, i.e. due to different analysis methods and foreground models. The statistical errors, due to limitations in the detector and the detected photon numbers, are much smaller than the shown systematic error bars. The overall increased flux in the $50$-month data is due to a different analysis strategy used in \cite{Ackerman2014} compared to the others. The figure shows that the bubbles have a typical \gray flux of $\sim 4\times 10^{-7}$ GeV \pcmsq \psec sr$^{-1}$ in the range of $1-100$ GeV. Complete integration in the $0.1-500$ GeV range produces a \gray luminosity of $\simeq 4 \times 10^{37}$ \ergps, for an assumed distance of $9.4$ kpc to the center of the FBs and a total size of $0.66$ sr \citep{Su2010, Ackerman2014}.

Before we go into further details of the spectrum and figure out what is the plausible emission mechanism that produces the observed \gray emission, we can study its main spectral features. As we noted earlier, the two relevant processes are IC emission and Hadronic emission (hereafter, the $p-p$ process).  For the IC channel, the required Lorentz Factor, $\Gamma$, to up-scatter a typical CMB photon to $1$ GeV is $\approx \sqrt{E_\gamma/E_{\rm cmb}} \approx 10^6$ (following $1:\Gamma:\Gamma^2$ rule of IC scattering; \citealt{Rybicki1986}). For a given population of CR electrons (CRe) with density distribution, $dN_{\rm CRe}/dE_{\rm CRe} \propto E_{\rm CRe}^{-p}$, the IC spectrum follows $E_\gamma dN_\gamma/dE_\gamma \propto E_\gamma^{-(p-1)/2}$, i.e. $E_\gamma^2 dN_\gamma/dE_\gamma \propto E_\gamma^{(3-p)/2}$. Such a spectrum is shown in Figure \ref{chap1:fig:FB-spectra} for a CRe population with the spectral index, $p=2.1$. It is evident that the \gray spectrum matches perfectly with IC emission at $E\lesssim 1$ GeV but deviates significantly above $1$ GeV. As we will see in section \ref{chap5:subsec:spectral-origin}, this deviation may represent IC  cooling losses and a high energy cut-off for the CR electrons. 

For the $p-p$ process, the required Lorentz Factor for the CR protons to produce $\sim$ GeV photon is $\sim 10$ GeV (energy of decayed $\pi^0$ is $E_{\pi^0}\sim (1/3) E_{\rm CRp}$ and then $\pi^0$ decays into two equal energy photons). Now, since the $p-p$ cooling time is $\sim 10$ Gyr at these energies, much longer than a typical injection and escape time of the CR protons, the resulting \gray spectrum simply follows the parent CRp spectrum, i.e. $E_p^{-p}$ \citep{Crocker&Aharonian2011}. An example of such a spectrum (for $p=2.1$) is shown by the horizontal dotted-gray line. Note that the downturn in the observed spectrum at $E_\gamma \lesssim 1$ GeV is due to the decayed $\pi^0$ mass approaching its rest mass ($m_{\pi^0} \approx 135$ MeV) which results in highly inefficient \gray production at $E_\gamma < m_{\pi^0}/2 \approx 70$ MeV (marked as the $\pi^0$ limit in figure \ref{chap1:fig:FB-spectra}) Note that the energy below the $\pi^0$ limit is produced mostly by bremsstrahlung and IC emission from secondary positrons and electrons generated from $\pi^+$ and $\pi^-$ decay, and is not included in the computation of the spectrum. Therefore, the $p-p$ spectrum looks consistent with the observed data once we account for a high energy cutoff of the CR protons at a few TeV. It is, therefore, unclear which is the main production channel for the observed \grays and we have to look into its signatures in other wavebands to fully understand the nature of the \grays as well as the source of the mechanical energy for these bubbles.

%% file: chap_2_multiwavelength_view.tex
\section{Multi-wavelength view toward the Galactic Center}
\label{chap2:multiwavelength-view}
Structures observed in different wavebands toward the GC can provide us with crucial information regarding the origin of the FBs. Over the last few decades, we have observed the sky through X-ray, UV, and radio telescopes. In this section, I provide a list of observations in the same part of the FB sky and provide a `first-look' conclusion about these observations and their possible relation to the FBs. 

\begin{figure*}
    \centering
    \includegraphics[width=0.9\textwidth, clip=True, trim={0cm 0cm 0cm 0cm}]{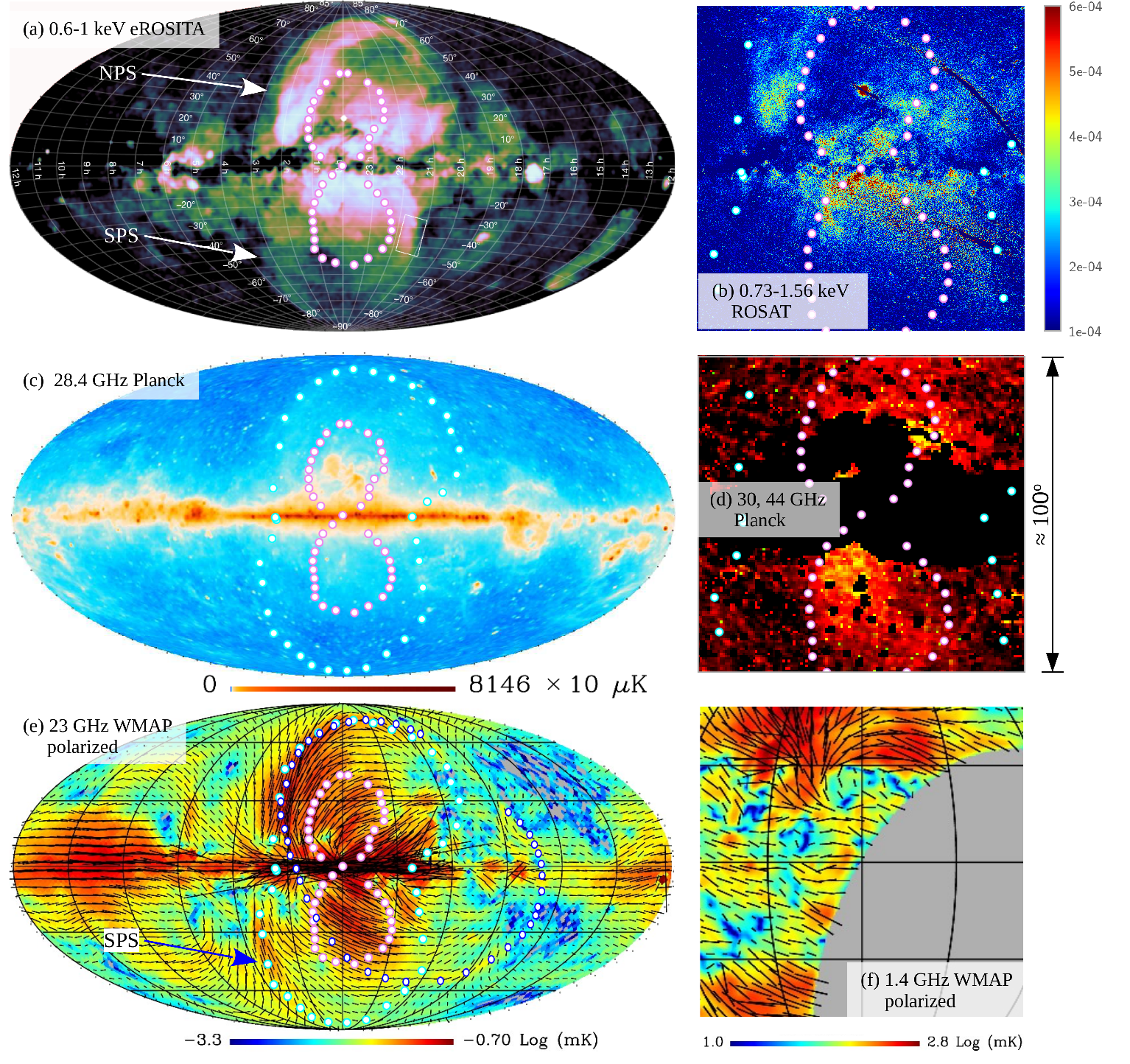}
    \caption{X-ray and radio sky maps. \textit{Panel (a)}: soft X-ray ($0.6-1$ keV) sky map from \erosita telescope highlighting the existence of two X-ray shells (\erosita bubbles) symmetric across the Galactic plane and the Galactic axis. The typical brightness of the shells is $\sim 3\times 10^{-3}$ cts \psec arcmin$^{-2}$. . Image reprinted with permission from \cite{Predehl2020} (copyright by the authors). \textit{Panel (b)}: \rosat sky map of the central $50^\circ$ of the Galaxy in the $0.73-1.56$ keV range \citep{Snowden1997} showing the central x-shaped X-ray emission coinciding with the outer edges of the FBs (magenta circles, \citealt{Su2010}). The color scale is in units of cts \psec arcmin$^{-2}$. The lower photon count in \rosat is due to the $\sim 10$ times smaller effective area compared to \erosita. \textit{Panel (c)}: Brightness temperature (in excess of the CMB) map at $28.4$ GHz. The edges of the \erosita bubbles and the FBs are shown using the magenta and cyan circles, respectively. The typical brightness inside the \erosita bubbles region is $\sim 50\: \mu$K. Image reprinted from \cite{PlanckCollaboration2016} (copyright by ESO). \textit{Panel (d)}: Central $50^\circ$ view of the sky at $30$ GHz (red) and $44$ GHz (yellow) after removing the Loop-I and the Galactic disk-related emission. Image reprinted from \citep{PlanckCollaboration2013}  with authors' permission (copyright by ESO). Typical brightness temperature is $\sim 20\:\mu$K. The dark patches represent the regions excluded from the template analysis due to high contamination by the disk emission. \textit{Panel (e)}: Polarization intensity map of \textit{WMAP} at $23$ GHz \citep{Vidal2015}. The small black lines represent the magnetic field orientation (perpendicular to the direction of the polarization vector). Overall alignment of the magnetic field can be seen along both the North Polar Spur (NPS) and South Polar Spur (SPS). Blue circles show the previously proposed full shape of Loop-I. \textit{Panel (f)}: \textit{WMAP} 1.4 GHz polarization map for the central $50^\circ$ showing strong depolarization near the Galactic disk \citep{Vidal2015}. The gray region was not observed in the survey. Images for the panels (e) and (f) are reprinted with permission from \cite{Vidal2015} (copyright by the authors).}
    \label{chap2:fig:xradio-map}
\end{figure*}

\subsection{North Polar Spur \textit{aka.} Loop-I \textit{aka.} \erosita bubbles}
\label{chap2:subsec:NPS}
Sky maps in the soft X-ray ($\sim 0.3-1.5$ keV) emission have been obtained by \rosat \citep{Snowden1997} and \erosita \citep{Predehl2020, Predehl2021}. One of the striking features in the \textit{ROSAT} all-sky survey (RASS) is a giant X-ray shell in the north-eastern sky extending from $b \approx 10^\circ$ to $\approx 75^\circ$ with a width of $\approx 20^\circ$, known as the Loop-I/North Polar Spur \citep[NPS;][]{Berkhuijsen1971, Haslam1974, Sofue1979}. The spur curves back toward the Galactic plane in the North-West quadrant, with a total East-West extent of $\Delta l \sim 90^\circ$, completely encompassing the FBs (Figure \ref{chap2:fig:xradio-map}). The outer edge of the NPS also falls precisely on the outer edge of the Haslam 408 MHz radio map \citep{Haslam1982}, $28.4$ GHz map \citep{PlanckCollaboration2016}, and the \gray residual map of the soft component (figure \ref{chap1:fig:Fermi-skymap}; \citealt{Ackerman2014}) indicating a common origin.

The question of whether the NPS/Loop-I is related to the FBs is connected to the question of whether it originates from the Galactic Center or somewhere in the Solar neighborhood (local). In the Galactic center origin scenario, the NPS is a $\sim 10$ kpc large bubble with a shell width of $\sim 2-4$ kpc and arises from an energetic event from the Galactic center \citep{Sofue1977, Sofue1994, Sofue2000, Bland-Hawthorn2003, Sofue2003, Kataoka2013}, whereas, the local origin scenario claims that the NPS is at a distance of $\sim 200$ pc from the Sun and is generated due to SNe explosions in a nearby OB association, Scorpio-Centaurus \citep{Berkhuijsen1971, Egger1995, Dickinson2018} or due to magnetic filaments lying along the sight-line between the Sun and the GC \citep{West2021}. 
One of the main arguments against the GC origin is that any GC event is expected to create symmetric signatures across the Galactic disk, i.e. we should expect a `South Polar Spur' in the southern hemisphere. The absence of such a counterpart of the NPS fueled further speculations about the local origin of this feature despite claimed signatures (although weak) of such counterparts in the \rosat and \textit{MAXI} data \citep{Snowden1997, Sofue2000} and inadequacy of SNe remnants from Sco-Cen OB association to produce such shells \citep{Shchekinov2018}. 
It was only very recently that the high-sensitivity X-ray telescopes, such as \erosita \citep{Predehl2021} and \textit{MAXI}, have been able to detect the southern counterpart of the NPS, known as the \erosita bubbles \citep{Predehl2020, Nakahira2020}. Although the southern counterpart is rather weak compared to the NPS, the size, shape, and symmetry across the Galactic disk of this newly discovered structure are in excellent agreement with the northern counterpart. The new X-ray observations suggest that the NPS/Loop-I is indeed produced by an energetic event at the GC and that each of the spurs (or \erosita bubbles) rises to $\sim 12-14$ kpc (for the assumed distance to the GC) from the Galactic disk which makes them one of the biggest structures in the Galaxy. 
Note that no such southern counterpart has been so far observed in the \gray maps despite the presence of two horn-shaped structures emanating from the Galactic disk resembling a mirror shape of the NPS in \gray (see figure \ref{chap1:fig:Fermi-skymap}). Whether these two \gray horns have a closed form in the southern hemisphere is for future analysis to determine. 

\subsubsection{Energetics}
Estimation of the energetics of the \erosita bubbles (post \erosita name for the North/South Polar Spurs) depends on their thermal and kinetic properties. \cite{Kataoka2013, Kataoka2015, LaRocca2020} analyzed spectra at different parts of the \erosita bubbles and inferred a plasma temperature, $T_s \simeq 3\times 10^6$ K, which is about $1.5$ times higher than the observed CGM temperature of $\approx 2\times 10^6$ K \citep[][see section \ref{chap1:subsec:cgm}]{Henley2010, Henley2013, Miller2015}. The observed temperature in the hot gas may be dominated by the inner CGM ($r\lesssim 20$ kpc) and may not represent the temperature of the entire CGM where the average temperature could be much lower \citep{Faerman2020}. However, for the purpose of this review, we are only worried about the conditions at the inner CGM. Now, assuming that the \erosita bubbles are created due to shock compression, the temperature suggests a Mach number, $\mathcal{M} \simeq 1.4$ which further implies a shock velocity of $\simeq 320$ \kmps (correcting for a weak shock in a $c_s (2\times10^6) \approx 210$ \kmps medium). The X-ray spectra also yield the density of the bubbles, $n_s = \sqrt{EM/d} \sim 0.004$ \pcc, where EM $ \simeq 0.05$ pc cm$^{-6}$ is the emission measure and  $d = 4$ kpc is the assumed line-of-sight width of the emitting region. Therefore, the energy density and the total energy in the \erosita bubbles are, $\epsilon = n_s k_B T_s/(\gamma -1) \simeq 2\times 10^{-12}$ erg \pcc and $E_{\rm th} \simeq 10^{56}$ erg, respectively \citep{Kataoka2013, Kataoka2015, Predehl2020}. Noting that the size of each of the bubbles is $\simeq 12$ kpc, the propagation time is estimated to be $\tau_{\rm dyn} \sim 20$ Myr (see eq \ref{chap1:eq:rs-agn}). Therefore, an energy release rate is $P_{\rm dyn} = E_{\rm th}/\tau_{\rm dyn} \sim 10^{41}$ \ergps. The exact propagation time scale, $\tau_{\rm dyn}$, depends on the assumed density profile and energy injection model (see section \ref{chap5:subsec:wind-model}), but the above rough estimation is expected to be close to a factor of unity.  

More recently, the \erosita bubbles temperature estimate has been challenged by \cite{AGupta2023} who analyzed \textit{Suzaku} spectrum based on a multi-temperature fit and found that the bubble spectra can be explained by two temperatures, one, $\simeq 0.2$ keV, consistent with the CGM temperature, and the second, $\sim 0.7$ keV. The authors note that the hotter component may not be particularly related to the bubbles, rather it may be an `extra-virial' component of the MW CGM itself since the $0.7$ keV component is also often found well outside the bubbles \citep{SDas2019a, SDas2019b}. \cite{AGupta2023} also find a super solar abundance of Ne/O and Mg/O in a few sight lines inside the bubbles (but not outside) indicating a possible presence of SNe ejected material in the bubble which further implicates a wind driven by SNe. If the estimated lower temperature for the bubbles is indeed true, it would mean that the shocks are undergoing radiative cooling, an interesting scenario for future exploration, especially since the radiative cooling time scale for the shock is $\tau_{\rm cool} \sim \epsilon/(n_s^2 \Lambda_{0.3 \mbox{keV}}) \sim 300$ Myr (assuming $\Lambda_{0.3 \mbox{keV}} \simeq 3\times 10^{-23}$ \ergps cm$^3$; \citealt{Gnat2007, Sarkar2021a}), much longer than the propagation time, $\tau_{\rm dyn}$ \footnote{Note that the cooling time-scale for the gas to cool from $0.3$ keV to $0.2$ keV (and not to zero) would only be $1/3$ of the $\tau_{\rm cool}$. The new cooling time ($\sim 100$ Myr) is still a factor of a few longer than the propagation time, $\sim 20$ Myr.}. The temperature estimate, however, has to be verified against different instruments and plasma models such as including the effects of non-equilibrium ionization in the shock \citep{Kafatos1973, Hamilton1983, Gnat2007, Sarkar2021b} and log-normal temperature distribution of the CGM \citep{Faerman2017, Vijayan2022}.
Interestingly, \cite{Yamamoto2022} find that the \textit{Suzaku} spectra of the NPS are consistent with a temperature of $0.5-0.6$ keV and that the spectra contain signatures of non-equilibrium ionization. Although a collisional equilibrium ionization interpretation of $T_s \approx 0.3$ keV plasma \citep[as was done in][]{Kataoka2013} cannot be ruled out, the non-equilibrium ionization spectrum is found to pass the statistical \textit{F-test}. It is, therefore, unclear how much (and which) of the non-standard plasma models (i.e. non-solar abundance or non-equilibrium ionization) is necessary and what is the exact temperature of the plasma. 
For this review, I will consider the NPS to be in collisional equilibrium and the temperature to be $0.3$ keV following \cite{Kataoka2013}.

\begin{figure*}
    \centering
    \includegraphics[width=0.9\textwidth, clip=True, trim={0cm 0cm 0cm 0cm}]{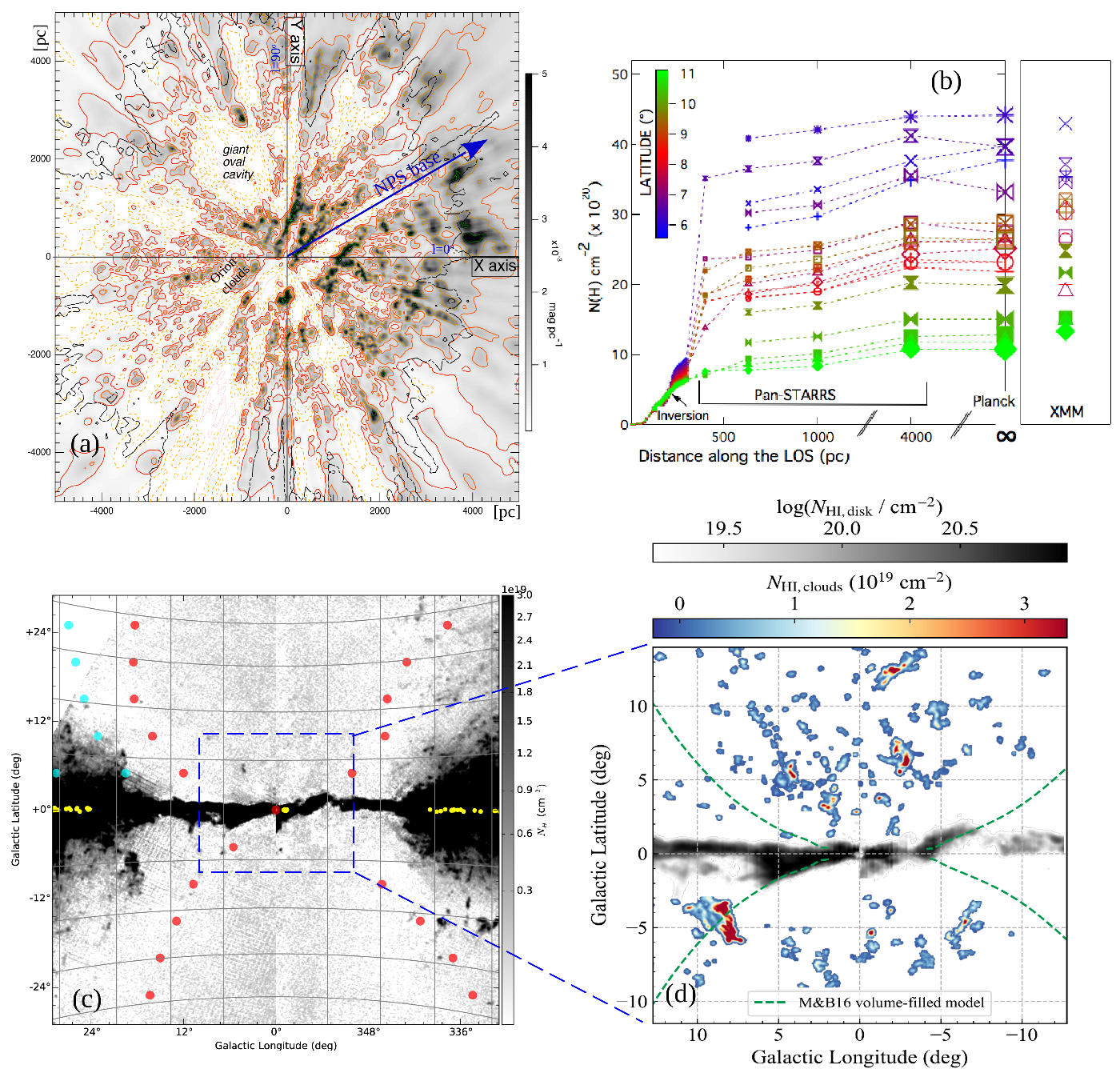}
    \caption{Panel \textit{a}: Spatial extinction map around the Sun obtained from a catalog of 
    $\approx 35$ Million stars. The direction to the NPS is shown using the blue arrow. Contour lines, starting from red to blue, represent iso-extinction levels of $10^{-5}$, $5\times 10^{-5}$, $10^{-4}$, $2\times 10^{-4}$, $5\times10^{-4}$, $10^{-3}$, $2\times10^{-3}$, $5\times10^{-3}$, $10^{-2}$, and $2\times 10^{-2}$ mag pc$^{-1}$. The image is reprinted with permission from \cite{Vergely2022} (copyright by the authors). Panel \textit{(b)}: Hydrogen column density (obtained from extinction) at different latitudes toward the North-East base of the NPS \citep{Green2015, Lallement2016}. The right part of the figure shows the column density required to explain the absorption features of XMM-Newton spectra along those directions. Image reprinted from \cite{Lallement2016} with authors' permission (copyright by ESO). Panel \textit{(c)}: \ion{H}{i} emission map at the GC, showing the presence of a \ion{H}{i} hole within the central $2.4$ kpc. The yellow circles show the H$\alpha$ regions (i.e. star-forming), the red circles show the edge of the FBs, and the cyan circles show the outline of an X-ray arc, known as the northern arc \citep{Su2010}. Image reprinted from \cite{Lockman2016} with the authors' permission (copyright by AAS). Panel \textit{(d)}: Outflowing \ion{H}{i} clouds (colored contours) in the GC. The Galactic disk contribution is shown by the gray scale. The dashed green line shows the approximate edge of the Fermi Bubbles \citep{Miller&Bregman2016}. Image credit - Enrico Di Teodoro. 
    }
    \label{chap2:fig:NHabs}
\end{figure*}

\subsubsection{Distance to the NPS}
From the morphological structure of the \erosita bubbles, it is almost certain that they originate from the GC. However, independent distance estimations of these bubbles have been confusing since they indicate a range of distances supporting both the local and the GC origin scenario. The most convincing evidence for the local origin of the NPS comes from the polarization of stellar light. Scattering across sharp edges, such as dust grains aligned with the magnetic field, can introduce polarization in the starlight. The polarization vector in such cases aligns with the sharp edge of the dust and thus indicates the direction of the magnetic field. On the other hand, the polarization vector in a pure synchrotron emission is perpendicular to the magnetic field. Therefore, in an ideal situation, the polarization vectors of optical light from stars behind a synchrotron emitting region are always perpendicular to the polarization vectors of the synchrotron radiation. Several authors found that this is indeed true for stars toward the NPS that are within a distance of $100-200$ pc \citep{Spoelstra1971, Spoelstra1972, Axon1976, Salter1983, Vidal2015}. While the orthogonality of the optical and radio polarization holds true for stars at $b\gtrsim 40^\circ$, the polarization vectors often swing around below this latitude to become parallel to the radio vectors. Additionally, the increase in polarization as a function of distance is not very different inside the NPS compared to the outside. It was, therefore, noted that the alignment of the optical and radio polarization could be due to a pure coincidence and this may arise from different volumes along the line of sight \citep{Spoelstra1972, Salter1983}. In fact, such a scenario is consistent with a poloidal magnetic field at the GC and a toroidal magnetic field in the Galactic disc.
\cite{Vidal2015} find that the overall magnetic field orientation toward the sky is consistent with a magnetic shell of radius $120$ pc and at a distance of $120$ pc toward ($320\deg,5\deg$) and probably represents a hot SNe bubble related to the Scorpio-Centaurus OB association \citep{Berkhuijsen1971, Wolleben2007, Dickinson2018}. It is, however, to be noted that such a local bubble would be indistinguishable, in terms of geometry on the sky, from a superbubble at the GC.

The GC origin scenario is mostly supported by the depolarization of the radio emission at $2.3$ GHz and X-ray emission toward the NPS. Low-frequency radio waves ($\sim$ GHz for the Galactic conditions) are quickly depolarized in the presence of random magnetic fields compared to the high-frequency waves (since the Faraday rotation is $\propto \lambda^2$). Such depolarization is clearly visible in the $2.3$ GHz \textit{WMAP} and \planck map at $b \lesssim 30\deg$ (see Panel \textsc{f} in figure \ref{chap2:fig:xradio-map}, and also \citealt{Carretti2013, Vidal2015, PlanckCollaboration2016}). Based on the polarization data, \citep{Sun2014, Sun2015} concluded that the depolarization implies that the radio emission is coming from a distance $\gtrsim 2$ kpc, at least toward $b\lesssim 40\deg$. At higher latitudes, the authors estimate smaller distances ($\sim 60-700$ pc) for the NPS based on the alignment of the radio and optical polarization data. However, they do not rule out the possibility that the NPS could be further away and that only its magnetic field is aligned with the local magnetic field of the Galaxy. It is also interesting to note in figure \ref{chap2:fig:xradio-map} (panel \textit{e}) that the South-East edge of the X-ray \erosita bubble perfectly matches with the SPS detected at $23$ GHz. The derived magnetic field direction in the SPS also aligns with the overall curvature of the southern \erosita bubble further enhancing the possibility that the radio arc and its polarization are related to the same shock wave originating from the GC.

Another distance constraint comes from the X-ray data. X-ray observations toward the NPS find that the spectra imply an absorbing Hydrogen column density of $\sim 3\times 10^{20}$ \pcmsq \citep{Willingale2003, Kataoka2013}, close to the Galactic values in those directions. The required absorbing column density for explaining the strong suppression of X-ray emission at $b\lesssim 10\deg$ is $\sim 5 \times 10^{21}$ \pcmsq is much higher than the total neutral gas content within $100$ pc of the Sun (assuming average ISM density of $1$ \pcc) unless there is a dense shell of gas in between \citep{Kataoka2013}. Although the 3D dust map of the local ISM (Panel \textit{a} of figure \ref{chap2:fig:NHabs}) shows the presence of such dense clumps towards the base of the NPS \citep{Lallement2014, Pushpitarini2014, Vergely2022}, the total gas column density, integrated to $4$ kpc, falls short of explaining the required column density (Panel \textit{b} of figure \ref{chap2:fig:NHabs}). The panel shows the Hydrogen column density (inverted from extinction estimates) toward the base of the NPS ($l\simeq 30\deg$, $b\simeq 5\deg-11\deg$) obtained from known stellar distances \citep{Lallement2016}. The right part of the panel shows the required column density to fit the X-ray absorption at those latitudes. It is apparent that the required Hydrogen column density falls short at every latitude, thus indicating a distance to the X-ray emitting NPS to be $\gtrsim 4$ kpc. Note that in a similar study, \citep{KDas2020} find that the extinction curve flattens out at $\sim 300$ pc and that the extinction required to produce the X-ray absorption is consistent with the flat value. This led them to conclude that NPS is $\lesssim 300$ pc away. However, I note that such a result only produces a lower limit (and not an upper limit as claimed in the paper) on the NPS distance. 

Further careful analysis of the \textit{Suzaku} and \textit{XMM-Newton} spectra towards the NPS ($l\simeq 20\deg-26\deg$, $b \simeq 20\deg-40\deg$) indicates absorption of \oviii Lyman series lines. The absorbing medium is characterized to have a temperature of $0.17-0.2$ keV (similar to the Galactic CGM; \citealt{Henley2010, Miller2015}) and a column density of $3-5 \times 10^{19}$ cm$^2$. The column density implies a distance of about $\sim 6-7$ kpc through the CGM \citep{Gu2016}. Additionally, the X-ray spectra indicate a metallicity of $0.2-0.7$ Z$_\odot$ \citep{Kataoka2013, Gu2016, Lallement2016}, consistent with the metallicity of the CGM \citep[$\simeq 0.5$ Z$_\odot$;][]{Miller2015, Faerman2017, Faerman2020} than the ISM \citep[$\simeq$ Z$_\odot$;][]{Maciel2010, DeCia2021}. See \cite{Lallement2022} for a more extended review of the NPS. 

While the GC origin of the NPS is promising, the absence of such a bright structure in the southern sky has been a point of concern even after the discovery of a fainter southern loop by \erosita. \cite{Sarkar2019} proposed that the brightness asymmetry could be due to a large scale ($\sim 10$ kpc) density inhomogeneity ($\approx 20\%$ less in the southern hemisphere) in the CGM which causes the southern loop to be much fainter than the northern one \footnote {Although the paper was originally intended to explain the absence of an SPS in the \rosat data, it nonetheless predicted a fainter southern structure.}. There is also an East-West asymmetry since the extent of the NPS/Loop-I is further out in the North-West quadrant than the North-East quadrant. The X-ray asymmetry implies an East-West density asymmetry, perhaps caused by a headwind in the CGM due to the motion of the Galaxy toward the Andromeda galaxy \citep{Sofue2019, Mou2023}. Nonetheless, the asymmetric features of the NPS/Loop-I/\erosita bubbles can be explained based on the asymmetries in the CGM properties and do not necessarily mean that the NPS is of a local origin. We, therefore, are in a stalemate where the optical polarization and X-ray emission observations point to different distances to the NPS. While the discovery of the \erosita bubbles makes it certain that there is at least one component of the NPS/SPS that is of GC origin, it is uncertain if there is another nearby component that is superimposed on the NPS. A more careful investigation of optical polarization and its connection to radio polarization must be done to resolve this issue.

\subsection{Other X-ray features}
\label{chap2:subsec:other-xray-features}
In addition to the NPS/SPS/\erosita bubbles, other X-ray features are also observed around the FBs. The most relevant feature is the X-shaped emission around the base of the FBs (at $|\Delta l| \lesssim 15\deg$, $|\Delta b| \lesssim 20\deg$) as noted by \cite{Bland-Hawthorn2003} in the ROSAT data (see panel b of figure \ref{chap2:fig:xradio-map}). The authors proposed that these X-shaped features are part of an open hypershell created by a central outflow. Interestingly, it appears that the FB edge and the X-ray edge of the open hypershell model are almost coincident with each other, suggesting that the FBs could represent a bi-conical shock produced from a GC event. However, the base for the X-shaped emission is centered around $l\approx +5\deg$ and has a half width of $\sim 5\deg$ ($\sim 0.5$ kpc), thus making it hard to pinpoint the origin of the emission, especially since the GC (\sgra) lies roughly at the Western edge of this X-shaped region. The energy requirement for the X-shaped region is $\sim 10^{55}$ erg \citep{Bland-Hawthorn2003}, about an order of magnitude smaller than the required energy of $\sim 10^{56}$ erg for the \erosita bubbles \citep{Sofue2000, Kataoka2013, Predehl2021}. 
\cite{Keshet2018} extended the analysis to study the presence of such a `shock edge' to higher latitudes and detected a brightness drop when going out from the FBs in the North-West and the South-East sections. The trend either reverses or vanishes in other sections of the FBs. The authors estimated that the brightness drop in different \rosat bands indicates a shock temperature of $\simeq 0.4$ keV and corresponding Mach number of $\simeq 4$ (considering projection effects along the line of sight), relatively higher than the values estimated for the NPS. The dynamical timescale and the average power, in such a case, would be $t_{\rm dyn} \sim 5$ Myr and $6\times 10^{41}$ \ergps. However, the power could be much larger if the energetic event was shorter in duration. As we will see in section \ref{chap5:origin-FBs}, the required power for the FBs varies from $\sim 10^{40}$\ergps to $\sim 10^{45}$ \ergps. 
The biggest support for a shock at the edge of the FBs comes from the central X-shaped feature. However, it is unclear if the X-shaped feature indeed represents hot shocked gas or represents a boundary layer between low-density hot outflowing gas and cold gas at the edge of the conical outflow. The interaction of hot and cold gas is known to produce X-ray emissions via charge exchange processes, as observed in the Solar wind, star-forming regions, and galaxy clusters \citep{Cravens1997, Snowden2009, Smith2012, Liu2012, Gu2023}. Unveiling the nature of these biconical X-ray features will require high-resolution spectrometric observations and careful analysis of the spectra. A proper understanding of this X-ray feature can provide crucial information on the origin of the FBs. 

Additional localized diffuse emission of size $\sim 30\deg$ is also observed using MAXI-SSC toward the North tip of the FBs. The emission is observed only in the $1.7-4.0$ keV band with no apparent counterpart in softer bands \citep{Tahara2015}. Further analysis of \textit{Suzaku} data in this region indicates the presence of a hot ($\simeq 0.7$ keV) component, similar to the findings by \cite{Yamamoto2022} in that region. So far, it is not yet clear if this feature is, in fact, related to the FBs. It would be an important piece of the puzzle if it is indeed related to the FBs.

\subsection{The \oviii/\ovii line ratio}
\label{chap2:subsec:oviii-ovii}
Absorption and emission line intensities of highly ionized Oxygen, namely \oviii and \ovii provide us with valuable information regarding the low-density gas inside and surrounding the FBs. Analysis of X-ray \ovii-K$\alpha$ absorption lines ($\approx 22$\AA) along ten's of line-of-sights indicates the presence of a $\sim 6$ kpc `hole' of hot gas at the GC \citep{Nicastro2016}, consistent with the size of the FBs. Emission line observations of \oviii ($18.97$ \AA) and \ovii ($21.62+21.8+22.12$\AA ~triplet) revealed that the region of the FBs has a relatively higher \oviii/\ovii line ratio compared to the ambient CGM \citep{Miller&Bregman2016}. The \oviii/\ovii line ratio is an excellent tracer of temperature for hot gas since the ratio is a steep function of the temperature. The observed line ratio implies that the temperature of the plasma is $\sim 5\times 10^6$, corresponding to a Mach number of $\simeq 2.3$, and expansion velocity $v_{\rm exp} \sim 500$ \kmps. This expansion velocity suggests that the age of the FBs is $\simeq 12$ Myr and that one would need $\sim 10^{42}$ \ergps to inflate the FBs \citep{Miller&Bregman2016}. However, the synthetic emission line ratios obtained from hydrodynamical simulations (including the effects of an extended CGM) suggest that the line ratios are consistent with a lower shock Mach number ($\sim 1.5$), therefore, require a power of $\sim 10^{40.5-41}$ \ergps \citep{Sarkar2017}. It is also to be noted that \cite{Miller&Bregman2016} did not consider the full set of data and excluded the NPS region since it was not clear at the time whether the NPS is related to the FBs or not. Probably an analysis of the full set of data, especially after the discovery of the \erosita bubbles, would bring the results from the toy models and the simulations much closer. Given the sensitive nature of the \oviii/\ovii line ratio to the temperature, it is an important check for any theoretical model aiming to explain the FBs or \erosita bubbles.

\subsection{Diffuse radio emission}
\label{chap2:subsec:Micro-Haze}
Large-scale radio emission in the sky is dominated by the CMB, free-free emission from thermal electrons, synchrotron emission from relativistic electrons in the Galactic magnetic field, and anomalous microwave emission due to spinning dust in the Galaxy. Out of these components, the properties of the CMB are known with excellent confidence and hence can be simply subtracted to obtain a clearer picture of the local sky. For the remaining components, one can consider emission templates and fit them together to obtain emissions from individual components, in a very similar manner to the \gray template analysis method discussed in section \ref{chap1:subsubsec:FB-instrument}. For example, the free-free emission template can be obtained from a combination of the H$\alpha$ map and radio recombination line maps, and the anomalous microwave emission template can be obtained from the thermal dust maps in the Galaxy \citep{PlanckCollaboration2016}.

\subsubsection{Loop I}
Figure \ref{chap2:fig:xradio-map} (panel c) shows the $28.4$ GHz sky map obtained from \textit{Planck} after the removal of the CMB emission \citep{PlanckCollaboration2016}. The most interesting feature, other than the emission from the disk, is the Loop-I structure extending from $l \simeq 45\deg$ to $-60\deg$ and $b \simeq 0$ to $80\deg$ in the Northern sky. This feature is much more prominent at lower frequencies as noticed in the Haslam $408$ MHz map \citep{Haslam1982} and \ion{H}{i} sky map \citep{Kalberla2005}. Interestingly, as shown in the figure, the outer edge of the radio Loop-I exactly follows the X-ray \erosita bubbles almost for the entire northern sky. This brings up a question - if the microwave Loop-I originated from the same event that also gave rise to the \erosita bubbles. The absence of a corresponding southern feature, however, poses a serious threat against this notion, repeating a very similar dilemma as was the case for the NPS in the \rosat map. The polarization map at $23$ GHz from WMAP shows the presence of further structures that originate from the disk at different longitudes and run almost parallel to the Loop-I edge in the northern sky \citep[][slso see panel e of figure \ref{chap2:fig:xradio-map}]{Vidal2015, Dickinson2018}. These features have arc-like shapes, bending towards the northwest side, and have a typical size of $\sim 20\deg-30\deg$. Similar structures are also noted in the southern sky. It has been argued that one of these smaller structures represents the southern part of the Loop-I (blue circles in fig \ref{chap2:fig:xradio-map}, panel e) and that the Loop-I is not a symmetric structure across the Galactic disk, and therefore, the Loop-I is a local structure \citep{Vidal2015, PlanckCollaboration2016, Dickinson2018}. However, the excellent morphological match of the southern \erosita bubble to another polarized structure, known as the South Polar Spur (SPS) suggests otherwise. Both the radio arc and the magnetic field directions have a surprising match with the \erosita bubble edge, as was also speculated in \cite{Sarkar2019}. It is, therefore, quite possible that the known SPS feature could be the southern counterpart of the Loop-I and that the previously thought southern boundary of the Loop-I belongs to another structure. 

Spectral properties of the Loop-I have been studied by using \textit{WMAP} data \citep{Bennett2003}. It is found that the spectral index, $\beta$,  (brightness temperature, $T_b \propto \nu^{-\beta}$) for the Loop-I is $\simeq 3$ \citep{Finkbeiner2004, Davies2006, Dobler2008}, indicating a CR electron index of $p \simeq 3$ at the high latitudes (following $p = 2\beta-3$ for synchrotron emission). The spectral index at lower latitudes is confusing due to other emissions from the ISM. The required Mach number for a shock to produce such electron spectral index can be found from classic Diffusive Shock Acceleration calculations, i.e. $p \simeq (\chi+2)/(\chi-1)$, where $\chi$ is the shock compression ratio \citep{Bell1978, Drury1983, Blandford1987}. Simple calculations of the Rankine-Hugnoit conditions across a shock suggest a Mach number $\mathcal{M} \simeq 2$ for producing the required spectral index \citep{Hoeft2007, Vink2015}. 
The required Mach number implies a shock velocity of either $\simeq 400$ \kmps in the inner CGM ($T_{\rm CGM} \simeq 2\times 10^6$ K) which is consistent with the X-ray spectra of the NPS or a shock velocity of $\simeq 30$ \kmps in the ISM ($T_{\rm ISM} \simeq 10^4$ K). If the Loop-I is indeed a local structure and is produced by the Sco-Cen OB association, then the required power to drive it can be estimated based on the wind-driven bubble calculation \citep{Weaver1977}. Considering that the bubble has to reach $\simeq 150$ pc and should have a Mach number of $\simeq 2$ (in a medium with a sound speed of $15$ \kmps), the required power and the age are $\simeq 2.5 \times 10^{38}$ \ergps and $\simeq 3$ Myr, respectively \citep[see eq 52 \& 53 of ][]{Dekel2019}. Assuming that the OB association has a typical IMF, the power can be produced by the stellar wind and SNe from a young cluster of mass $\sim 10^4$ \msun \citep{Leitherer1999}. The required mass is, in fact, quite consistent with the mass estimated for the Sco-Cen association \citep{Damiani2019, Zerjal2023}. Therefore, it seems that both the GC origin scenario and the local origin scenario for the NPS/Loop-I can produce the observed microwave emission and its spectral properties. 

\subsubsection{Microwave Haze}
\label{chap2:subsubsec:haze}
Removal of large-scale diffuse emission reveals further substructures that are even more interesting, at least in the context of the current review. \cite{Finkbeiner2004} subtracted the free-free emission (correlated with H$\alpha$ emission), thermal dust emission, and soft synchrotron emission (correlated with the Haslam $408$ MHz feature) from the \textit{WMAP} data and discovered another diffuse structure at $|b| \lesssim 30\deg$ toward the GC, naming it `Haze'. It was initially thought that it was another un-subtracted component of the free-free emission. The haze was later found to be of synchrotron origin, independent of the free-free emission template, with a spectral index ($T_b \propto \nu^{-\beta}$) of $\beta = 2.4-2.7$ \citep{Dobler2008, Su2010}. The presence of the haze was confirmed by \cite{PlanckCollaboration2013} who estimated the spectral index to be $\beta \simeq 2.56\pm0.05$. The required spectral index for the cosmic ray electrons is $p \approx 2.1$ (following $p=2\beta-3$), suggesting that the CR electrons are produced by very strong shocks with Mach, $\mathcal{M} \gtrsim 5$, following the diffusive shock acceleration theory (see previous discussion). 

Figure \ref{chap2:fig:xradio-map} (panel d) shows the close-up view of the microwave haze within the central $50\deg$. The panel also shows the outer edge of the FBs in the magenta circles. Although the haze emission spills slightly out of FB edge, the emission is almost contained within. This brings up an interesting question - whether the haze and the FBs are connected in some way. Given that both the microwave haze and the FBs require the same electron spectral index (see section \ref{chap1:subsubsec:FB-spectrum}) and occupy the same spatial volume, it is highly likely that both the radio and \gray structures are generated from the same population of electrons \citep{Su2010}.  Further analysis of the microwave spectrum showed that the typical magnetic field in the haze is, $B \sim 8\: \mu$G \citep{Su2010, Dobler2012, Ackerman2014, Narayanan2017}. As will be discussed later, the haze spectrum is of supreme importance in determining the spectral origin of the FBs. 

Interestingly, the haze does not show any polarization signatures (at least in $\gtrsim 23$ GHz maps), indicating that the magnetic field inside the haze is probably randomly oriented. Observations of the haze at $2.3$ GHz using the Parkes Radio Telescope, however, revealed that the emission is significantly polarized ($\sim 20-30\%$) and has a steeper spectral index, $\beta \simeq 3-3.2$ in the $2.3-23$ GHz range, i.e. electron spectral index of $p \simeq 3-3.4$ \citep{Carretti2013}. The spatial extent of the $2.3$ GHz polarized emission also extends $\sim 10\deg$ beyond the boundary of the FBs toward the North-West and South-West sides and almost touches the \erosita bubble edges. The magnetic field is estimated to be $B \sim 6\: \mu$G  (for a volume-filling emission) or $12\: \mu$G (for a shell-like emission). The presence of such a steep spectral index indicates the cooling of the electron population and possibly represents an older population that was advected out of the GC. Now for a relativistic electron to produce synchrotron emission at a frequency of $\nu$, the required Lorentz factor is \citep{Rybicki1986}
\begin{equation}
    \Gamma \sim 9\times 10^3 \, \nu_{2.3GHz}^{1/2}\: B_{10 \mu G}^{-1/2}
    \label{chap2:eq:Gamma_sync}
\end{equation}
and the synchrotron cooling time is 
\begin{eqnarray}
    t_{\rm syn, cool} \sim \frac{\Gamma m_e c^2}{P_{\rm sync}} &\approx& 27 \mbox {  Myr  } \nu_{2.3 GHz}^{-1/2}\: B_{10\mu G}^{-3/2} \nonumber \\
    &\approx& 24 \mbox{ Myr  } \Gamma_4^{-1}\: B_{10\mu G}^{-2}
    \label{chap2:eq:t_sync_cool}
\end{eqnarray}
where, $P_{\rm syn} = (4/3) \sigma_T c \Gamma^2 (B^2/8\pi)$ is the synchrotron power generated from the electron, and $\Gamma_4 = \Gamma/10^4$. 
The above estimate indicates that the polarized microwave bubbles are $\gtrsim 25$ Myr older \citep[][supplements]{Carretti2013}. The relatively sharp change in the spectral index at $\gtrsim 23$ GHz (inside the FBs) indicates that the higher energy electrons are probably being accelerated \textit{in situ} within the FBs 
and, therefore, maintain a flat spectral index of $p\approx 2.1$. This is corroborated by the fact that the $\gtrsim 23$ GHz emission does not show any polarization signatures, suggesting a different (most probably tangled) magnetic field structure inside the FBs. Fast advection of CR (within $t\lesssim t_{\rm syn,cool}$) from the GC or acceleration at the forward shock (assuming FB represents a forward shock) would show some alignment in the magnetic field and hence produce polarization signatures.

Alternatively, the CRe could be diffused out from the freshly accelerated CRe electron population inside the FBs \citep{Crocker2015, Sarkar2015b}. Since the diffusion coefficient for the CR is given as \citep{Gabici2007}\footnote{Ideally the diffusion coefficient in lower density CGM can be $\sim 10^{2-3}$ higher than the classical value \citep{Chan2019, SJi2020} but simultaneously anisotropic diffusion can reduce the diffusion perpendicular to the field by a factor of $\sim 10^{2-3}$ \citep{Shalchi2010}.}
\begin{equation}
    D \sim 4\times 10^{27} \mbox{ cm}^2 \mbox{s}^{-1} \,\, \Gamma_4^{1/2}\,B_{10\mu G}^{-1/2}\,,
\end{equation}
the diffusion length for the CRe before it cools via synchrotron emission is
\begin{eqnarray}
    S \sim \sqrt{6\: D\: t_{\rm syn, cool}} &\approx& 1.4 \mbox{ kpc } \, \Gamma_4^{-1/4}\: B_{10 \mu G}^{-5/4}\, \nonumber \\
    &\approx& 1.4 \mbox{ kpc } \, \nu_{2.3GHz}^{-1/8}\: B_{10 \mu G}^{-9/8}\,.
    \label{chap2:eq:diffusion-sync}
\end{eqnarray}
Therefore, one would expect that the low-energy electrons would diffuse beyond the acceleration zone by $\sim 10\deg$ (assuming a distance of $\sim 8$ kpc to the emitting region), very close to the observed extension of the 2.3 GHz polarized emission around the FBs \citep{Carretti2013}. However, the insensitivity of the diffusion length to the emitting frequency means that the diffusion would be only slightly less for a $23$ GHz emitting electron compared to a $2.3$ GHz emitting electron, in contradiction to the clear scale separation at these two frequencies. Therefore, either the diffusion has steep dependence on the electron energy (perhaps due to how it `leaks' from the FBs; \citealt{Crocker2015}) or the diffusion does not play a role at all, in which case, the $2.3$ GHz emitting electrons are simply advected out from the Galaxy over a time-scale of $t_{\rm sync, cool} \sim 25$ Myr. Accurate numerical models of the outflow and CR propagation are required to understand the origin of this extended polarized emission.

\subsubsection{HI hole at the Galactic Center} 
Energetic activities such as SNe or SMBH accretion events at the GC are expected to heat the surrounding gas via shocks or radiation, and may even create outflows. The existence of such a heated region at the GC was first discovered by \cite{Lockman1984} in the \ion{H}{i} emission map. The author found that the \ion{H}{i} layer of the Galactic disk is much thinner at Galactocentric radius, $R\lesssim 3$ kpc. The \ion{H}{i} scale height was found to drop sharply by almost a factor of $3-4$ as one goes toward the GC. The result was confirmed by an updated data set from the Parkes Galactic All-Sky Survey \citep{McClure-Griffiths2009, Kalberla2010} and including the effect of non-circular motions such as the one due to the Galactic bar \citep{Lockman2016}. It was found that the \ion{H}{i} thickness drops sharply at $R\simeq 2.4$ kpc, just outside at the edge of the FBs (see fig \ref{chap2:fig:NHabs}, panel c). The sharp drop at $R \simeq 2.4$ kpc confirms the idea that the absence of \ion{H}{i} is not due to the modeling of an unknown component of the Galactic rotation curve. Moreover, the anti-correlation of the \ion{H}{i} column density and the \gray flux from the FBs supports the idea that the \ion{H}{i} hole is indeed due to either gas removal from the region or heated by an outflow from the center. It was also found that the central $2.4$ kpc region is highly deficient of H$\alpha$ emission (except at the very center), indicating suppression of star formation due to an energetic event at the GC. 
 
\subsection{High Velocity clouds}
High velocity ($\sim 100$ \kmps) warm/cold clouds are often detected in external galaxies that have an active outflow \citep{Lynds1963, Heckman1990, Cecil2001, Strickland2004}. Efforts to detect such an outflow from our Galaxy did not become mainstream mostly due to the lack of evidence that the Galaxy has an outflow, despite the claims by \cite{Sofue1977, Sofue1984, Bland-Hawthorn2003}. The discovery of the FBs opened a new window to study a galactic wind in our own Galaxy. Currently, we have a number of observations, both in emission and in absorption, detecting high-velocity gas clouds outflowing from the GC, enabling us to directly infer the properties of the nuclear outflow.

\begin{figure*}
    \centering
    \includegraphics[width=0.7\textwidth, clip=true, trim={0cm 0cm 0.1cm 0cm}]{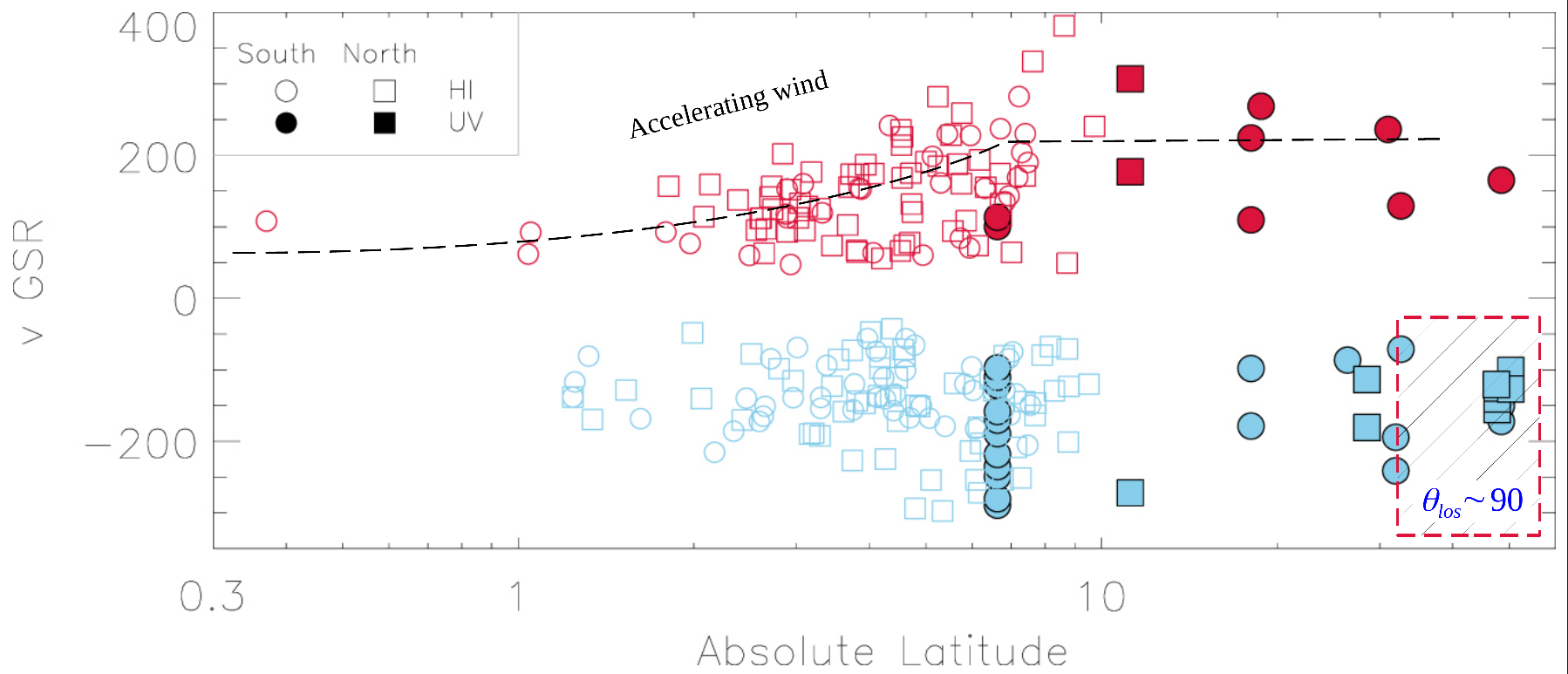}
    \caption{Warm/cold cloud velocities (in the Galactic Standard of Rest frame) toward the FBs \citep{Ashley2020}. The dashed black line shows the typical behavior of the clouds, indicating acceleration at $|b|\lesssim 7\deg$ and then flattening out at higher latitudes. The cyan and red points show the blue-shifted and the redshifted clouds. The red-hatched box is the region where the GSR velocity of the clouds should be zero since the radial velocity vector becomes perpendicular to the line-of-sight direction (if the clouds are within the FBs). Image reprinted from \cite{Ashley2020} with authors' permission (copyright by AAS).}
    \label{chap2:fig:vGSR}
\end{figure*}

\subsubsection{In emission}
\label{chap2:subsubsec:HI-clouds}
\cite{McClure-Griffiths2013} detected about $86$ \ion{H}{i} clouds with an LSR (local standard of rest) velocity ranging from $\sim -200$ to $+200$  \kmps within the central $5\deg$ of the GC. The clouds have a typical size in the range of $5-30$ pc, a mass range of $\sim 30-10^3$ \msun (total mass $\sim 4\times 10^4$ \msun), and an estimated density of about $\sim 1-10$ \pcc. The Galactic Standard of Rest (GSR) velocity for the clouds was found to be consistent with an outflow velocity of $\sim 200$ \kmps and a full wind opening angle of $\sim 135\deg$. The authors estimated that the total required power (including the hot wind phase) to drive such an outflow is $\sim 2\times 10^{39}$ \ergps over the dynamical time of $\sim 2$ Myr. 
Observations using the Green Bank Telescope revealed many more \ion{H}{i} clouds within the central $\sim 10\deg$, outflowing at a GSR velocity of $\sim 330$ \kmps and containing a total mass of $\sim 10^6$ \msun \citep{DiTeodoro2018} (see figure \ref{chap2:fig:NHabs}, panel d). The mechanical power of the wind driving such an outflow is estimated to be $\sim 3-5\times 10^{40}$ \ergps over a dynamical time-scale of $\sim 10$ Myr. The estimated high outflow velocity for the clouds suggests that these clouds are being entrained by a diffuse and high-velocity wind from the GC. Some of these clouds are even detectable in CO($2\rightarrow 1$) line emission and show a sign of acceleration as the height increases, indicating a longer dynamical time \citep{DiTeodoro2020, Lockman2020, Noon2023}.  
Optical line emissions such as H$\alpha$ and [\ion{N}{ii}] are also detected inside the FBs at a distance of $\sim 6.5$ kpc toward $(l,b) \simeq 10\deg.4-11\deg.2$ \citep{Krishanarao2020}. The optical emission lines indicate the presence of a warm ($\sim 10^4$ K) gas with a Hydrogen column density, $N(H^+) \approx 3\times 10^{18}$ \pcmsq and electron density of $\approx 2$ \pcc. Molecular clouds are also detected at the edge of the \ion{H}{i} hole \citep{Lockman2016} at Galactocentric radii of $2.6-3.1$ kpc, i.e. $l,|b| = 12\deg-26\deg, \lesssim 5\deg$ \citep{YSu2022}. 
The molecular clouds are found to have a head-tail structure, with the tails pointing away from the disk suggesting an entrainment by the outflow.  The dynamical time and the mass of the clouds are estimated to be $\sim 3$ Myr and $\sim 10^6$ \msun. Therefore, the power of the molecular structures is estimated to be $4\times 10^{39}$ \ergps, consistent with the previous estimations (for the cold cloud only). 
The authors, however, claim that the required power could be $\sim 10$ times more if one also assumes molecular clouds at $110 \lesssim |z| \lesssim 260$ pc, in which case, the required total wind power to drive such an expansion/outflow of the molecular gas would be $\sim 4\times 10^{41}$ \ergps \footnote{Considering another factor of $10$ to get the diffuse wind power}, much higher than previous estimates. 

\subsubsection{In absorption}
\label{chap2:subsubsec:warm-clouds}
Warm/cold clouds have also been observed at higher latitudes ($|b| \gtrsim 10\deg$) using absorption lines against background stars or AGNs. The combination of a distant star/AGN and a foreground star is used to locate the clouds inside the Galactic outflow \citep{Keeney2006, Zech2008, Savage2017, Cashman2021}. Such observations have detected metal ions, such as \ion{C}{iii}, \ion{C}{iv}, \ion{Si}{ii}, and \ion{Si}{iv}, with Local Standard of Rest (LSR ) velocity, $v_{\rm LSR}$, ranging from $-100$ \kmps to $+170$ \kmps or Galactic Standard of Rest (GSR) velocity, $v_{\rm GSR} \sim -140$ to $+140$ \kmps. The clouds are estimated to be at a height of $5-12$ kpc from the disk. A single `explosion' event of $\gtrsim 50$ Myr in age can explain the dynamics of the clouds and are consistent with the cloud motions in a Galactic fountain model \citep{Shapiro1976, Bregman1980}. Based on the density and temperature estimation from the detected lines, it is found that the pressure of the clouds is $p/k_B \sim 10^5$ K \pcc, much higher than the ISM ($\sim 10^4$ K \pcc; \citealt{McKee1977, Cox2005, Draine2011}) but consistent with the pressure of Galactic winds \citep{Strickland1997, Strickland2004, Sarkar2015a, Fielding2018, Schneider2020, Sarkar2022b}, implying a compression of the clouds by the wind. 

Metal absorption lines are often detected against background AGNs passing through the FBs. Such absorption lines also have LSR velocities ranging from $-250$ \kmps to $+250$ \kmps on both sides of the Galactic disk and are estimated to be at a distance of $\sim 2-7$ kpc from the disk. The clouds show signs of deceleration with increasing latitude \citep{Fox2015, Bordoloi2017, Karim2018}. The kinematics of the clouds appear to be consistent with a momentum-driven outflow from the GC at a launching velocity of $\sim 1000$ \kmps. The estimated dynamical time for the clouds to reach their current location is estimated to be $6-9$ Myr. However, more recent observations by \cite{Ashley2020} show that the clouds only accelerate to $|b| \sim 7\deg$ and either decelerate or remain at a constant velocity outside (figure \ref{chap2:fig:vGSR}). \cite{Ashley2020} also reveal the existence of clouds at $b\gtrsim 30\deg$ that have non-zero $v_{\rm LSR}$ unlike expected. If the clouds are indeed inside the FBs at $|b|\gtrsim 30\deg$ (and  $|l|\lesssim 10\deg$), their radial velocity vector makes $\sim 90\deg$ angle to the line-of-sight and hence should not produce any observable velocity (see figure \ref{chap2:fig:vGSR}). Therefore, such clouds are either outside the FBs or have non-radial velocity vectors, possibly due to an in-situ formation of the clouds rather than being entrained from the GC. In a later paper, \cite{Ashley2022} discovered that many clouds inside the FBs have metallicities ranging from $\sim 20\%$ to $300\%$ Solar, indicating that the low metallicity clouds could be part of the CGM or could form due to the mixing of outflow and CGM material at the FB edge \citep{Sarkar2015b}.

%% file: chap_4_GC_activity.tex
\section{Activity at the Galactic center}
\label{chap4:GC-activity}
It is without question that the large-scale structures that we discussed in the previous sections are related to the activities at the Galactic Center. Therefore, to understand the origin of these structures, one has to look into the past and present activities at the GC that are/were producing significant energy. While the emission from \sgra and star-forming regions inform us about the current accretion rate or the star-formation rate, `\textit{relics}' provide a good understanding of the energy sources that are not directly observable or were present in the past. In this section, we look into some of these features that are probably connected to the large-scale Fermi/eROSITA bubbles.

\begin{figure*}
    \centering
    \includegraphics[width=0.9\textwidth] {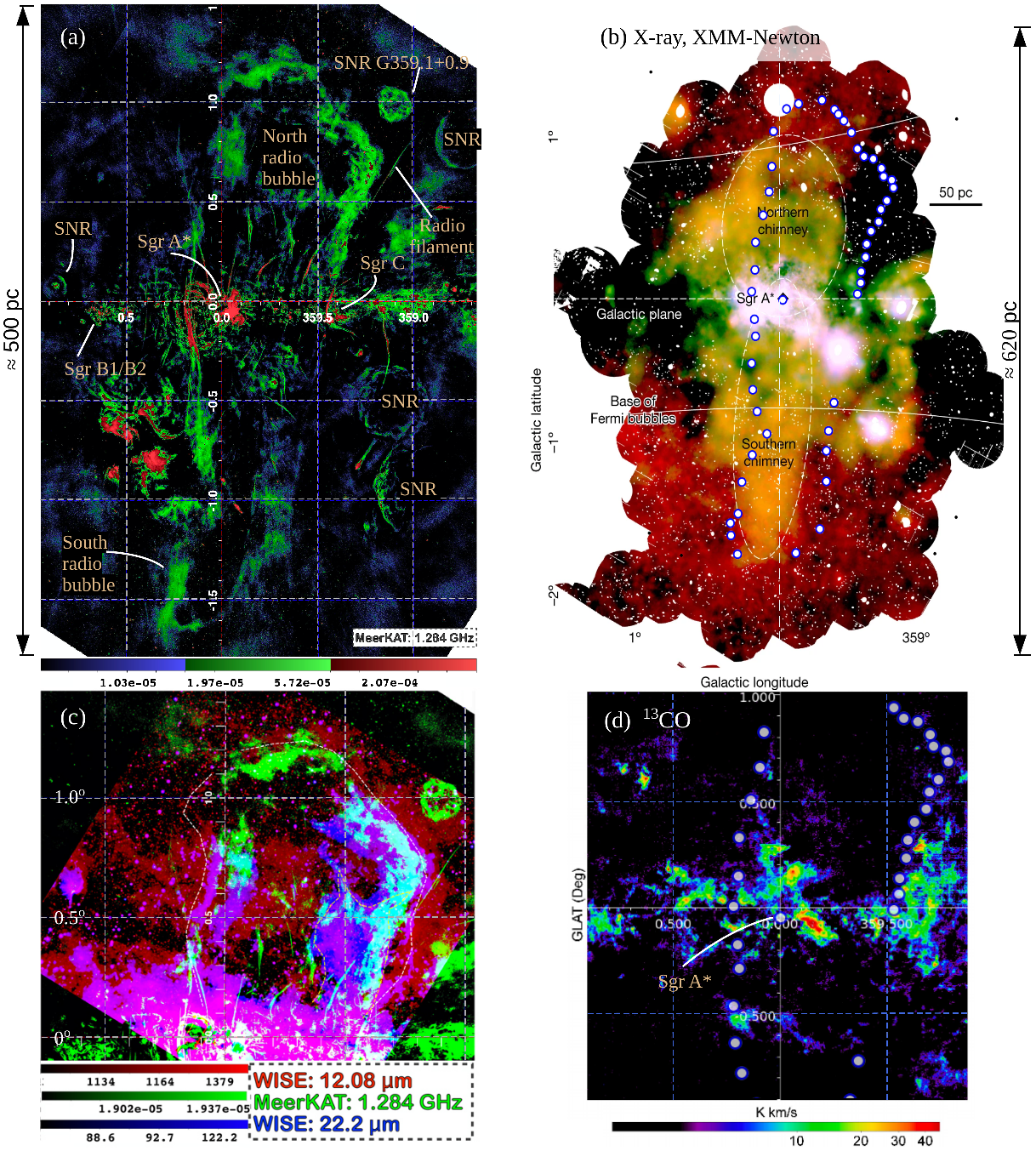}
    \caption{Panel (a): MeerKAT $1.284$ GHz map ($1\deg \approx 150$ pc) showing radio bubbles and filaments at the GC \citep{Ponti2021}. Two vertical radio loops of total size $\sim 450$ pc are clearly visible on the map. The image is reprinted from \cite{Ponti2021} with authors' permission (copyright by ESO). Panel (b): X-ray emission from $\sim 300$ pc region of the GC and observed using XMM-Newton \citep{Ponti2019}. The red shows emission in the $1.5-2.6$ keV energy band, the green shows the \ion{S}{xv} line emission ($2.35-2.56$ keV), and the purple-blue shows the continuum emission in the $2.7-2.97$ keV. The blue circles show the outline of the $450$ pc radio loop of the MeerKAT map. Image is reprinted from \cite{Ponti2019} with permission (copyright by the authors). Panel (c): Warm dust emission at $12\mu$m and $22\mu$m bands from \cite{Ponti2021} showing that the 12$\mu$m emission extends beyond the northern radio bubble. Each grid box is $0.5\deg$. Panel (d): $^{13}$CO emission map ($-3.5 < v < +3.5$ \kmps) showing a similar bubble/shell of molecular gas \citep{Veena2023}. The blue circles show the outer edge of the radio bubbles. The central blue dot shows the location of the \sgra. Image is reprinted from \cite{Veena2023} with permission (copyright by the authors).}
    \label{chap4:fig:meerkat}
\end{figure*}

\subsection{Structures at the GC}
\label{chap4:subsec:GC-structures}
\subsubsection{Galactic center bubbles} \label{chap4:subsubsec:GC-bubbles}
Numerous Spurs, filaments, and bubbles are observed within a few degrees of the GC. A well-studied feature is the $\Omega$-shaped radio lobe, known as the Galactic Center Lobe, that extends $\simeq 1.2\deg$ above (north) the Galactic plane. The Lobe is connected to the Galactic disk via two vertical spurs at $l\approx +0.2\deg$ and $l\approx -0.6\deg$, first discovered by \cite{Sofue&Handa1984} in the $10$ GHz map and has since been investigated several times \citep[e.g.][]{Yusef-Zadeh1987, Chevalier1992, Law2009, Law2010}. A southern counterpart of the Lobe, with a similar extent, was also discovered more recently using $1.28$ GHz MeerKAT observations \citep{Heywood2019, Heywood2022}, revealing a symmetric bipolar bubble/loop structure for the Galactic Center Lobe (see figure \ref{chap4:fig:meerkat}, panel a). The radio bubbles together have a size of $\simeq 75$ pc in width and $\simeq 450$ pc in height, assuming a Galactic Center distance of $8.5$ kpc \citep{Eisenhauer2003, Ghez2008, Gillessen2009, Gravity2019}. 
Due to its flat spectral index \citep{Sofue&Handa1984} and the correlation of the radio continuum with the radio recombination lines \citep{Law2009}, the origin of the radio emission was initially thought to be a free-free emission from ionized gas. The spectral indices of the radio continuum, however, have been revised to be much steeper ($F_\nu \propto \nu^{-0.5\mbox{ to } -1.0}$, i.e. $p = 2-3$), consistent with a synchrotron origin \citep{Law2010}. The radio intensity and the spectral index indicate that the magnetic field in the Lobe is $\sim 40-100 \mu$G, similar to the magnetic field estimated in supernovae bubbles elsewhere in the disk region \citep{Reich1987, Yusef-Zadeh2022a}. Assuming energy equipartition in the radio bubbles, the energy density and the total energy are, $u = 2\times B^2/(8\pi) \sim (1-3)\times 10^6$ K \pcc and $E_{450{\rm pc}} = u \times$ volume $\sim (1-5)\times 10^{52}$ erg, respectively (the factor $2$ is to account for the equivalent thermal energy in equipartition). The numbers are consistent with the estimations from the radio recombination lines by \cite{Law2009} who assumed an ionized gas density ($\sim 10^3$ \pcc) and temperature ($\sim 4,000$ K). The estimated energy density is almost $2$ orders of magnitude higher than that of the typical ISM toward the GC and signifies that the GC is actively driving an outflow. 
Based on the radio recombination lines observations, the line-of-sight expansion velocity of the radio loops is found to be $\sim 10$ \kmps \citep{Law2009}, which implies a dynamical time-scale, $t_{\rm dyn} \sim 4\, (\Delta R/40\mbox{ pc})\:(10 \mbox{ \kmps}/v)$ Myr, where $\Delta R$ is the cylindrical radius of the radio bubbles. Surprisingly, the dynamical time is slightly higher but close to the synchrotron cooling time, $1-2$ Myr, in the radio bubbles \citep{Heywood2019}. This implies that a significant part of the CR electrons must be accelerated \textit{in-situ}, which further proposes that the radio loops possibly indicate the location of shocks expanding from the GC. This idea of the shock is indeed supported by mid-IR observations at $8 \mu$m, $12 \mu$m, and $22 \mu$m \citep{Uchida1985, Shibata1987, Bland-Hawthorn2003, Ponti2021}. 
Additionally, \cite{Ponti2021} find that the mid-IR emission follows the radio emission very closely and that the $12\mu$m emission extends about $\sim 0.1\deg$ beyond the radio and the $22 \mu$m emission (panel c in figure \ref{chap4:fig:meerkat}). Given that both the mid-IR bands detect warm dust emission ($12\mu$m is warmer), the separation of the scale in these two bands can only be explained if the radiation from a shock creates a precursor region \citep{Draine2011, Sutherland2017} ahead of the shock. Depending on the shock velocity, such a photon precursor can sufficiently heat the dust to emit in the $12 \mu$m band \citep{Draine2007, Draine2011} while the dust inside the shock is heated by collisional processes, and therefore, emits in both $12\mu$m and $22 \mu$m bands \citep{Ponti2021}. 

Recently a molecular bubble was also observed within the central degrees, coinciding with the shape of the radio bubbles \citep[panel d in figure \ref{chap4:fig:meerkat};][]{Veena2023} and with the Polar Arc \citep{Bally1988}. The molecular edge in the north-western side encapsulates the radio emission, suggesting that the molecular gas could have been advected upwards by a shock \citep{Walch2015} or advected up and then slid down along the edge of the shock surrounding a super-bubble \citep{Suchkov1994, MacLow1999, ARoy2016}. The Eastern edge of the molecular bubble, on the other hand, deviates from the radio loop and moves toward negative longitude, past the \sgra. Both the Eastern and Western edges of the molecular bubble show signs of convergence at $l,b \simeq -0.4\deg, -0.2\deg$, significantly away from the \sgra, indicating an origin other than the central black hole. The shell expansion velocity lies within $-3.5< v_{\rm los} < 3.5$ \kmps, significantly slower than the \ion{H}{i}/molecular outflows velocities ($-180 \lesssim v_{\rm los} < 280$ \kmps) observed at higher latitudes ($2.5\deg \lesssim b \lesssim 10\deg$; \citealt{Dickinson2018, DiTeodoro2020}). It is, therefore, possible that the cold outflows at different scales originate from different sources \citep{Veena2023}.

The radio loops have also been observed in X-ray emission at the $1.5-3.0$ keV band. \cite{Ponti2019} detected a bipolar X-ray chimney originating close to the GC ($\Delta l = \pm 0.4\deg (\pm 60)$ pc) and extending till $b\simeq 1\deg$ ($\approx 150$ pc) in the north and till $1.5\deg$ ($\approx 220$ pc) in the south of the Galactic plane (see Panel b in figure \ref{chap4:fig:meerkat}). The X-ray emission seems to be roughly confined within the radio loop (more so in the south than in the north), suggesting that the X-ray emitting hot gas may have originated from the same events as the radio loops and the X-ray emission traces the hot gas inside a superbubble.   
The density and the temperature of the hot gas are estimated to be $k_B T \approx 0.7-0.8$ keV (corresponding sound speed, $c_s \approx 450$ \kmps) and $\sim 0.2$ \pcc, respectively. Indeed, the thermal pressure, $p/k_B \sim 2\times 10^6$ K \pcc, and the total energy, $E_{450{\rm pc}} \sim 4\times 10^{52}$ erg, of the hot gas are very similar to the estimates based on the magnetic energy density and the radio recombination lines at the radio loops which further suggest a common origin \citep{Ponti2019}. A direct velocity estimate for the hot gas is currently difficult due to the lack of spectral resolution of the X-ray telescopes. A lower limit on the expansion time scale can, however, be estimated based on the sound speed of the hot gas. This lower limit on the expansion time is $t_{\rm exp, min} \sim 150\,{\rm pc}/450$ \kmps $\approx 3\times 10^5$ yr. As mentioned, this time scale only gives a lower limit. The dynamical time-scale of the actual expansion would depend on the speed of the shock driven by this hot gas and can be much lower \citep{Castor1975, Weaver1977}. 

The estimated energy and expansion time-scale for the radio loops implies a power, $L_{450 {\rm pc}} \sim 5\times 10^{52} \mbox{ erg}/4$ Myr $ \sim 4\times 10^{38}$ \ergps. The power can be produced by the central SMBH given an accretion rate of $\dot{M} \sim 10^{-7}$ \mpy (assuming, an efficiency factor of $0.1$) or a total of $\sim 500$ SNe over the dynamical time (equivalent to a star-formation rate of $\sim 0.01$ \mpy) at the GC \citep{Chevalier1992, Law2010, MZhang2021}. The presence of a $\sim 30$ pc radio bubble and x-ray emission around the Quintuplet super-cluster system, east of the \sgra, provides an example of how the $450$-pc radio bubbles could be powered by star-forming regions in the disk \citep{Yusef-Zadeh1987, Sofue2003, Ponti2015}. Some authors also claim that the $450$-pc radio loops could be also the result of a MHD instability in the disk \citep{Sofue&Handa1984, Shibata1987, Heyvaerts1988}. 

Molecular and X-ray emissions are also observed at even smaller scales ($\sim 0.1\deg$). \cite{Hsieh2015, Hsieh2016} observed an Hour-Glass shaped CS($2\rightarrow 1$) features, each of size $\sim 15$ pc, originating from the \sgra. From the typical expansion velocities ($v_{\rm los} \sim 100$ \kmps) of nearby molecular clouds at a distance of $\sim 34$ pc from the \sgra, the dynamical time-scale for the energetic event can be estimated to be $\sim 6\times 10^5$ yr (correcting for an acceleration). Excess x-ray emission is also observed within the Molecular Hour-Glass indicating a similar structure are the $450$ pc radio loop but much smaller. The x-ray emitting hot gas is found to have a density of $2$ \pcc and a temperature of $0.7-1$ keV. Therefore, the pressure and the total energy are $\sim 2\times 10^7$ K \pcc and $E_{15 {\rm pc}} \sim 6\times 10^{50}$ erg, respectively \citep{Muno2004, Ponti2019}. The expansion time scale from the sound speed of the hot gas is $\sim 3\times 10^4$ yr, implying a lower limit on the age of this Hour-Glass feature. The required power for the $15$-pc lobes is estimated to be $L_{15pc} \sim 6\times 10^{50}$ erg$/10^5$ yr $\approx 3\times 10^{37}$ \ergps ($\lesssim 6\times 10^{38}$ \ergps, in case of an upper limit). The fact that the power for the $15$-pc and $450$-pc lobes are very similar despite their large difference in scales means that both these structures could be powered by similar events happening at different epochs. Although the power for these structures falls short of the required power for the FBs (as we will see in section \ref{chap5:origin-FBs}), these structures could be the weaker version of the `chimneys' that powered the Fermi/eROSITA bubbles \footnote{Note that the central $\sim 20$ pc also hosts a $\sim 7$ keV diffuse plasma \citep{Koyama1989, Muno2004}, the nature of which is still unknown. Recent claims suggest that this emission could be due to unresolved stars at the GC that have super Solar Fe abundances \citep{Anastasopoulou2023, Hua2023} and may not be a diffuse component of the plasma.}.

\subsubsection{Central Molecular Zone (CMZ)}
\label{chap4:subsubsec:CMZ}
The densest part of our Galaxy lies at the center of the Galaxy and is called the central molecular zone (CMZ). The CMZ is a disk-like region of dense ($\gtrsim 10^3$ \pcc) gas within $+1.7\deg \lesssim l \lesssim -1\deg$ ($R\simeq 200$ pc, assuming a Galactocentric radii of $8.5$ kpc), $|b|\lesssim 0.5\deg$ ($H\simeq 75$ pc), and $|v_{\rm rot}| \lesssim 150$ \kmps \citep[Panel a of Fig \ref{chap4:fig:GC}; ][]{Sofue1995, Morris1996, Molinari2011, Henshaw2016, Henshaw2022, Sormani2022}. 
The total molecular mass within the CMZ is $M_{\rm cmz} \sim 3-5\times 10^7$ \msun \citep{Dahmen1998, Pierce-Price2000, Molinari2011} which is about $5-10\%$ of the total molecular gas of the whole Galaxy ($\sim 6\times 10^8$ \msun; \citealt{Roman-Duval2016}). The total gas surface density in the CMZ is $\sim 10^3$ \msun pc$^{-2}$ \citep{Henshaw2022} which is almost 30 times higher than the Solar neighborhood \citep{Spilker2021}. 
A detailed analysis of the far-IR dust emission suggests that the CMZ consists of a twisted elliptical ring of cold-dense gas with a semi-major axis, $a = 100$ pc, and a semi-minor axis, $b = 60$ pc, and is extended only to Sgr B2 in the east and Sgr C in the West \citep{Molinari2011}. According to this model, \sgra is significantly offset (toward the West) from the center of the ellipse and lies closer to the front side of the ring.

The CMZ is thought to originate as a result of the gas dynamics within the Galactic bar. Stable orbits in the bar potential ($x_1$ orbits) become intersecting when the orbital energy reaches a critical value \citep{Binney2008}. While an intersecting set of orbits can remain stable in a collisionless system, the gas undergoes a shock at the intersection and loses angular momentum. The gas then flows towards the center and finally settles into a ring-like structure closely following the $x_2$ orbits and thus forming the CMZ-like rings in galaxies \citep{Binney1991, Athanassoula1992, Sormani2015}. 
Gas transiting from the $x_1$ to the $x_2$ orbit, however, does not form stars because of its high-velocity dispersion due to the shear in a flat rotation curve. The gas settles down to the $x_2$ orbit and starts forming stars only after reaching a solid body rotation regime where the shear-driven turbulence and angular momentum transport is suppressed \citep{Krumholz2015}. 
Detailed dynamical modeling suggests that the gas accumulation time for the Milky-Way CMZ is about $\sim 10-50$ Myr which represents the expected time scale between two star-formation events, with each event lasting for a dynamical time of a few Myr at the CMZ \citep{Kruijssen2014, Krumholz2017, Armillotta2019, Armillotta2020, Sormani2020a}. The models also predict the star formation rate and the gas depletion time in such a dynamical system to be $\sim 0.03-1$ \mpy and $10^{8-9}$ yr, respectively.  For a detailed discussion of the individual features of the CMZ and its dynamics, the readers are referred to a recent review by \cite{Henshaw2022}.

The CMZ is known to be coincident (both spatially and kinematically) with a stellar ring, known as the \textbf{Nuclear Stellar Disk (NSD)}. The disk extends from $R\simeq 30$ pc to $\simeq 300$ pc and dominates the gravity due to its high stellar mass content, $\sim 0.7-1.5\times 10^8$ \msun \citep{Launhardt2002, Sormani2020b}. The stellar disc is thought to have formed from multiple episodes of star formation at the CMZ itself  \citep{Armillotta2019, Sormani2020a}. Understanding the star-formation history of the NSD can provide us with crucial information regarding the SNe energy production rate at the GC.

\subsubsection{Nuclear interstellar matter}
\label{chap4:subsubsec:nuclear-ism}
While the CMZ sets the overall star formation within the central $\sim 100$ pc of the GC, the transfer of gas from the CMZ (at $\sim 100$ pc) to \sgra (at $10^{-6}$ pc) is still poorly understood. Direct observations of gaseous and stellar structures close to the \sgra can provide us with a better view of the activities at the \sgra. A detailed review of such nuclear structures is given in \citep{Genzel2010, Bryant2021}. In the current review, I provide only a summary of these structures for the sake of completeness. 

\begin{figure*}
    \centering
    \includegraphics[width=0.95\textwidth, clip=true, trim={0cm 0cm 0cm 0cm}]{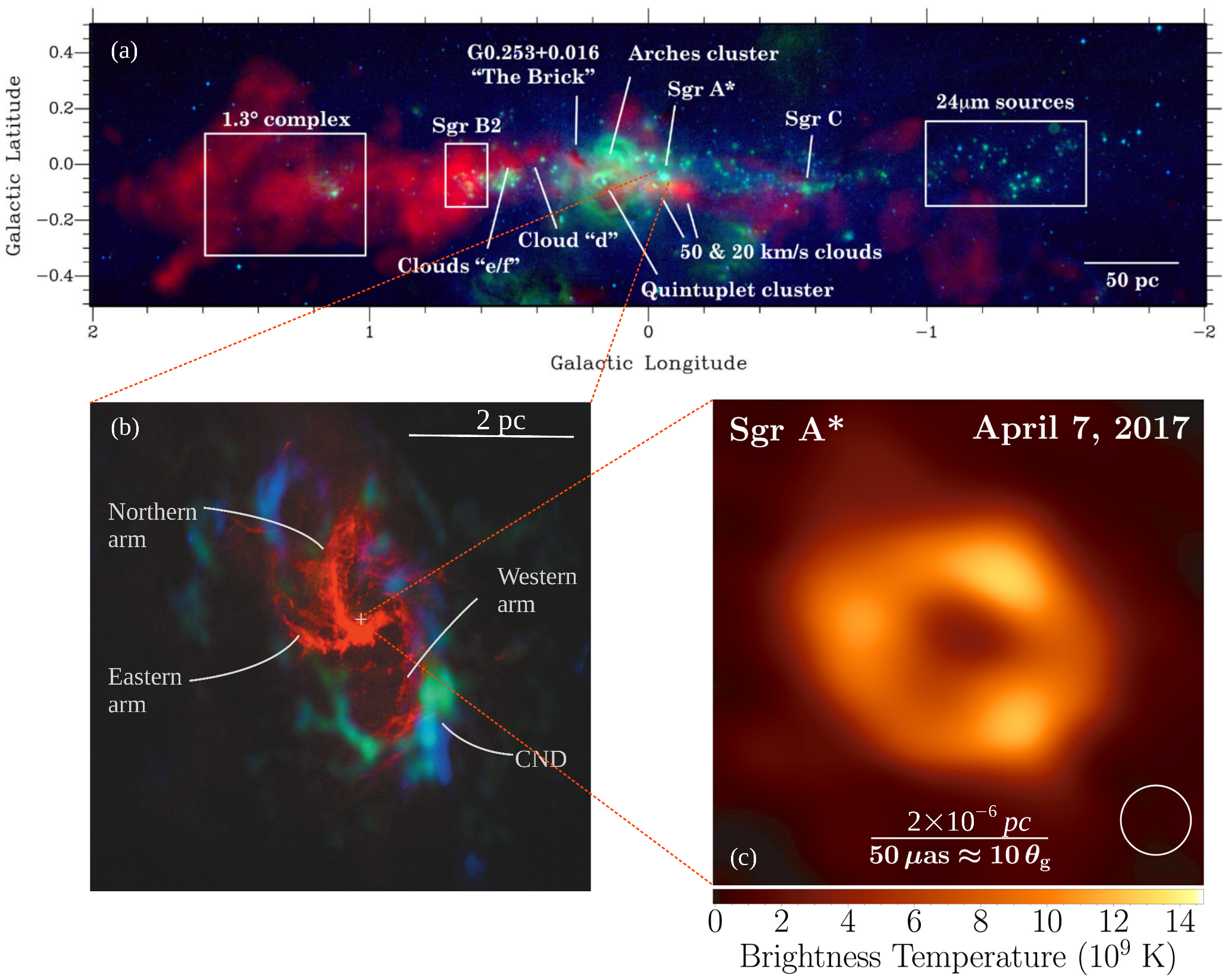}
    \caption{Galactic center at different scales. Panel (a) ($1\deg \approx 150$ pc): multi-wavelength emission from the central molecular zone. The red shows NH$_3$ emission and represents dense ($\gtrsim 10^3$ \pcc) gas. The green and blue shows the $21.3\mu$m and $8.28 \mu$m emission, respectively, and represent the PAH emission from cloud edges, young stellar objects, and evolved stars. The image is reprinted with permission from \citep{Kruijssen2014} (copyright by the authors). Panel (b): Interstellar matter in the central few pc of \sgra showing the circumnuclear disk (CND) in green and blue (HCN emission), and the ionized streams in red (6 cm radio continuum emission). The location of the \sgra is marked with a `+' sign. Image is reprinted from \cite{Genzel2010} with authors' permission (copyright by the APS). Panel (c): Event Horizon Telescope observation of the accretion disk around the \sgra at $1.3$ mm (adapted from \citep{EventHorizon2022_I}). Here, $\theta_g = G M_{\rm bh}/(c^2 D)$ is the gravitational radius of the central SMBH with mass $M_{\rm bh}$, and $D$ is the distance from Earth. The white circle shows the beam size used for the image reconstruction. The image is reprinted with permission from \citep{EventHorizon2022_I} (copyright by the authors).}
    \label{chap4:fig:GC}
\end{figure*}

\textbf{Circumnuclear disk (CND)} is a ring like gaseous structure at a distance of $1.5-4$ pc from \sgra (panel b, Fig \ref{chap4:fig:GC}). The CND has a total mass of $10^{5-6}$ \msun and hosts several clumps with densities of $\sim 10^{6-8}$ \pcc and gas temperatures of $\sim 50$ K \citep{Becklin1982, Brown1984, Genzel1985, Serabyn1986, Genzel2010, Hsieh2021}. Most of these clumps are compact enough to overcome the tidal force from \sgra and may either remain in rough equilibrium or form stars. \citep{Shukla2004, Christopher2005, Hsieh2021}. The critical density to be tidally stable within the central $1$ pc is $\sim 10^8$ \pcc and, therefore, clouds are tidally disrupted inside of this radius as is evident from the shape of the ionized streams (`Eastern arm', `Northern arm' and the `bar') inside this radius (figure \ref{chap4:fig:GC}). The central $1-1.5$ pc radius can be described as the `ionized cavity' ($T\sim 5-7\times 10^3$K) that is affected by the UV radiation either from \sgra or from the nuclear star-cluster \citep{Becklin1982, Guesten1987, Shukla2004, Christopher2005}. Within the ionized cavity, different streams spiral in toward \sgra with the closest approach at $\sim 0.04$ pc where the mass inflow rate is estimated to be $\sim 10^{-3}$ \mpy \citep{Genzel1994}. The `Western arm' is shown to be in a circular orbit and most probably represents an ionized surface of the CND. These ionized streams bear evidence of how gas from $\sim$ pc scale can be fed to the central black hole. It is, however, unclear how the gas from $\sim 100$ pc scale reaches the central few pc, i.e. the CND. There is some evidence of molecular filaments connecting the CND to a molecular cloud at $\sim 20$ pc away, suggesting that the CND could have formed during a close passage of such a cloud \citep{Hsieh2017, Hsieh2018}. The CND (inclination $\approx 70\deg$) and ionized streams (inclination $\approx 50\deg$) are significantly misaligned with the large scale Galactic plane (inclination $=90\deg$) \citep{Jackson1993, Paumard2004, Zhao2009, Genzel2010}, an important issue I will come back to below.

\subsubsection{Nuclear Star Cluster (NSC)}
\label{chap4:subsubsec:NSC}
The central parsec of the GC is known to host a compact star cluster ($M_\star \simeq 10^6$ \msun) where the majority of stars are old (age $\gtrsim 1$ Gyr) that have an average rotation vector in the same direction as the Galaxy itself \citep{Becklin1978, Launhardt2002, Genzel1996, Trippe2008, Schodel2009, VonFellenberg2022}. A younger generation of stars (age $\approx 6\pm2$ Myr) is also observed within the central $0.5$ pc \citep{Paumard2006, Bartko2009}. Almost 2/3-rd of the young stars are seen to have a well-defined clockwise component ($j_z >0$) and the rest are observed to have a counterclockwise rotation. The masses of the components are estimated to be $\approx 10^4$ \msun and $\approx 5\times 10^3$ \msun for the clockwise and counterclockwise rotating components, respectively. Additionally, the rotation vectors of individual stars within the clockwise system slightly vary from each other but, overall, remain at an inclination of $122\deg\pm7\deg$ (about $\sim 30\deg$ away from the Galactic rotation axis) \citep{Paumard2006, Bartko2009, Lu2009, Genzel2010}. The counterclockwise systems, although less well defined, have an inclination of $60\deg\pm15\deg$. Therefore, these young stellar disks are even more misaligned (compared to the CND) from the Galactic rotation axis and lie in between the inclination of the CND and the Northern arm. 

The origin of the nuclear young stellar disks has been explained in terms of star-formation in an accretion disk around \sgra \citep{Levin2003, Bonnell2008, Nayakshin2018, Generozov2022}. The accretion event in question happened during the passage of a (or two) cloud of mass $\sim 10^5$ \msun and an initial velocity of $40-80$ \kmps toward the central SMBH. Numerical simulations suggest that during this event, about $10-30\%$ gas was directly accreted onto the black hole over a viscous time-scale of $\sim 10^7$ yr with an accretion rate of $\sim 10^{-4}-10^{-3}$ \mpy. The accretion, therefore, could have produced about $0.1\%-1\%$ of Eddington luminosity, i.e. $5\times 10^{41-42}$ \ergps \citep{Bonnell2008, Hobbs2009}. Given the timeline for the accretion event ($6\pm2$ Myr) being consistent with the time-scale to generate the $450$ pc radio bubbles seen in MeerKAT images (section \ref{chap4:subsec:GC-structures}), it is possible that both of these features are connected to the same activity. In such a case, the central SMBH would have had a much lower luminosity ($\sim 4\times 10^{38}$ \ergps) than the expected power from the accreted gas.

\subsection{Star formation rate and history}
\label{chap4:subsec:SFR}
One of the engines that can power the Fermi/eROSITA bubbles is stellar feedback from the GC. Massive ($\gtrsim 8$ \msun) stars generate mechanical energy via stellar winds and SNe \citep{Leitherer1995, Leitherer1999, Sternberg2003}. The generation of the stellar energy is directly related to the star formation rate (SFR) within the last $\lesssim 40$ Myr (age of a $8$ \msun star). Therefore, estimating the current star formation at the GC within the last $\sim 10$ Myr is crucial in determining whether star formation can power the Fermi/eROSITA bubbles. Several different methods have been used to estimate the `current' SFR within the central $\sim 100$ pc (the CMZ) and it is found to lie in a range of $0.04-0.8$  \mpy. A detailed discussion of different methods and star formation locations is given in a recent review by \cite{Henshaw2022}. Therefore, I only provide a brief overview. 

One of the direct methods to estimate the SFR in the CMZ is to count individual sources, such as young stellar objects (YSO), \ion{H}{ii} regions, masers, and supernovae remnants (SNR). YSOs are proto-stars (age $\lesssim 1$ Myr) that are still accreting colder material from the surroundings and thus produce excess infrared (IR) emission. \citep{Yusef-Zadeh2009} counted the number of YSOs within $|l|, |b| < 1.3\deg, 0.17\deg$ and estimated that the current SFR is $\simeq 0.15$ \mpy for the last $\sim 0.1$ Myr. This value was later updated to $0.05-0.08$ \mpy due to contamination by main sequence stars \citep{An2011, Immer2012, Koepferl2015, NandaKumar2018}. Such YSOs are often embedded inside dusty molecular clouds and can produce powerful masers that can hint at the current star formation activity. Water masers indicate an SFR of $\sim 0.04$ \mpy \citep{Lu2019a}. Counting of SNRs can also produce a star formation rate in the very recent past ($\sim 10-40$ kyr at the CMZ densities). 
Based on approximately $15$ X-ray and radio SNRs, \cite{Ponti2015} estimated an SFR of $\sim 0.04-0.15$ \mpy. More recently, \cite{Ponti2021} detected 4 more SNRs in the CMZ which updates the estimated SFR to be $0.05-0.2$ \mpy in the last $10-40$ kyr. New IR sources have also been detected with \textit{SOFIA} \citep{Hankins2020} which will probably affect the estimated SFR. Integrated IR emission of star-forming regions (resolved or unresolved) in the CMZ indicates an SFR of $\sim 0.1$ \mpy \citep{Yusef-Zadeh2009, Barnes2017}. 
Massive stars also emit H-ionizing radiation that creates \ion{H}{ii} regions around the stars. Free-free emission or radio recombination lines from these \ion{H}{ii} regions have been used to estimate the SFR to be $\sim 0.015$ \mpy \citep{Longmore2013}. Such an estimation, however, only provides a lower limit on the SFR since many known star-forming regions are also not visible in such radio emissions. \cite{Henshaw2022} note that the average SFR in the CMZ over the past $\sim 5$ Myr is $\sim 0.07$ \mpy, and over the last $5-100$ Myr is $\sim 0.09$ \mpy and is almost an order of magnitude smaller than the expected SFR inside the CMZ based on the gas surface density \citep{Longmore2013, Barnes2017, Henshaw2022} or dynamical modeling of the CMZ \citep{Krumholz2017, Armillotta2019}. 

More recently, the GALACTICNUCLEUS survey \citep{Nogueras-Lara2019} has provided us with near-IR color-magnitude information of about $700,000$ stars. The color-magnitude data reveals that most of the stars ($\gtrsim 80\%$) within the nuclear stellar disc formed about $7-8$ Gyr ago and about $5\%$ were formed at another star-formation period ($\sim 100$ Myr long) $1$ Gyr ago \citep{Nogueras-Lara2020}. The authors also note that although the CMZ is forming stars at a slow pace for the last $\sim 500$ Myr, it had a jump in star formation in the last $\sim 30$ Myr with an SFR of $\approx 0.2-0.8$ \mpy. 
This is also supported by the discovery of new young stars in different star-forming regions \citep{Nogueras-Lara2022}. Note that this rate of star formation and the time scale are consistent with the expected amplitude and periodicity of star formation at the CMZ \citep{Krumholz2017, Armillotta2019, Armillotta2020, Sormani2022}. The large discrepancy of the estimated SFR at the CMZ using different methods produces a challenge to estimate the `true' SFR at the GC. Probably higher sensitivity surveys (such as SOFIA; \citealt{Hankins2020}) could help in reducing the large discrepancy between the source counting and stellar light modeling. 

Although other star formation indicators based on Mira variables recover the major star formation episode at $\sim 8$ Gyr ago, they fail to reproduce the star formation history of \cite{NoguerasLara2019} at $\lesssim 1$ Gyr time scales \citep{Sanders2023}. This is mainly because of the unreliability of the age-period relation for Mira variables at such a young age \citep{Trabucchi2019, Sanders2023}.

\begin{table*}[]
    \centering
    \begin{tabular}{p{0.7in} c c c c p{1.3in}}
    \hline \\
    Feature          & Scale [kpc]   & Velocity/Mach      & Age [Myr] & Power [\ergps] & Comments \\
    \hline\hline\\
    eROSITA bubbles  & $\simeq 12$ & Mach $\approx 1.5$ & $\sim 20$ & $\sim 10^{41}$ & Thermal, $T\sim 0.3$ keV, sec \ref{chap2:subsec:NPS} \\
    Fermi Bubbles   & $\simeq 8$ & - & - & - & Hadronic/IC,  CR spectrum $p \approx 2.1$, sec \ref{chap1:subsec:FBs}\\
    Microwave Haze  & $\simeq 6-7$ & - & - & - & Synchrotron,  $p \approx 2.1$, $B \approx 5-10 \mu$G, sec \ref{chap2:subsec:Micro-Haze} \\
    Warm clouds & $\sim 2-7$ & $v_{\rm los} \sim \pm 250$ \kmps & $6-9$ & - & Model dependent velocity, sec \ref{chap2:subsubsec:warm-clouds} \\
    \ion{H}{i} clouds & $\sim 1.5$ & $v_{\rm los} \sim \pm 200$ \kmps & $\sim 1-10$ & $10^{39.3}-10^{41.6}$ & sec \ref{chap2:subsubsec:HI-clouds}\\
    $450$-pc Radio bubble & $\sim 0.2$ & $v_{\rm los} \sim \pm 10$ \kmps & $\sim 4^\dagger$ & $\sim 10^{38.6}$ & Synchrotron, $p\approx 2-3$, $B \approx 40-100 \mu$G, sec \ref{chap4:subsubsec:GC-bubbles} \\
    X-ray chimney & $\sim 0.15$ & $c_s \sim \pm 450$ \kmps & $\sim 0.3$ & $\lesssim 10^{39.6}$ & Thermal, $T\sim 0.7$ keV, sec \ref{chap4:subsubsec:GC-bubbles} \\
    15-pc lobes & $\sim 0.015$ & $v_{\rm los} \sim 100$ \kmps & $\sim 0.6$ & $10^{38.8}$ & sec \ref{chap4:subsubsec:GC-bubbles} \\
    Nuclear young star cluster & $0.0005$ & - & $6\pm2$ & - & $M_\star \sim 10^4$ \msun, Sec \ref{chap4:subsubsec:NSC}\\
    \hline\\
    \end{tabular}
    \begin{tabular}{p{1in} c c c p{1.5in}}
    \hline\\
        Activity & Scale & Rates & Power [\ergps] & Comments \\
        \hline\hline\\
        Star formation &  $\sim 100$ pc & $0.07$, $0.2-0.8$ \mpy & $10^{40.3}$, $10^{40.9-41.5}$& over $\sim 5$ Myr, $\sim 30$ Myr, sec \ref{chap4:subsec:SFR}\\
        AGN (Current) & $\sim 10^{-6}$ pc & $10^{-8}$ \mpy & $10^{38-38.7}$  & Sec \ref{chap4:subsec:MWBH-acc-rate}\\
        AGN ($\sim 100$ yr ago) & $\sim 10^{-6}$ pc & $\sim 10^{-5 \mbox{ to }-4}$ \mpy & $10^{41-42}$ &  for $\Delta t\sim 1-10$yr, Sec \ref{chap4:subsec:MWBH-acc-rate} \\
        AGN ($\sim$ 1-3 Myr ago) & $\sim 10^{-6}$ pc & - & $10^{43.7-44.7}$ &  for $\Delta t\gtrsim 4$ kyr, Sec \ref{chap4:subsec:MWBH-acc-rate} \\
        \hline
    \end{tabular}
    \caption{Top Table: Summary of the observational multi-wavelength features toward the GC. Parameters, such as age and power are model-dependent quantities. Bottom Table: Summary of GC activities as observed/estimated from different tracers. $^\dagger$ For the $450-pc$ radio bubble a cylindrical radius of $40$ pc and $v_{\rm los}$ are used to obtain the age. }
    \label{chap4:tab:sum-parameters}
\end{table*}

\begin{figure*}
    \centering
    \includegraphics[width=0.8\textwidth, clip=True, trim={0.5cm 0cm 1cm 1cm}]{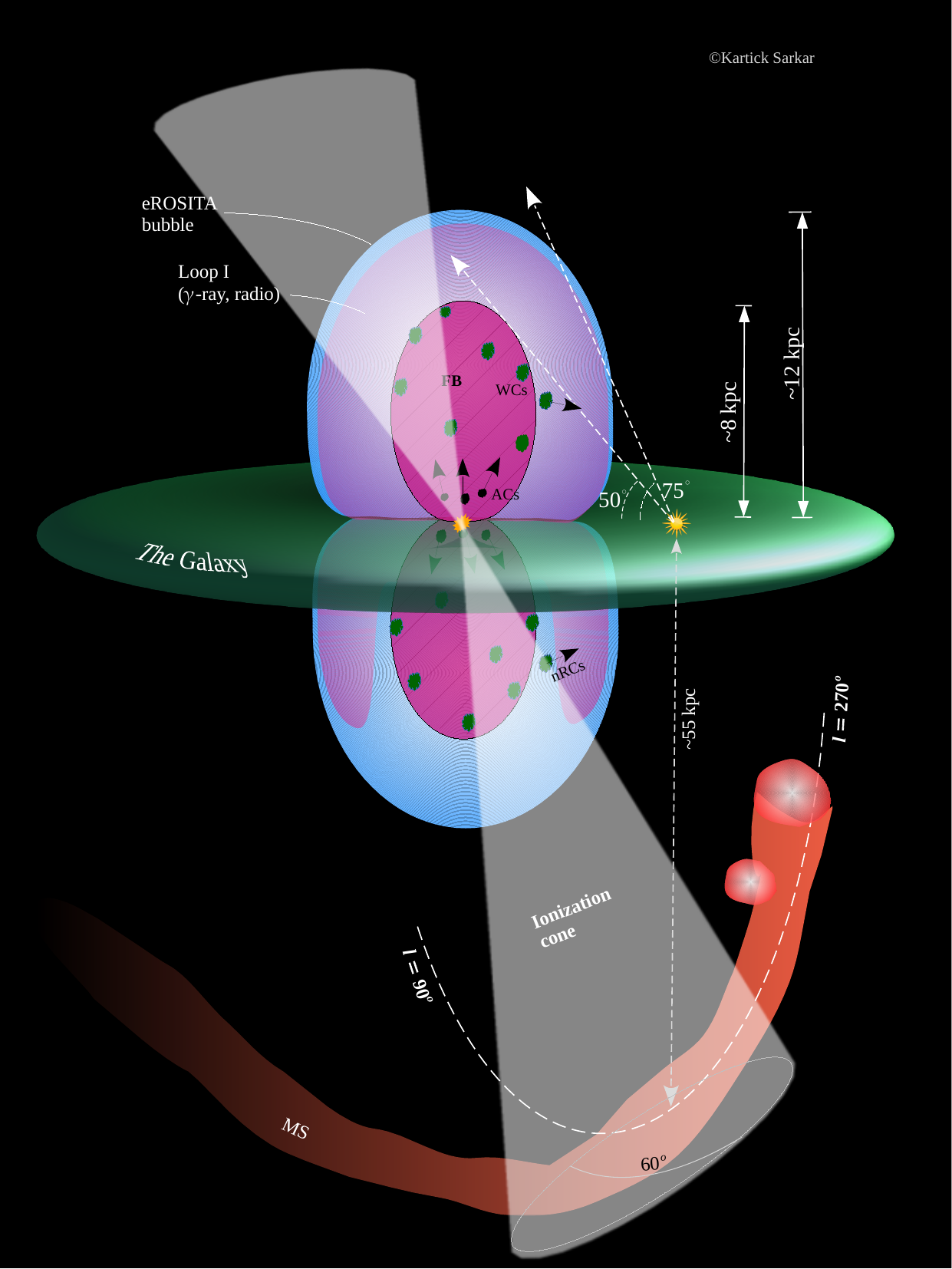}
    \caption{Multi-Wavelength representation of the Fermi/eROSITA bubbles. The cyan shows the eROSITA bubbles (x-ray; \citealt{Predehl2020}), the light purple shows the $\gamma$-ray \citep{Ackerman2014} and radio Loop-I/NPS \citep{PlanckCollaboration2016}. The slightly different sizes in $\gamma$-ray/radio and x-ray bubbles are just to show the different colors. In practice, their edges almost exactly coincide with each other. The deep purple bubbles are the Fermi Bubbles (FBs; \citealt{Su2010}). The black hatched region represents the radio Haze emission \citep{PlanckCollaboration2013}. Green clouds are warm clouds (WCs) detected in UV absorption lines \citep{Fox2015, Bordoloi2017}. Green clouds with arrows are non-radial clouds (nRCs; \citealt{Ashley2020}). The black outflowing clouds are atomic clouds (ACs) detected in \ion{H}{i} emission \citep{DiTeodoro2018}. The white light cone shows the $\sim 3$ Myr old ionization cone \citep{Bland-Hawthorn2019}.  The ionization cone intersects the Magellanic Stream (MS) in the southern hemisphere. Note that the Magellanic Stream lies at $l\approx 90\deg/270\deg$ plane and extends from $(l,b) \approx (270\deg,-30\deg)$ to $(90\deg,-30\deg)$ \citep{Putman2003}, thus implying a shape that lies perpendicular to the image plane.}
    \label{chap4:fig:obs-cartoon}
\end{figure*}

\subsection{AGN activitiy}
\label{chap4:subsec:MWBH-acc-rate}
Our Galaxy is known to contain a supermassive black hole (SMBH) of mass $M_{\rm bh} \approx 4.14\times 10^6$ \msun at its center, \sgra, at a distance of $D = 8.127$ kpc \citep{Genzel1996, Schodel2002, Ghez2003, Genzel2010, Gravity2019, EventHorizon2022_IV}. The characteristic length, time, and angular scales are $r_g \equiv G M_{\rm bh}/c^2\approx 2\times 10^{-7}$ pc, $t_g \equiv G M_{\rm bh}/c^3 \approx 20.4$ s, and $\theta_g \equiv  G M_{\rm bh}/(c^2 D) \approx 5$ micro arcsec, respectively. The presence of the SMBH at the GC means that gas accretion onto the SMBH can result in mechanical power that could be responsible for providing the required energy to the Fermi/eROSITA bubbles. The produced power depends on the rate of accretion onto the BH following $L_{\rm bh} \sim 0.1 \dot{M}_{\rm acc} c^2$, as noted in section \ref{chap1:subsubsec:agn-outflow}. 
Observations of polarization and its variability in mm, sub-mm, and near IR wavelengths within the central $0.1$ pc of the \sgra have been instrumental in determining the mass accretion rate of the SMBH \citep{Aitken2000, Marrone2007, Eckert2012, Brinkerink2016, Bower2018, Do2019, Michail2023}. The observation of polarized emission indicates a low Faraday depth and hence a low mass accretion rate. Different theoretical models for the accretion flow\footnote{See \cite{Yuan2014} for a recent review on the different accretion flow models abound black holes.} produce an accretion rate in the range of $10^{-9}-10^{-8}$ \mpy \citep{Agol2000, Quataert2000b, Yuan2004, Marrone2007, PSharma2007}. More sophisticated numerical simulations employing state-of-the-art methods for solving the General Relativistic Magneto-Hydrodynamics (GRMHD) equations also agree on the estimated accretion rates from the simpler accretion flow models \citep{Dexter2009, Drappeau2013, Ressler2020, Ressler2023}. 
Very recently, it has been possible to resolve the central $\sim 10^{-6}$ pc ($\sim 5 r_g$) using the Event Horizon Telescope (EHT; \citealt{EventHorizon2022_I}) that uses interferometry at mm wavelengths (panel c of figure \ref{chap4:fig:GC}). The observation shows a clear sign of an accretion disk and the shadow of the central SMBH. Detailed modeling of the reconstructed image reveals a `best-bet region' for crucial parameters for the black hole, \textit{viz.} accretion rate $\dot{M}_{\rm acc} \approx (0.5-1)\times 10^{-8}$ \mpy, bolometric luminosity, $L_{\rm bol} \approx (0.7-1)\times 10^{36}$ \ergps, and the mechanical luminosity, $L_{\rm bh} \approx (1-5)\times 10^{38}$ \ergps. The spin of the SMBH and the inclination of the accretion disk are estimated to be $(a, i) = (0.5, 30\deg)$ or $(0.94, 10\deg)$. A low inclination means that the accretion disk is not aligned with any of the larger scale structures, namely, central molecular zone ($\sim 100$ pc, $i\sim 90\deg$), circumnuclear disk ($\sim 2$ pc, $i\sim 66\deg$), ionized streams ($\sim 0.1-1$ pc, $i\sim 45\deg-60\deg$), or the nuclear young stellar disks ($\sim 0.5$ pc, $i\sim 60\deg, 122\deg$), and rather faces toward the Earth. The current accretion rate at the \sgra is, therefore, $\sim 10^{-7}$ times the Eddington accretion rate, $\dot{M}_{\rm edd} \approx 0.1$ \mpy. 

X-ray observations of the central $0.06$ pc indicate the presence of a hot ($\sim 1.3$ keV) plasma with density $n_e \sim 25$ \pcc which further suggests that the accretion rate at the Bondi radius ($r_B \sim 10^5 r_g$) is about $10^{-6}$ \mpy \citep{Baganoff2003}. This accretion rate is found to be consistent with the accretion of stellar wind material from nearby Wolf-Rayet stars \citep{Loeb2004, Moscibroka2006, Ressler2020}. The apparent discrepancy of the accretion rate at $\sim 10^5 r_g$ and at $\sim 10 r_g$  poses a challenge to the understanding of how gas flows from large scale to being finally accreted at the black hole.
Multi-scale simulations covering the whole range of $\approx 1 r_g$ to $6\times 10^6 r_g$ suggest that even though the wind feeds the SMBH at a higher rate (at $\sim 10^6 r_g$ scale), the accreted material close to the black hole quickly becomes `magnetically arrested' and gas accretion to the black hole happens only via a thin disk. Such highly magnetized accretion flow produces jets/outflows that restrict the final accretion to the black hole to only $\sim 10^{-8}$ \mpy \citep{Ressler2020, Ressler2023}.  

Direct evidence of a past jet at the GC  has been difficult to pinpoint because of the confusion with the other sources. \cite{Cecil2021} discuss indirect evidence of an intermittent jet from \sgra that may have caused the $\sim 15$-pc cavity around \sgra and claim that the jet direction must be $\sim 10\deg$ away from the Galactic rotation axis to produce such vertical bubbles in the last $\sim$ Myr. While it is a possible scenario, similar bubbles have also been noted in SNe simulations around \sgra \citep{MZhang2021}. The direct evidence of jet-like features is seen in mm/sub-mm recombination lines in ALMA/VLA images in the form of radial streaks that converge toward the \sgra \citep{Yusef-Zadeh2020}. The `jet-like' features extend to $\sim 0.1$ pc and lie in the Galactic plane at a position angle of $60\deg$ with a jet opening angle $\sim 30\deg$, in contrast to the claims of a vertical jet. Analysis of MeerKAT images also revealed that, statistically, short filaments are converging toward \sgra and lying on the Galactic plane \citep{Yusef-Zadeh2023}. Based on the velocity and distance of a molecular cloud, Sgr A E, along the radio filaments, the authors estimate that the event that produced these radio filaments must be related to \sgra $\sim 6$ Myr ago. The estimated time scale is similar to the dynamical time scale of the $450$-pc radio bubbles and the age of the young nuclear stellar disc. Could they be of the same origin?

Although the current accretion rate onto \sgra is $\sim 10^{-8}$ \mpy, several observations suggest that the central SMBH was about $\sim 10^3$ time more luminous only about $\sim 100$ yr ago. Observations of fluorescent Fe K$\alpha$ line (at $\approx 6.4$ keV) in several molecular clouds (including at Sgr B2 and Sgr C cloud complexes) around \sgra indicates an X-ray luminosity of $L_X \sim 10^{39}$ \ergps (mechanical luminosity of $10^{41-42}$ \ergps) \citep{Kayoma1996, Murakami2001, Inui2009}. This means that \sgra was active not too long ago. Considering the light travel time to the clouds, and that we do not see such brightness in \sgra, the event most probably happened $\sim 100$ yr ago. Variation of Fe K$\alpha$ line intensity across the clouds as well as during repeated observations indicate that \sgra is variable on a $\sim (1-10)$ yr time-scale \citep{Muno2007, Chuard2018}. Detailed modeling of the X-ray line intensity variation across molecular clouds suggests that there could have been two events in the last $\sim 300$ yr, one at $\sim 240$ yr ago, lasting for $\sim 10$ yr and another at $\sim 100$ yr ago, lasting for a few yr \citep{Clavel2013, Churazov2017, Chuard2018, Marin2023}. Both these events probably had an X-ray luminosity of $\sim 10^{39}$ \ergps. Although these activities were enhanced by a factor of $10^3-10^4$ compared to the current activity, the average mechanical luminosity of \sgra over the last $\sim 100$ yr is only $10-100$ times higher once we consider the short variability time-scale of such eruptions. 

Signatures of a past accretion event have also been detected in terms of excess ionization in the Large Magellanic Stream (MS) compared to local high-velocity clouds. The excess ionization in the MS indicates an ionization source at the center of the Galaxy \citep{Bland-Hawthorn2013, Bland-Hawthorn2019, Fox2020}. The excess ionization in H$\alpha$ and metal lines, such as \ion{C}{IV} and \ion{Si}{IV}, is mainly observed within a half opening angle of $25\deg-30\deg$. 
The recombination time, $\sim 0.6-2.9$ Myr, of the ionized plasma and to-and-fro light travel time, $\sim 0.3$ Myr (at the MS distance of $50$ kpc) means that the event causing the excess ionization occurred about $1-3$ Myr ago. The required photon flux ($\sim 10^{53}$ photons \psec) at the GC is only possible through an accretion activity of $0.1-1$ times the Eddington rate at the central SMBH \citep{Bland-Hawthorn2013, Bland-Hawthorn2019}. 
For the MW central SMBH, such an accretion rate means a power of $5\times 10^{43-44}$ \ergps. This AGN time-scale at the GC surprisingly fits well with the formation time of the nuclear young stellar disks and the $450$ pc radio bubbles (although the radio bubble power is much lower). 
The duration of the event is, however, uncertain. One can estimate a lower limit on the accretion duration to be the ionization time scale of the plasma and is  $\sim 4$ kyr at a density of $\sim 0.1$ \pcc. Given that the ionization time scale is, much shorter than the other time scales, the Eddington accretion phase at the \sgra could even be as short as $\sim 4$ kyr to produce such an ionization cone.


%% file: chap_5_proposed_origin.tex
\section{Proposed origin of the Fermi/eROSITA Bubbles} 
\label{chap5:origin-FBs}
A successful theoretical model for the Fermi and eROSITA bubbles (FEBs) requires one to simultaneously explain the multi-wavelength and multi-scale observations that are discussed in the previous sections. Although not all the features may be relevant for the FEBs, they paint an overall picture of the outflow from our Galaxy, of which FEBs are part.
The specific question of the origin of the FEBs can be divided into two parts, one, what provides the mechanical energy to inflate these bubbles against the pressure of the ambient CGM? and second, what is the origin of the \gray emission? 

For the mechanical origin of the bubbles, it is evident from the observed/estimated activities at the GC (sec \ref{chap4:GC-activity} and table \ref{chap4:tab:sum-parameters}) that the FEBs may be powered either by a continuous energy source (SNe at regular intervals or intermittent AGNs) or a single outburst from the central SMBH. 
For the spectral origin, the \grays may be produced either via the hadronic ($p-p$) channel or via IC scattering of ambient light such as CMB, IR, and the interstellar radiation field (ISRF). There are also some possible scenarios where the \grays originate from dark matter annihilation or pulsars \citep[see][and references therein]{Hooper2013}. Such emission, however, can only explain the low-latitude \grays (depending on the assumed interaction cross section) and are clouded by the huge uncertainty in the nature of dark matter itself. 
In the following, I will only focus on the established channels of the \gray production, i.e. Hadronic, and Leptonic. Due to the history of this topic, the Fermi Bubbles (FBs) and the eROSITA bubbles (in earlier convention, NPS/Loop-I) were not thought to be related, and hence in most of the cases, the models explain only the Fermi Bubbles. Therefore, a distinction in the notation, i.e. FBs or FEBs, must be noted carefully in the following discussion.

\subsection{Simple outflow models} 
\label{chap5:subsec:wind-model}
Propagation of shock waves due to a continuous or instantaneous energy source in a background medium has been discussed extensively in the literature (see \citealt{Ostriker1988} for a review, and also \citealt{Sedov1946, Taylor1950, Sedov1959, Cox1972, Weaver1977, Shapiro1979}). The general solution for a Blast-Wave (BW) in an isotropic medium is self-similar for which the radius and the velocity of the shock front, $r_s$, at any time, $t$, is given as simple power laws, $r_s(t) \propto t^\eta$ and $v_s(t) = \eta r_s/t$. A more accurate expression of the shock radius is given by \cite{Ostriker1988} as
\begin{equation}
    r_s(t) = \left[ \frac{\xi E_b}{\bar{\rho}(r_s)}\right]^{1/5} t^{2/5}
\end{equation}
where, $E_b$ is the total energy of the BW, $\bar{\rho}(r_s)$ is the average density within $r_s$, and $\xi = \frac{3}{4\pi \eta^2 \sigma}$ with $\sigma = E_b/(M v_s^2) \approx 0.74$, which is generally a constant with time but has a weak dependence on the density profile. In the above equation, the radius of the self-similar blast wave is always related to the total energy and the average density ($\bar{\rho}(r_s)$) inside the shock radius even if the total energy and the density profiles are time and/or radius dependent \citep{Ostriker1988}. For a given power-law density profile, $\rho(r) = \rho_0\:(r/r_0)^{-\alpha}$, the shock radius and velocity are then written as 
\begin{eqnarray}
    r_s(t) &=& \left[ \frac{3-\alpha}{3} \frac{\xi E_b}{\rho_0 r_0^\alpha}\right]^{1/(5-\alpha)}\: t^{2/(5-\alpha)}, \quad\quad \mbox{    and    } \nonumber \\
    v_s(t) &=& \frac{2}{5-\alpha}\, \frac{r_s}{t} \quad \nonumber \\
    &=& \frac{2}{5-\alpha}\,\left[ \frac{3-\alpha}{3} \frac{\xi E_b}{\rho_0 r_0^\alpha}\right]^{1/(5-\alpha)}\: t^{(\alpha-3)/(5-\alpha)} \,,
    \label{chap5:eq:r_s}
\end{eqnarray}
implying $\eta = 2/(5-\alpha)$. The above equations boil down to the solutions noted earlier in equation \ref{chap1:eq:rs-agn}. The shock temperature is given as $T_s = \frac{3}{16} \frac{\mu m_p}{k_B}\: v_s^2$ for a strong shock ($\mu \approx 0.6$ is the mean molecular weight for a fully ionized plasma at Solar metallicity). For a weak shock (Mach $\lesssim 3$, as is the case for the FEBs), one has to consider the Mach number, $\mathcal{M}$, of the shock while calculating the temperature. The shock temperature is then given as 
\begin{equation}
    T_s = \left[\frac{1}{\gamma \chi \mathcal{M}^2} + \frac{\chi-1}{\chi^2}\right]\: \frac{\mu m_p}{k_B}\: v_s^2
    \label{chap5:eq:T_s}
\end{equation}
where, $\chi = \frac{\gamma+1}{(\gamma-1)+2/\mathcal{M}^2}$ is the density jump across the shock.

\begin{figure*}
    \centering
    \includegraphics[width=0.9\textwidth]{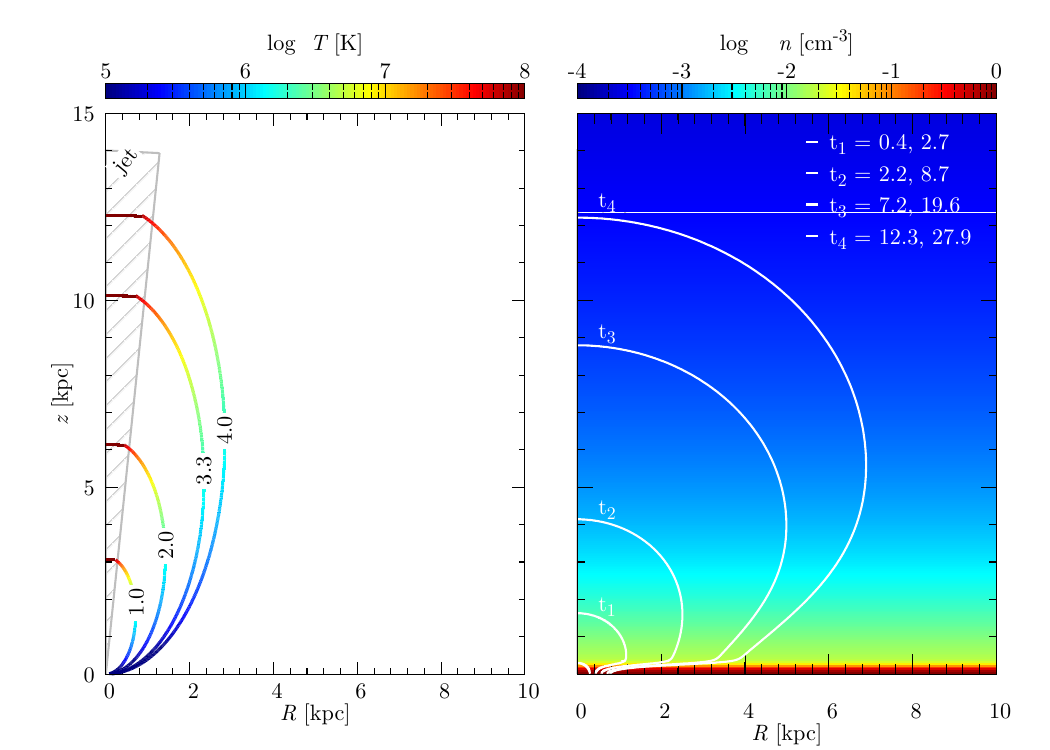}
    \caption{Analytically estimated propagation of shock waves for different energy injection methods. \textit{Left Panel}: Forward shock from a ballistic jet with $v_j = 0.01 c$ and opening angle $\theta_j = 4\deg$ \citep[see][for details]{Mondal2022}. The color of each line represents the shock temperature. \textit{Right Panel}: Shock wave from an energetic explosion ($E=10^{56}$ erg) and a continuous wind injection ($L_{\rm wind} = 10^{40.7}$ \ergps) at the GC. The solution is based on the Kompaneets approximation \citep{Kompaneets1960} of shock propagation on a given density profile (shown in color; $n \propto z^{-1.5}$). Each curve represents the shock surface at a given dimensionless time. The real times for the (burst, wind) cases are noted in the legend in units of Myr.}
    \label{chap5:fig:shock-propagation}
\end{figure*}

Shocks driven by a constant energy source (namely, winds) are also self-similar and can be represented by the equation \ref{chap5:eq:r_s} with the modification $E_b = L\:t$, where $L$ is the power of the energy source \citep{Castor1975, Weaver1977}. The solution for a power law density profile practically remains the same except that $\eta = 3/(5-\alpha)$ and $\sigma \approx 1.2$ \citep{Ostriker1988}. The higher value of $\sigma$ compared to a blast wave case is due to a higher fraction of thermal energy stored in wind-driven bubbles \citep{Weaver1977}. As noted in section \ref{chap1:subsec:cgm}, the inner CGM of the Milky Way is represented by a power law density profile with $\rho_0 \approx 1.3\times 10^{-2}$ \mpcc and $\alpha = 1.5$. Therefore, for the Milky Way, the radius and velocities of a blast wave and a wind-driven shock can be written down as 
\begin{eqnarray}
    \label{chap5:eq:rs_bw_dumb}
    &r_s& \approx 6.1\, \mbox{  kpc   }\, E_{56}^{2/7}\, t_{10}^{4/7}\,, \nonumber \\
    &v_s& \approx 344\,\mbox{  km s}^{-1}\,\, E_{56}^{2/7}\, t_{10}^{-3/7} \quad\quad \mbox{For blast-waves, and }\\
    \label{chap5:eq:rs_wind_dumb}
    &r_s& \approx 3.2 \mbox{ kpc }\, L_{41}^{2/7}\, t_{10}^{6/7}\,, \nonumber \\
    &v_s& \approx 273\,{\rm km s}^{-1}\, L_{41}^{2/7}\, t_{10}^{-1/7} \quad \mbox{ For wind-driven shocks}
\end{eqnarray}
where, $L = 10^{41} L_{41}$ \ergps is the mechanical luminosity of the wind. In practice, the shock will be a bit faster along the Galactic axis than the horizontal direction because of the slightly disk-like nature of the Milky-Way CGM in the inner $\sim 10$ kpc owing to the gravity of the stellar disk and the CGM rotation \citep{Hodges-Kluck2016}. Equations \ref{chap5:eq:rs_bw_dumb} and \ref{chap5:eq:rs_wind_dumb} only represent the typical location of the shock in each case.

The main difference between the blast wave-type and a wind-type solution is that the wind bubble consists of further structures, such as the reverse (internal) shock and the contact discontinuity. For a conical wind, the bubble can have two strong internal shocks, one, a collimation shock due to the pressure of the ambient medium that tries to collimate the wind to narrower geometry and, second, a reverse shock due to the high-velocity wind-catching up to the heavy and slowly moving shocked CGM shell. These shocks can easily reach Mach $\gtrsim 10$ and, therefore, can be ideal locations for accelerating cosmic ray (CR) particles other than accelerating the particles at the forward shock-front itself. On the other hand, a lack of such strong shocks inside a blast wave means that the CR particles either have to be accelerated at the forward shock or the GC and then diffused/advected to the current location. 

Solutions for BW in an axisymmetric media (such as planar atmospheres or axisymmetric CGM) have been obtained using the Kompaneets approximation \citep{Kompaneets1960} where the propagation of the shock wave is solved based on the local shock velocity directions \citep{Maciejewski1999, Olano2009, Irwin2019}. Figure \ref{chap5:fig:shock-propagation} shows a few such solutions for the (blast wave, wind) case. Note that the shock in the case of a burst reaches $z\approx 12$ kpc much faster than the wind case. This is, essentially, because the blast wave stores more kinetic energy than the wind bubbles, so the BW can propagate much faster. Therefore, even if an energy burst and a wind have the same shock locations, the shock velocity and hence the shock temperature are expected to be higher in the burst case. Therefore, observation of the temperature of the eROSITA bubbles can differentiate between these two scenarios. 
The figure also shows a simplified shock propagation in a `ballistic jet' scenario where the jet deposits a fraction of the energy per unit length at every height. This energy then expands horizontally to finally form the surface of the shock \citep[see][for details]{Mondal2022}. Although the left panel of the figure shows a simplistic shock surface for the ballistic jet, the actual structures of a jet-driven bubble are more complicated to obtain since it requires proper consideration of the collimation process due to the ambient medium \citep{Bromberg2011, Harrison2018}.

\subsection{Spectral origin}
\label{chap5:subsec:spectral-origin}
A successful theory for the spectral origin of the FBs has to also explain the \gray production mechanism, sharp edges, uniform surface brightness across the FBs, and the microwave haze.  
As mentioned, the most plausible ways to produce \grays in the FBs are the Hadronic ($p-p$) channel and the Leptonic channel (IC scattering of low energy photons).

\subsubsection{Hadronic channel}
\label{chap5:subsubsec:hadronic}
\begin{figure}
    \centering
    \includegraphics[width=0.45\textwidth]{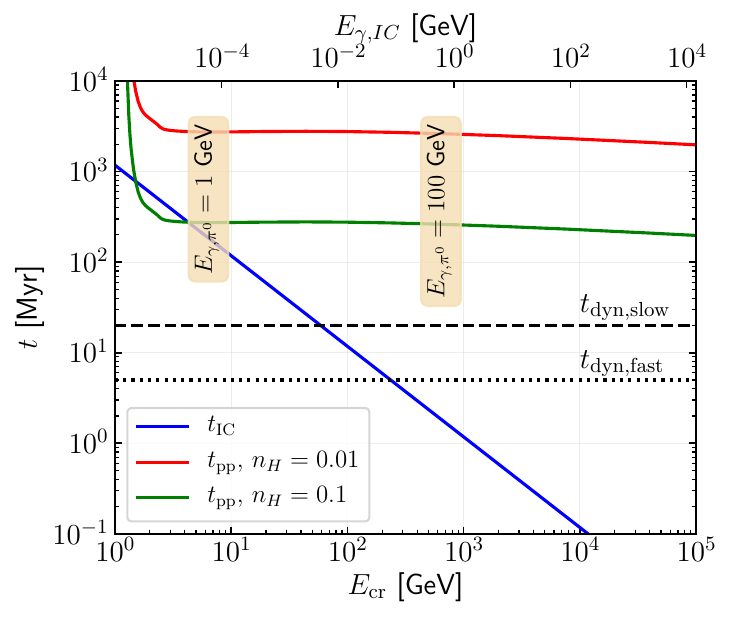}
    \caption{Left: Different time scales in the FB-spectrum as a function of CRe or CRp energy. The IC cooling time, $t_{\rm IC, cool}$, due to the CMB is shown using the blue solid line. The Hadronic ($p-p$) cooling times, $t_{pp}$, for $n_H = 0.01$ and $0.1$ \pcc are shown using the red and green solid lines, respectively. The high $t_{pp}$ at $E_{\rm cr}\lesssim 2$ GeV is because of ineffective inelastic collisions near the rest mass of the proton ($0.938$ GeV). Pion production below $\sim 2$ GeV happens instead via resonant processes \citep{Kafexhiou2014}. The top x-axis shows the energy of \gray photons produced (from CMB) in the IC process corresponding to the CRe energy in the lower x-axis. The vertical labels indicate the typical CRp energy for producing \gray photons at $1$ GeV and $100$ GeV via $\pi^0$ decay. The horizontal black lines represent the dynamical time scales corresponding to a slowly expanding and a fast expanding FB. }
    \label{chap5:fig:tcool}
\end{figure}

In the Hadronic channel, a CR proton (or heavier nuclei, together they are called the proton) collides with a gas phase proton and primarily produces pions which then further decay down to \grays and other particles \citep{Longair1981}. 
\begin{eqnarray}
    p + p &\rightarrow& \pi^+ + \pi^- + \pi^0 \nonumber \\
    \pi^+ &\rightarrow& \nu_\mu + \mu^+ \rightarrow \nu_\mu + e^+ + \nu_e + \bar{\nu}_\mu \nonumber \\
    \pi^- &\rightarrow& \bar{\nu}_\mu + \mu^- \rightarrow \bar{\nu}_\mu + e^- + \bar{\nu}_e + \nu_\mu \nonumber \\
    \pi^0 &\rightarrow& 2\:\gamma 
    \label{chap5:eq:pp-chain}
\end{eqnarray}
The time scale for the production of the \grays and other particles in this channel is dominated by the $p-p$ collision times, $t_{pp} \sim 1/(n_H\:\sigma_{pp}\:c)$, where, $n_H$ is the gas density and $\sigma_{pp}$ is the $p-p$ scattering cross-section. The $p-p$ time scale is shown in figure \ref{chap5:fig:tcool} as a function of CRp energy where I assumed the value of $\sigma_{pp}$ from \cite{Kafexhiou2014} and $n_H = 0.01, 0.1$ \pcc. Therefore, for a Hadronic emission from low-density ($\sim 0.01$ \pcc) gas, the bubbles have to be $\lesssim 3$ Gyr old (unless constantly replenishing the CR population) \citep{Crocker&Aharonian2011, Crocker2012}. The age of the bubbles can be much lower ($\lesssim 300$ Myr) if the $p-p$ interaction occurs in denser clouds ($\sim 0.1$ \pcc), probably condensed out of the low-density medium, inside the bubbles or at the contact discontinuity \citep{deBoer2014, Crocker2015, Mou2015}. 

For a given CRp with energy, $E_p$, a typical \gray photon produced in the $p-p$ channel has an energy of $\sim E_p/6$ (the CRp energy is divided almost equally into $\pi^+$, $\pi^-$, and $\pi^0$ and then $\pi^0$ decays into two equal energy photons). 
For a given CR energy spectrum, $N(E)$ (number density per unit volume, per unit energy), the \gray spectrum (volume emissivity) can be calculated following 
\begin{equation}
    E_\gamma^2\: \frac{dN_\gamma}{dE_\gamma} = E_\gamma^2 \int \frac{d\sigma_\gamma(E_\gamma, T_p)}{dE_\gamma}\: n_H\: c\: N(E_p)\: dT_p
    \label{chap5:eq:hadronic-spec}
\end{equation}
where, $E_p$ is the energy of the CRp, $T_p = E_p - m_p c^2$ is the kinetic energy of the CRp, $\frac{d\sigma_\gamma(E_\gamma, T_p)}{dE_\gamma}$ is the differential cross section for a proton with kinetic energy, $T_p$, to create a \gray photon of energy $E_\gamma$. Figure \ref{chap1:fig:FB-spectra} shows an example of such a \gray spectrum (gray dashed line; arbitrarily scaled), calculated for an assumed CR spectrum, $N(E) \propto E^{-2.1} \exp{[-E/(2\mbox{ TeV})]}$. Here we consider the approximate differential cross-sections given in \cite{Kafexhiou2014}. The figure shows that the \gray spectrum follows the CR spectrum in the range of $\sim 1-100$ GeV, as expected. The \gray spectrum then falls off due to the exponential cut-off of the CR spectrum. The lower \gray production at $E_p \lesssim 2$ GeV is expected due to the inefficiency of the inelastic collision as $E_p$ reaches the rest mass energy of the proton. At this energy, the \gray production is primarily via resonant processes \citep{Kafexhiou2014} and through IC scattering of the secondary leptonic particles (equation \ref{chap5:eq:pp-chain}) not included in the above equation. Although the $p-p$ channel can successfully explain the \gray spectrum from the FBs, the synchrotron emission from the secondary leptonic particles fails to explain the spectrum or the intensity of the microwave haze emission \citep{Ackerman2014, Crocker2015, Cheng2015a}. A complete spectral model of the FBs using the Hadronic emission, therefore, requires another leptonic component, probably CRe, that has a similar spectrum as the CRp. 

The origin of the CR inside the bubbles remains another aspect of the question. As discussed in section \ref{chap1:subsubsec:FB-spectrum} (also figure \ref{chap1:fig:FB-spectra}), the observed \gray spectrum requires a CRp spectral index of $p\approx 2.1$. It is understood that high energy CR can be injected into the bubbles from the Galactic center as well as accelerated at fluid shocks \citep{Fujita2013, Thoudam2013, Taylor2017}. These models explore the idea that the FB edges represent a strong fluid shock where CRe are accelerated and then diffused inwards to fill the bubble. The sharp edges of the shocks can naturally explain the sharp edges of the FBs. One, however, requires a strong shock with Mach, $\mathcal{M}\gtrsim 5$ to produce CR with a spectral index of $p\approx 2.1$, following a diffusive shock acceleration mechanism where ($p \approx (\chi+2)/(\chi-1)$, with $\chi$ as the density compression ratio; \citealt{Bell1978, Drury1983, Blandford1987, Fujita2013, Fujita2014}). Such strong shocks would imply a shock velocity, $v_s \gtrsim 10^3$ \kmps and a shock temperature, $T_s \gtrsim 1.2$ keV, much higher than the temperature observed in the eROSITA bubbles \citep{Kataoka2013} or inferred from FB-edges \citep[][also see section \ref{chap2:subsec:NPS}]{Keshet2018}. 

A tale-tell sign of the Hadronic processes would be neutrinos from the FBs. Initial estimates show that neutrinos from the FBs should be detectable at $E\gtrsim 30$ TeV where the atmospheric CR events are less dominant \citep{Lunardini2012, Lunardini2014, KM3NET2013}. Using the ANTARES detector, \cite{Adrian-Martinez2014} found no statistically significant neutrino emission toward the FBs. Neutrino fluxes from IceCube also remain inconclusive due to the emission from larger scales ($\sim 100$ kpc) in the Galactic CGM and one would require a factor of 10 more counts to obtain a reasonable signal from the FBs region \citep{Taylor2014, Sherf2017}. High energy neutrinos ($\sim$ PeV) were detected by IceCube which indicates a CRp spectrum with a cut-off of $E_{\rm cut} \sim 3$ PeV, inconsistent with the FB \gray spectrum \citep{Razzaque2018}. Additional confusion with the Galactic disk at low latitudes cannot be completely ignored in their result. 
A solid detection of the secondary particles from the FBs is still awaited and would provide us with a great amount of information regarding the particle processes at the FBs.

It is to be noted that although the Hadronic channel is successful in reproducing the spectral shape of the \gray emission, it fails to reproduce the observed intensity due to the low gas density inside the FBs. \cite{Crocker2015} bypassed this issue by assuming condensation of denser clouds from the low-density bubbles, whereas, \cite{Mou2015} achieved the gamma-ray intensity due to a thicker contact discontinuity. However, simulations including radiative cooling and realistic CGM find that neither the contact discontinuity nor the condensed clouds have enough mass to produce sufficient \gray intensity \citep{Sarkar2015b}, therefore, ruling out the hadronic channel.

\subsubsection{Leptonic channel}
\label{chap5:subsubsec:leptonic}
The leptonic channel of \gray production is via IC scattering of high energy CR electrons (CRe) with low energy background photons, such as the CMB, and Galactic star-light/IR emission. A low-energy photon can be scattered to higher energies provided the photon energy $h\nu\ll m_e c^2$ such that classical Thompson scattering is applicable in the rest frame of the electron. Close to this limit ($h\nu \sim m_e c^2$), relativistic corrections need to be considered and the IC scattering cross-section decreases significantly \citep{Klein1929, Rybicki1986}. In the classical limit, the upscattering follows the $1:\Gamma:\Gamma^2$ rule such that the required energy for an electron to upscatter a photon of energy, $E_b$, to energy $E_\gamma$ is
\begin{equation}
    E_e = m_e c^2 \sqrt{\frac{E_\gamma}{E_{b}}} = 16 \left(\frac{E_\gamma}{\rm GeV}\right)^{1/2}\: \left( \frac{E_b}{eV}\right)^{-1/2} \quad\quad \mbox{GeV}\,.
\end{equation}
Therefore, the required energies to produce \gray photon in the range of $1-100$ GeV are $0.5-5$ TeV for CMB photons ($E_b \sim 10^{-3}$ eV), $50-500$ GeV for IR photons ($E_b \sim 0.1$ eV), and $16-160$ GeV for star light ($E_b \sim 1$ eV). 
The CRe also lose energy due to the IC scattering with the low-energy photons such that the IC cooling time is
\begin{equation}
    t_{\rm IC, cool} \sim \frac{3 m_e c}{4\sigma_T U_{\rm ph} \Gamma} \approx 0.6 \mbox{ Myr  } \left( \frac{\mbox{eV cm}^{-3}}{U_{\rm ph}}\right)\: \left( \frac{10^6}{\Gamma}\right)
    \label{chap5:eq:t_IC_cool}
\end{equation}
where, $U_{\rm ph}$ is the energy density of the low-energy photons. For black-body radiation, $U_{\rm ph} = 0.26\: (T/T_{\rm cmb})^4$ eV cm$^{-3}$. Figure \ref{chap5:fig:tcool} shows the IC cooling time due to the CMB. One important aspect in the context of the FBs is that $t_{\rm IC, cool} \sim 1$ Myr for electrons that produce $\sim 1$ GeV \grays. The short cooling time means that either the FBs are younger than $\sim$ Myr or that the electrons are constantly replenished to maintain a steady source for the \grays.

\begin{figure}
    \centering
    \includegraphics[width=0.45\textwidth]{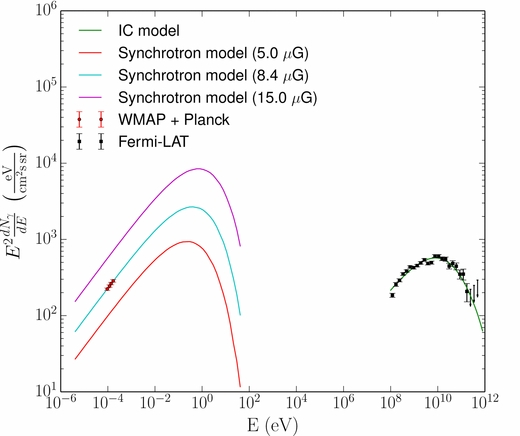}
    \caption{A combined fit of the \gray and microwave emission for a given CRe population, $N \propto \Gamma^{-2.2} \exp(-\Gamma/2.5\times 10^6)$. Different colors show the synchrotron emission for different magnetic field values. Image is reprinted from \cite{Ackerman2014} with authors' permission (copyright by AAS).}
    \label{chap5:fig:combined-spectrum}
\end{figure}

The spectrum (volume emissivity) for the IC scattering of CRe is written as \citep{Rybicki1986}
\begin{equation}
E_\gamma^2 \frac{dN_\gamma}{dE_\gamma} = \frac{3 c \sigma_T}{4}\, \int d\epsilon \: \left(\frac{E_\gamma}{\epsilon}\right)\: \mathbb{N}(\epsilon) \, \int_{\Gamma_1}^{\Gamma_2} d\Gamma \frac{N(\Gamma)}{\Gamma^2} f\left(\frac{E_\gamma}{4\Gamma^2 \epsilon} \right) 
\label{chap5:eq:IC_spectrum}
\end{equation}
where, $\mathbb{N}(\epsilon)$ is the photon density. For a black-body of temperature, $T$, the photon density is 
\begin{equation}
\mathbb{N}(\epsilon) = \frac{8\pi}{h^3 c^3}\, \frac{\epsilon^2}{\exp(\epsilon/k_B T)-1}\,,
\end{equation}
$N(\Gamma)$ is the number density of CR electrons with Lorentz factor, $\Gamma$, and the function, $f(x) = 2x\: \log(x) + 1+x-2 x^2$. A full spectrum (arbitrarily scaled) is shown in figure \ref{chap1:fig:FB-spectra} (black solid line) for an electron spectrum of $N(\Gamma) \propto \Gamma^{-2.1}$ at $\Gamma<10^6$ and $\propto \Gamma^{-3} \exp(-(\Gamma/10^7)^2)$ at $\Gamma > 10^6$ interacting with CMB photons. The CMB requires $E_e\sim$ TeV ($\Gamma\sim 10^6$) to produce $\sim$ GeV photons. The IC cooling time scale at this energy is $t_{\rm IC, cool} \sim 2$ Myr (figure \ref{chap5:fig:tcool}), indicating that the CRe population suffers from a cooling break at $\Gamma \sim 10^6$ which justifies a steeper CRe spectral index ($\Delta p \approx 1$) above $\Gamma >10^6$. 
A general fit assuming a CRe population containing a simple power law and an exponential cutoff, i.e. $N \propto \Gamma^{-2.2} \exp(-\Gamma/2.5\times 10^6)$, has also been shown to fit well with the FB spectra \citep{Su2010, Ackerman2014} indicating that the steepening of the \gray spectrum could simply be due to the lack of CRe acceleration at $E_e \gtrsim$ TeV. The \gray spectrum has also been explained based on IC contributions from CMB + IR + optical light where the $1$ GeV break is simply where the IR and starlight photons take over \citep{Yang2017}. One big advantage of the leptonic origin of the \gray is that the same CRe population can also explain the spectrum and intensity of the microwave haze \citep{Su2010, PlanckCollaboration2013, Ackerman2014}. One only requires a magnetic field of $B \sim 5-10 \mu$G in the FBs (figure \ref{chap5:fig:combined-spectrum}).

The age of the FBs is crucially dependent on the cooling time of the CRe and its origin. One simple solution is that the CRe are injected by an AGN-jet activity at the base of the FBs from where it advects and diffuses throughout the FB. The short cooling time of the CRe requires the age of the FBs, $t_{\rm fb} \lesssim t_{\rm IC, cool} \approx 2$ Myr \citep{Guo2012a, Guo2012, Yang2012}. Additionally, CRe can also be accelerated at the forward shock which can provide a high CRe density in a shell such that the projected \gray surface brightness appears to be flat across the FBs and latitude independent \citep{Yang2017}. As usual, the CRe are confined by the shock that generates them and guarantees a sharp edge of the FBs. In all these cases, the age of the FBs remains to be $\lesssim 2$ Myr, and therefore, requiring a Mach $\gtrsim 5$ flow which is in contradiction to the estimated temperature of the NPS/eROSITA bubbles. The age constraint can be overcome if the CRe are efficiently accelerated \textit{in-situ}  inside the FBs due to internal shocks or turbulence \citep{Su2010, Mertsch2011, Cheng2011, Cheng2012, Petrosian2012, Lacki2014, Cheng2015b, Sasaki2015, Sarkar2015b, Mertsch2019}. Depending on the type of outflow driving mechanism, the age of the bubble can be in the range of $\sim 8-30$ Myr. In these models, the FBs represent either the reverse shock or the contact discontinuity where the CR can remain confined due to the presence of a draped magnetic field outside the bubbles \citep{Dursi2008, Ruszkowski2008} thus generating a sharp boundary for the \grays. Interested readers are encouraged to consult \cite{Ruszkowski2023} for a detailed description of CR physics.

Although the \textit{in-situ} acceleration mechanism appears promising in terms of the cooling time, the existing models suffer from limitations in producing a flat surface brightness for the entire bubble. Additionally, it is unclear if internal shocks in the outflow can produce CRe with energies are high as a $\sim$ TeV. A complete model with a proper \textit{in-situ} CRe acceleration mechanism and its interaction with the magnetic field, such as synchrotron cooling and magnetic confinement would provide a definitive answer.

\begin{figure*}
    \centering
    \includegraphics[width=\textwidth, clip=true, trim={2cm 1cm 0cm 1cm}]{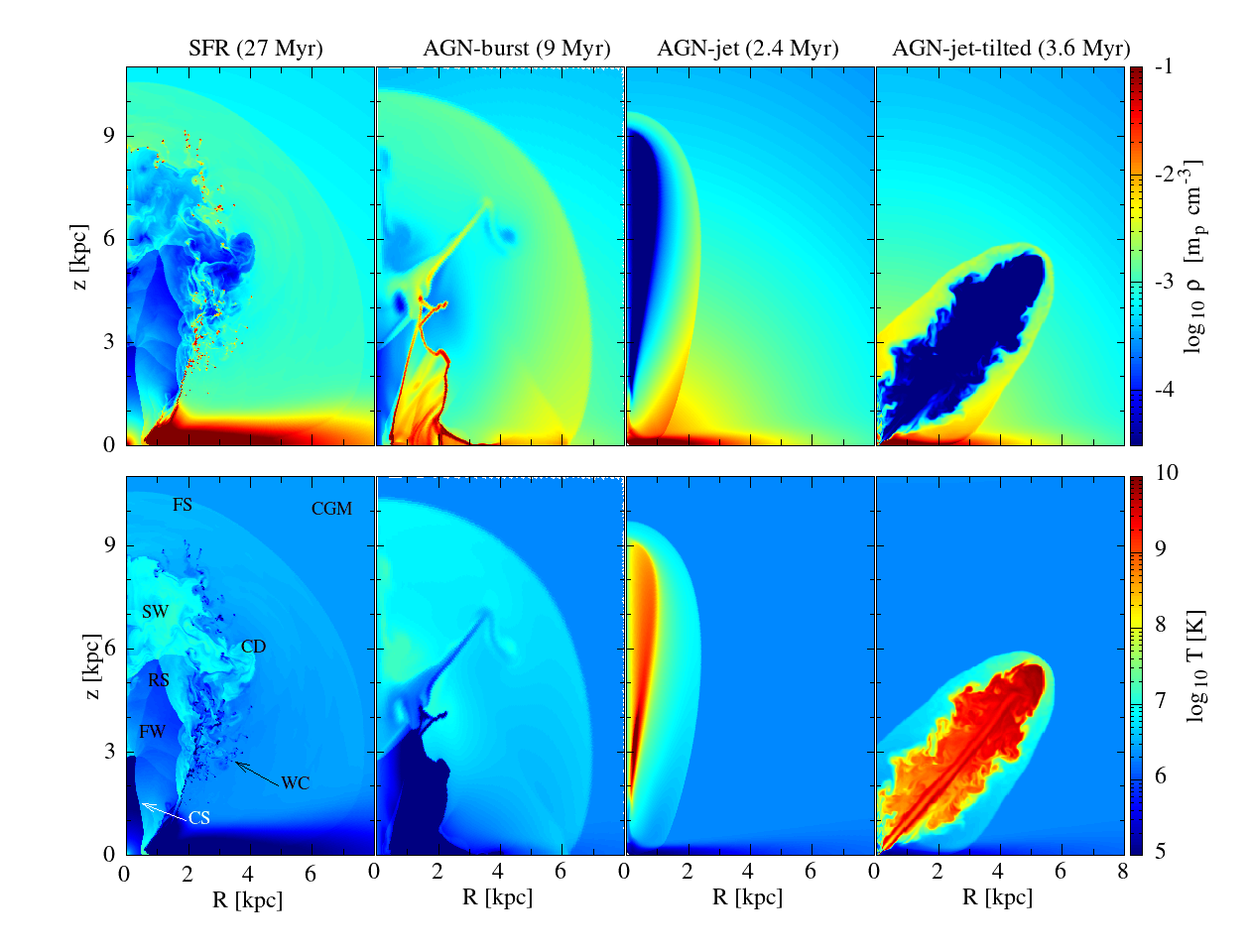}
    \caption{Density (top row) and temperature (bottom row) color maps of hydrodynamical simulations of Fermi/eROSITA bubbles for different scenarios. First column: a star-formation-driven wind with SFR $= 0.5$ \mpy ($L_{\rm wind} = 5\times 10^{40}$ \ergps) at $t=27$ Myr \citep{Sarkar2015b}. Different regions of the wind are marked as - CGM: circumgalactic medium, FS: forward shock, CD: contact discontinuity, SW: shocked wind, RS: reverse shock, FW: free wind, CS: collimation shock, and WC: warm clouds. This scenario assumes that the FBs are represented by the contact discontinuity and the eROSITA bubbles (NPS) are represented by the shocked CGM (material between FS and CD). Second column: an AGN burst scenario where a total of $10^{56}$ erg thermal energy is deposited within $40$ pc during the first $t_{\rm inj} = 0.01$ Myr \citep{Mondal2022}. The snapshot is taken at $9$ Myr. In this scenario, the eROSITA bubbles are represented by the shock, and the FBs are represented by the lower-density bubble inside the shock \citep{Yang2022}. Third column: an AGN jet scenario, where a jet with $L_{\rm jet} = 2\times 10^{43}$ \ergps is turned on for $t_{\rm inj} = 0.04$ Myr ($E_{\rm tot} = 2.5\times 10^{55}$ erg) \citep{Mondal2022}. This scenario explains the FBs as the forward shock with $\mathcal{M} \sim 4-5$ and assumes that the eROSITA bubbles originated from a separate event unrelated to the FBs. Fourth Column: a misaligned AGN jet scenario where the jet with $L_{\rm jet} = 2\times 10^{41}$ \ergps is continuously injected at an angle of $45\deg$ from the Galactic rotation axis \citep{Sarkar2023}. The snapshot is taken at $t = 3.6$ Myr, i.e. $E_{\rm tot} = 2.3\times 10^{55}$ erg. The panel shows that misaligned jets do not naturally produce axisymmetric and hemispherical symmetric bubbles unless the jets are dissipated within the ISM. Such dissipation could occur either due to interactions with molecular clumps \citep{DMukherjee2018} or via magnetic kink-instability \citep{Tchekhovskoy2016}.}
    \label{chap5:fig:HD-simulations}
\end{figure*}

\subsection{AGN driven models} 
\label{chap5:subsec:AGN-models}
Energy from the central SMBH is an obvious source of the mechanical energy for the Fermi/eROSITA bubbles. The required SMBH power can vary by many orders of magnitude leaving different signatures on the resultant bubbles. Below I summarize different types of AGN-driven models that have been discussed in the context of the Fermi/eROSITA bubbles.

\subsubsection{AGN jet-driven}
\label{chap5:subsubsec:AGN-jet-bubbles}
The nature of the output mechanical energy from an SMBH depends on the accretion rate, $\dot{M}_{\rm acc}$. If $\dot{M}_{\rm acc}/\dot{M}_{\rm edd} \lesssim 10^{-3}$, then the density in the accretion disk around the SMBH is low enough to form a non-radiative accretion flow which can lead to formation of collimated outflows such as jets \citep{Narayan1995, Yuan2012, Yuan2014, Guistini2019}. Additionally, relativistic jets can also be produced in the presence of a rotating black hole where the accreted gas extracts the rotational energy of the BH \citep{Blandford1977, Tchekhovskoy2011, Kwan2023}. 

Jet-driven bubbles have been simulated in the context of the FBs where the jet is injected for a short period of time ($\sim 0.1-0.5$ Myr) with jet power, $L_{\rm jet} \sim (1-3)\times 10^{44}$ \ergps i.e., $0.2-0.5 L_{\rm edd}$ and total energy $\sim 10^{57}$ erg \citep{Guo2012, Guo2012a, Yang2012, Yang2013, Yang2017, Yang2022}. The FBs, in these models, are represented by the contact discontinuity at an age of $t_{\rm FB} \sim 1-6$. Although the contact discontinuity is expected to suffer from the shear-driven Kelvin-Helmholtz instability (KHI), it is found that the presence of CR diffusion and fluid viscosity can suppress such instabilities and produce the sharp edge for the FBs \citep{Guo2012a}. For younger bubbles ($\sim 1$ Myr), the KHI time scale is longer than the age of the bubbles so that the bubble edges remain sharp. Additionally, the magnetic draping of the ambient magnetic fields can suppress such instabilities as well as CR diffusion across the bubbles, and thereby, maintain a smooth and sharp edge \citep{Yang2012}.
\cite{Yang2013} found that the decline of the Haze emission at $|b| \gtrsim 35\deg$ can be explained if a second jet from \sgra is present. Several authors also modeled the FBs assuming that the edges of the FBs represent a forward shock of Mach$\sim 4-5$ similar to the X-ray analysis of \cite{Keshet2018}. 
\cite{Zhang2020, Mondal2022} performed numerical simulations of jets that last for $0.01-1$ Myr and have energy $\sim 3\times 10^{55}-3\times 10^{56}$ erg, i.e. jet power $\sim 10^{42-45}$ \ergps. In all the jet-driven scenarios, the estimated age of the FBs is $\sim 1-6$ Myr. These age estimations coincide with the age of the nuclear young star cluster ($6\pm2$ Myr) which indicates the possibility that the FBs were produced by the same accretion event.

While the jet-driven models are successful in reproducing the shape and spectrum of the FBs (due to their young age), they suffer from a few drawbacks.
\begin{itemize}
    \item  The jets are expected only in sub-Eddington accretion rate ($\dot{M}_{\rm acc}/\dot{M}_{\rm edd} \lesssim 10^{-3}$) and therefore, are not expected have very high jet power as assumed.
    
    \item The young FBs also produce a shock that is very hot ($\gtrsim 1$ keV) which is in contradiction with the observed temperature of the NPS ($\sim 0.3$ keV; section \ref{chap2:subsec:NPS}). One possible solution could be that the electrons and protons do not have enough time to be in equilibrium, so the observed temperature is much smaller than the actual shock temperature in such fast shocks. The electron-proton time scale, given as the coulomb interaction time scale \citep{Berginskii1965},
    \begin{equation}
    t_{\rm ep} \sim 2 \mbox{ Myr } \left( \frac{T_s}{4\times 10^6}\right)^{3/2}\: \left(\frac{10^{-3}}{n} \right) 
    \end{equation}
    is indeed comparable to the dynamical time of the FBs in the AGN jet scenario. In practice, this equilibrium time scale for $M$ is much lower than the above value due to the initial temperature of the electron ($\sim 0.2$ keV for the CGM; \citealt{Miller2015}) and heating due to shock compression \citep{Vink2015}. \cite{Sarkar2023} showed that for the Milky Way CGM profile, $t_{\rm ep} \sim 1$ Myr and it is independent of the shock Mach number. Any bubbles older than $\sim 1$ Myr should not suffer from $e-p$ non-equilibrium. Even in the cases, where $t_{\rm dyn} \sim t_{\rm eq}$, the electron temperature can reach up to $\sim 70\%$ of the equilibrium value. Therefore, the high Mach number as found in the jet models may be in contradiction with the NPS observations. In the cases, where the FBs are assumed to be forward shocks \citep{Zhang2020, Mondal2022} the temperature constraint is more relaxed due to an estimated higher temperature of the shock ($\sim 0.4$ keV; \citealt{Keshet2018}).

    \item The jets are assumed to be perpendicular to the Galaxy plane. As we see in section \ref{chap4:subsec:GC-structures}, structures such as the circum-nuclear disk ($\sim 1.5-4$ pc), nuclear young stellar cluster ($\sim 0.5$ pc) and the EHT accretion disk ($\sim 10^{-6}$ pc) are all significantly misaligned ($\sim 20\deg-80\deg$) with the Galactic rotation axis \citep{Genzel2010}. 
    Therefore, if there was indeed a jet from \sgra in the past, the jet would follow the angular momentum of the accretion disk, i.e. the gas surrounding it, and would be significantly misaligned with the Galactic rotation axis. The most convincing evidence of the jets at the GC is probably a number of observed radio streaks near the GC \citep{Yusef-Zadeh2020, Yusef-Zadeh2023}. These `jets' also are almost aligned with the Galactic plane instead of pointing along the Galactic rotation axis. Even from the galaxy formation point of view, the angular momentum of the gas at $\sim 1$ pc from the SMBH can be at $\sim 0\deg-60\deg$ angle with the large scale ($\sim$ kpc) gas flows in a galaxy \citep{Angles-Alcazar2021}. 
    Therefore, although it can be argued that a past jet could be pointing toward the Galaxy rotation axis, the probability is very low. A misaligned jet in the Milky Way would maintain its initial jet direction and would not be able to produce the axisymmetry and hemispherical symmetry of the FEBs \citep{Sarkar2023}. One possible solution could be to dissipate the jet energy inside the ISM such that the subsequent evolution of the bubbles is then determined by the density profile of the ISM and CGM. 
    Possible mechanisms of jet dissipation are i) magnetic kink instability \citep{Tchekhovskoy2016}, or ii) interaction with dense clouds \citep{DMukherjee2018, Cecil2021}. However, for such dissipation to happen within the ISM, the jets have to be weak. For magnetic kink-instability to occur within the ISM (scale height, $H$ pc), one must have $L_{\rm jet} < L_{\rm MKI} = 4\times 10^{41}$ \ergps $n_{\rm ism}\: H_{200pc}^2$, where $n_{\rm ism}$ is the ISM particle density \citep{Tchekhovskoy2016, Sarkar2023}. The criteria for jet dissipation via interaction with molecular clouds is uncertain and most probably depends on the local density inhomogeneity and the jet power. Recent study on the exact criteria of jet dissipation found that MW ISM is not clumpy enough for a jet to be dissipated \citep{RDutta2024}. Therefore, re-orientation of the bubbles along the Galactic rotation axis is not possible via jet-clump interaction in our MW. However, there remains a possibility if the jet directly hits the CMZ ring. In that case, the jet could be dissipated and produce the symmetric bubbles Fermi/eROSITA bubbles \citep{RDutta2024}. Additionally, if the accretion disk itself if precessing fast enough to distribute the jet energy isotropically, the symmetric bubbles could be achieved. The detailed modeling of these scenarios are left for future exploration.
    Very short duration jet bursts could in principle still produce a symmetric set of bubbles about the Galactic plane provided that the jet power is $\gtrsim 5\times 10^{44}$ \ergps $\approx L_{\rm edd}$, i.e. the jet duration of $\lesssim 0.01$ Myr (since the total energy of the FEBs is $10^{56}$ erg). Such short-duration jets would choke before they can escape the ISM (scale height $\sim 200$ pc), therefore, effectively creating a blast-wave-like evolution of the forward shock \citep{Yang2022, Sarkar2023}. The problem of the high shock temperature still remains for such blast wave bubbles since they produce a shock temperature (for the NPS) to be $\approx 0.4$ keV compared to the observed $\approx 0.3$ keV \citep{Kataoka2013, Miller&Bregman2016}.
\end{itemize}

\subsubsection{AGN wind-driven}
\label{chap5:subsubsec:AGN-wind-bubbles}
In highly accreting black holes ($\dot{M}_{\rm acc}/\dot{M}_{\rm edd} \gtrsim 10^{-2}$), the output energy is mostly in the form of radiation \citep{Churazov2005, Guistini2019} which can further couple with the gas via Compton scattering to drive a wide-angle outflow from the SMBH \citep{King&Pounds2003, Tombesi2013}. Alternatively,  wide-angle winds can also be generated due to magnetic field-mediated centrifugal force or simple buoyancy in the accretion disk in such highly accreting systems \citep{Yuan2012, YFJiang2014, YFJiang2019b}. The advantage of the wind-driving is that the wind is wide-angled compared to the jets that are highly collimated ($\theta_{\rm jet} \lesssim 10\deg$), and therefore, the winds can easily couple with the ISM. The flow of the wind can then be easily modified by the CMZ or the large-scale Galactic ISM to produce bubbles/shocks that are symmetrical about the Galactic plane and the rotation axis. 

\cite{Zubovas2011, Zubovas2012} considered such quasi-spherical outflows from \sgra at $\sim 6$ Myr ago, with a wind power , $L_{\rm wind} \approx 2.5\times 10^{43}$ \ergps lasting for $\sim 0.05-1$ Myr. The FBs were explained as the low-density bubbles behind the forward shock. The FBs can also be explained by lower power winds ($L_{\rm wind} \sim 10^{41-42}$ \ergps; \citealt{Mou2014, Mou2015, Mou2023}) where the winds are continuously blown out from the GC for $t_{\rm fb} \sim 7-15$ Myr and just stopped $\sim 100$ yr ago. The assumed wind power is consistent with the X-ray reflection nebulae constraints (section \ref{chap4:subsec:MWBH-acc-rate}; \citealt{Kayoma1996, Murakami2001, Inui2009}). 
The obtained \gray surface brightness produces slightly edge-brightened FBs but is consistent with flat surface brightness \citep{Mou2015}. The shock temperature for the low-power bubbles (at least for the $L_{\rm wind} \sim 10^{41}$ \ergps case) are also consistent with the NPS temperature as well as the \ion{O}{viii}/\ion{O}{vii} intensity ratios outside the NPs \citep{Sarkar2017}. More recently, \cite{Fujita2023} showed that wind-driven bubbles with power, $L_{\rm wind} \sim 10^{42}$ \ergps) are also consistent with the estimated non-equilibrium temperature ($\sim 0.7$ keV) from the \textit{Suzaku} telescope \citep{Yamamoto2022}. However, note that the estimation of the temperature in the assumed non-equilibrium plasma is still unclear. See section \ref{chap2:subsec:NPS}.

\subsubsection{TDEs}
\label{chap5:AGN-TDE-bubbles}
Apart from the accretion of typical ISM gas, SMBHs can also produce energy via accreting stars through tidal disruption events (TDEs; \citealt{Hills1975}). Typically, about half of the tidally disrupted stellar material falls onto the SMBH and the rest can remain unbound to form a quasi-spherical outflow. The bound material forms an accretion disk around the SMBH which can produce further winds as well as collimated jets \citep{Rees1988, Piran2015}. The typical time-scale, luminosity, and total energy of such an event is $t_{\rm tde} \sim$ yr, $L_{\rm tde} \sim L_{\rm edd}$, and $E_{\rm tde} \sim 10^{51} M_{{\rm bh},6}^{2/3}$ erg, where $M_{{\rm bh},6} = M_{\rm bh}/10^6$ \msun \citep{Rees1988, Piran2015, Stone2016}. Although the observed rate of TDEs in the local universe is $\sim 1-6\times 10^{-5}$ yr$^{-1}$ galaxy$^{-1}$ \citep{Donley2002, Gezari2008, Yao2023}, the expected rate for MW type SMBH is $\sim 0.3-4 \times 10^{-4}$ yr$^{-1}$ galaxy$^{-1}$ \citep{Magorrian1999, JWang2004, Gezari2008, Stone2016}. Particularly for the \sgra, the TDE rate is estimated to be $\approx 0.7-2.5 \times 10^{-4}$ yr$^{-1}$ \citep{Generozov2018}. Therefore, a MW type SMBH ($M_{\rm bh} \approx 4\times 10^6$ \msun) is likely to produce a mechanical power of $\sim 10^{39.5-40.5}$ \ergps which is slightly lower than but close to the required $\sim 10^{40.5-41}$ \ergps to explain the X-ray spectrum or \ion{O}{viii}/\ion{O}{vii} line ratios of the FEBs (section \ref{chap2:subsec:NPS}). 

Formation of the FBs by TDEs is considered by some authors \citep{Cheng2011, Ko2020} who showed that this energy injection mechanism also could explain the sizes and the shape of the FBs. The required TDE-driven power was estimated to be $\sim 10^{41}$ \ergps based on the assumption of $E_{\rm tde} \sim 3\times 10^{52}$ erg (applicable for highly penetrating TDEs; \citealt{Alexander2005}). The estimated age of the FBs is then $t_{\rm FB} \sim 30$ Myr (to produce $E_{\rm FEB} \sim 10^{56}$ erg). On the other hand, one requires an enhanced TDE rate by a factor of $10$, maybe due to the different geometry of the NSC, in the past $\sim 20-30$ Myr for the typical TDE energy of $\sim 10^{51} M_{bh,6}^{2/3}$ erg to obtain the same power (Tsvi Piran, \textit{private communication}). It can be shown that the individual TDEs effectively behave as SNe-like explosions because the distance traveled by the TDE ejecta within its lifetime is only $z_{\rm tde} \sim (E_{\rm tde} t_{\rm tde}^2/\rho_{\rm ISM})^{1/5} \approx 1$ pc, which is similar to the thermalization scale of the SNe ejecta in the ISM. Additionally, it can also be shown that even if the TDE produces a jet, the jet would only propagate a distance of $z_{\rm tde} \sim v_{h,jet} t_{\rm tde} \sim 1$ pc, very similar to the simple wind-like outflow from the TDE  \citep{Sarkar2023}. Here, $v_{h,jet}$ is the velocity of shock at the jet head.

\subsubsection{AGN-burst}
\label{chap5:subsubsec:AGN-burst}
As noted in section \ref{chap5:subsubsec:AGN-jet-bubbles}, very short duration ($\lesssim 0.01$ Myr) and powerful ($L_{\rm jet} \gtrsim 5\times 10^{44}$ \ergps) jets in our Galaxy would effectively behave as a blast-wave. The solutions for such blast waves are discussed in \ref{chap5:subsec:wind-model}. Long before the discovery of the symmetric X-ray eROSITA bubbles, \cite{Sofue1977, Sofue1984, Sofue1994, Sofue2016} explored the idea of a `giant Galactic center explosion' to explain the radio and X-ray NPS\footnote{The reason for categorizing \citeauthor{Sofue1977}'s models in AGN-burst is because of his assumption of an instantaneous energy explosion. The energy from a starburst is released continuously over a period of $\approx 40$ Myr even if all the stars form instantaneously \citep{Leitherer1995, Leitherer1999}.}. The required energy and dynamical time for the `Bipolar Hyper-Shell' were found to be $\sim 3\times 10^{56}$ erg and $\sim 15$ Myr, very close to the currently accepted values for the eROSITA bubbles \citep{Kataoka2013, Predehl2020}. \cite{Sofue2016} further point out that such an energy and dynamical time would not be achievable by star-formation since the required SFR at the GC is $\sim 1$ \mpy for the last $\sim 15$ Myr, about $1/3$rd of the SFR in the whole Galaxy. The authors, therefore, argue for a case where the central SMBH is producing the energy. The shock temperature of such a high-energy explosion is, however, inconsistent with the \ion{O}{viii}/\ion{O}{ii} line ratio observations \citep{Sarkar2023}. It remains to be seen if any particular CGM density profile in combination with such an explosion is still consistent with the shock temperature data.

\subsection{Star formation driven}
\label{chap5:subsec:SFR-models}
As we saw in section \ref{chap4:subsec:SFR}, the star formation rate in the CMZ is estimated to be in the range of $0.07-0.8$ \mpy, which means that the wind power is $L_{\rm wind}\approx 10^{40-41}$ \ergps. The lower power of the wind means that the shock reaches $\sim 10$ kpc at $t_{\rm dyn} \sim 40-80$ Myr (equation \ref{chap5:eq:rs_wind_dumb})\footnote{Note that equation \ref{chap5:eq:rs_wind_dumb} should be modified in case of a weak shock. Therefore, even if the current version of the equation produces subsonic flow for the $L_{\rm wind}\sim 10^{40-41}$ \ergps case, the corrected equation produces a $\mathcal{M}\approx 1.5$ flow, as is seen in simulations}. Such a long time scale for the FEBs could be argued based on the following facts \citep{Crocker2015}.
\begin{itemize}
    \item The FBs and the $2.3$ GHz polarized bubbles seem to be tilted by $\sim 30\deg$ toward the west in both the southern and northern hemispheres \citep{Su2010, Carretti2013}. Such a tilt can be a result of a westward CGM wind due to the relative motion ($v_{\rm cgm-wind}\sim 50$ \kmps) of the MW in the local galaxy group medium \citep{VanderMarel2012}. Such a motion would tilt the expanding bubbles if the expansion/shock speed, $v_s \sim v_{\rm cgm-wind}/\tan(30\deg) \sim 80$ \kmps. The dynamical time in such a case is $t_{\rm dyn} \sim 110$ Myr \citep{Crocker2015}. More recently, \cite{Mou2023} simulated such a CGM wind and concluded that $v_{\rm cgm-wind} \approx 200$ \kmps and $t_{\rm dyn} \sim 20$ Myr is required to explain the tilted/asymmetric features of the NPS and the $2.3$ GHz emission. Nonetheless, the bending features indicate a longer time scale for the bubbles. A similar bending of the polarized radio ridges near the waist of the FBs also indicates a slow rise of these ridges over a time scale of $\gtrsim 10$ Myr \citep{Crocker2015}. Contributions from the orientation of the local magnetic fields cannot, however, be ruled out in determining the shape of the bubbles and may be important for explaining the tilt in a much younger bubble.
    \item A steep electron spectrum at $2.3-23$ GHz in the polarized emission \citep{Carretti2013} indicates a synchrotron cooling time (and hence a dynamical time) of $t_{\rm dyn} \sim 25$ Myr in the polarized bubbles (section \ref{chap2:subsec:Micro-Haze}).
\end{itemize}
Based on the above arguments, \cite{Crocker2014, Crocker2015} proposed a star-formation-driven model for the FBs where the bubbles are produced by a constant star formation at the GC. The model requires an SFR $\approx 0.1$ \mpy (consistent with the SFR estimates from source counting; section \ref{chap4:subsec:SFR}) over a time-scale of $t_{dyn} \sim 300$ Myr. The longer lifetime of the bubbles allows the low-density bubble ($n \sim 0.01$ \pcc) material to radiatively cool down ($t_{\rm rad, cool} \sim 500$ Myr) and condense into denser clouds such that the $p-p$ channel can produce the observed \gray emission.

\cite{Lacki2014} proposed that a SFR $\approx 0.1$ \mpy at the CMZ could inflate the FBs over $t_{\rm dyn} \sim 10$ Myr provided the thermalization efficiency of the SNe is $\approx 0.75$ (higher than the assumed value of $0.3$ in this review). The \gray emission is assumed to arise from the CR electrons freshly accelerated at the reverse/termination shock of the wind. However, one requires SFR$\sim 0.2$ \mpy to explain the \gray intensity in this model. The first hydrodynamical simulations of the star-formation-driven wind model for the FBs were performed by \cite{Sarkar2015b} who find that FBs could be explained if the SNe-driven wind power, $L_{\rm wind} \approx 5\times 10^{40}$ \ergps (SFR $\approx 0.5$ \mpy\footnote{In the paper, the luminosity to SFR conversion was assumed to be $L_{\rm wind} \approx 3\times10^{41}$ \ergps $SFR/$(\mpy), a factor of $2$ lower than the assumed conversion factor in this review. For the same $L_{\rm wind}$, equation \ref{chap1:eq:Edot-sfr} produces a SFR of $0.3$ \mpy}) over a time period of $t_{\rm dyn} \approx 27$ Myr (figure \ref{chap5:fig:HD-simulations}). The simulations obtain a forward shock velocity of $\sim 300$ \kmps ($\mathcal{M} \approx 1.5$) consistent with the estimated temperature of the eROSITA bubbles (back then, the NPS). The authors also noted that if the forward shock is assumed to be coincident with the NPS/Loop-I feature, the size and shape of the contact discontinuity (CD) automatically represent the size and shape of the FBs. 
The fact that \textit{the recovery of one of the features (either the Fermi Bubbles or the NPS/Loop-I) self-consistently reproduces the other feature strongly suggests that both the Fermi and eROSITA bubbles were generated by the same event}. The ratio of the sizes i.e. the ratio between the forward shock location and the CD location, therefore, is a strong constraint for models trying to reproduce the FEBs from a single event. The model also recovers the kinematics of the warm absorption clouds ($v_{\rm los} \sim \pm 200$ \kmps) as discussed in section \ref{chap2:subsubsec:warm-clouds} \citep[also][]{Fox2015, Bordoloi2017}. Because of the expansion through the ISM and then the CGM, the CD also contains a significant amount of the ISM gas as it expands. At the CD, some of the denser ISM gas directly condensates into smaller clouds while some of it mixes with the lower metallicity CGM gas and then condensates into clouds. 
The warm clouds form at the CD, therefore, can have a variety of metallicities ranging from the ISM to the CGM values (as observed by \cite{Ashley2022} and also significant non-radial velocities. The clouds are almost co-moving with the CD, i.e. they have a non-radial expansion, in addition to having local turbulent motion around the CD. The non-radial velocity and a local turbulent motion for the clouds are key to explaining the $v_{\rm los} \sim \pm 200$ \kmps even though the overall expansion speed is only $\sim 300$ \kmps. The \gray and microwave emission in this model is assumed to be from CR electrons (with $p\approx 2$) that are confined within the CD, therefore, producing a sharp edge for the FBs. The simulations also show that the hadronic processes only produce $\sim 1$\% of the observed \gray intensity, thereby, ruling out the hadronic scenario. Although the star-formation-driven wind model of \cite{Sarkar2015b} has been quite successful in explaining different features of the FEBs, there are certain drawbacks to the model.
\begin{itemize}
    \item The required star-formation rate ($\sim 0.5$ \mpy) seems to be a factor of a few higher than the observed rates by source counting ($\sim 0.07-0.1$ \mpy; section \ref{chap4:subsec:SFR}). The required SFR and the duration of the event are, however, in excellent agreement with the recent estimations of SFR $\sim 0.2-0.8$ \mpy over the last $\sim 30$ Myr \citep{Nogueras-Lara2019}. Given that the range of the SFR in \citeauthor{Nogueras-Lara2019}'s estimates is due to different stellar spectral models, the statistical uncertainty of this estimate is significant. At this point, it is still not clear what the true star formation rate at the CMZ is. 
    \item The simulations only solve hydrodynamical equations and the \gray and microwave emission is obtained based on \textit{ad-hoc} prescriptions pf the CR energy density and spectral index. A complete numerical simulation for the star-formation-driven model with accurate CR propagation is still awaited.
\end{itemize}  

\cite{Mertsch2019} solved the kinetic equations for the CR propagation in a star-formation-driven wind ($v_{\rm wind} \sim 300$ \kmps) and concluded that the \grays and microwave emissions, and the morphological features of the FBs can be explained by the IC and synchrotron emission from CR electrons that are accelerated \textit{in situ} in the turbulence. The total energy required to produce the FBs is estimated to be $\sim 6\times 10^{54}$ erg over a duration of $24$ Myr i.e., $L_{\rm wind} \sim 10^{39}$ \ergps, and, therefore, requires only $\sim 1$\% of the energy available in star-formation. The only explains the FBs and does not attempt to explain the eROSITA bubbles. More recently, \cite{Tourmente2023} solved the CR transport equations in a `Galactic breeze' setup where a subsonic wind ($v_{\rm wind} \sim 200$ \kmps) is injected from the GC. The wind achieves a steady-state bi-polar CR structure over a time scale of $\sim 300$ Myr and produces a hadronic \gray emission that is consistent with the observations. The model, however, does not reproduce the observed sharp edge of the FBs.


%% file: chap_6_summary.tex
\section{Summary and concluding remarks}
\label{chap6:Summary}

\subsection{Further implications}
\label{chap6:subsec:implications}
While investigating the origin of the FEBs is worthy in its own right, it also has implications for the larger context of galaxy formation and evolution. 

\subsubsection{Using Fermi/eROSITA bubbles to constrain MW CGM}
The shape and brightness of the NPS/eROSITA bubbles provide us with important information regarding the CGM density profile in our Galaxy, irrespective of its SNe or AGN origin. As discussed in section \ref{chap5:subsec:wind-model}, the propagation of the shock in a background medium largely depends on 3 parameters, $\rho_0 r_0^\alpha$, $E_b$, and $\alpha$ (for an assumed power-law density profile). These parameters can be solved from a given shock radius, $r_s$ ($\propto (\rho_0 r_0^\alpha)^{1/(5-\alpha)}$), shock temperature, $T_s$ ($\propto v_s^2$), and an emission measure ($\propto n_{\rm shock}^2 dl$; where $n_{\rm shock}$ is the post-shock density, and $dl$ is the integration length). The fourth parameter, $t$, requires further assumptions. If one assumes that the FBs represent the contact discontinuity, then the ratio of the contact discontinuity and the forward shock can be used to obtain the dynamical time \citep[as was done in][]{Sarkar2015b}. Therefore, the characteristics of the FEBs can, in principle, be used to put constraints on the CGM density profile of the inner $\sim 10-12$ kpc of our Galaxy. Practically, the computation of the shock shape in a general axisymmetric density distribution (dependent on cylindrical radius and height) requires a more sophisticated method \citep[e.g.,][]{Sakashita1971} to solve for the partial differential equation arising from the Kompaneets approximation than the ones used to generate figure \ref{chap5:fig:shock-propagation}.  

Several authors have used the shape and brightness of the eROSITA bubbles to put constraints on the CGM conditions. \cite{Sarkar2019} proposed that the absence of the south polar spur in the ROSAT X-ray maps could be due to a density asymmetry between the northern and southern hemispheres. The model shows that only a $20$\% lower density in the southern hemisphere compared to the northern one could make the southern X-ray bubble significantly faint, owing to the projection effects at the Solar vantage point and a density$^2$ effect on the emissivity. A possible sign of such an asymmetry could also explain the relatively larger ($3\deg-5\deg$) southern Fermi Bubble. Although eROSITA discovered a southern X-ray bubble, the relatively lower brightness of the southern bubbles is consistent with the North-South density asymmetry. \cite{Sofue2019} used the shape of the NPS to conclude that a spherical symmetric $\beta$-profile does not explain the shape of the NPS, rather, one needs a `semi-exponential' disk-like ambient medium where the disk scale height increases rapidly with cylindrical radius, consistent with the MW density profiles inferred by \cite{Nakashima2018}. Apart from the North-South asymmetry, the NPS also shows asymmetry in the East-West direction (figure \ref{chap2:fig:xradio-map}). Several authors proposed that the East-West asymmetry in the NPS could be caused by an East-West density asymmetry, probably due to our Galaxy's motion ($\sim 55$ \kmps) through the intra-group medium toward the M31 galaxy \citep{Sofue2000, Mou2018, Sarkar2019, Sofue2019, Mou2023}. Therefore, the shape, temperature, and brightness of the eROSITA bubbles can provide us with useful information about the inner CGM ($\sim 10$ kpc) of the MW.

\subsubsection{\gray bubbles in other galaxies}
\label{chap6:subsubsec:gray-other-gals}
An obvious question is - can we detect FEB-like features in external galaxies? \cite{Owen2022a, Owen2022b} performed detailed MHD simulations including CRe spectral evolution for an AGN-jet scenario ($E_{\rm tot} \sim 3\times 10^{57}$ erg) in a general galactic CGM setup. The simulations showed that the shock structure for the jet activity in an external galaxy could be visible in radio ($\sim 10$ GHz) and \grays ($1-100$ GeV). Although the radio emission from such an activity should be visible at $\gtrsim 7$ Myr (at $\sim$ Mpc distance), the \gray emission disappears after an IC cooling time of $\sim 1$ Myr. The \gray bubbles, therefore, are only visible to $\sim 5-10$ kpc ($\sim 0.4\deg$ at $1$ Mpc), whereas, the radio emission can be visible to $\sim 40$ kpc  ($\sim 2\deg$ at $1$ Mpc) due to the long synchrotron cooling time at the relevant CRe energies. Although detecting the \gray bubbles using \textit{Fermi}-LAT telescope would be difficult due to its limited sensitivity and spatial resolution ($\Delta\theta\sim 0.15\deg$ at $10$ GeV), radio emission should be detectable using telescopes such as Square Kilometer Array (mid) and Green Bank Telescope. Interestingly, \cite{Pshirkov2016} detected excess \gray emission around M31 (distance $\approx 760$ kpc) and attributed them to two large ($\sim0.5\deg \equiv 6-7$ kpc) bubbles of \gray emission. Given the lower angular resolution of \textit{Fermi}-LAT and the uncertainty of the contribution from a halo-like emission surrounding M31, it is not certain if these are indeed \gray bubbles. The aspect of detecting such bubbles around nearby galaxies, however, remains very interesting. 

Similar to the \gray and radio emission, large-scale outflows can also have detectable shocked shells (like the eROSITA bubbles) in external galaxies. For example, \cite{JTLi2019} find a shell-like emission in hard X-ray ($>2$ keV) corresponding to radio and optical emission in galaxy NGC3079. Cosmological galaxy formation simulations that include some form of AGN feedback notice the presence of large $\sim 10-50$ kpc X-ray emitting bubbles in Milky-Way type galaxies \citep{Pillepich2021}. The estimated X-ray brightness ($10^{-9}$ \ergps \pcmsq sr$^{-1}$) is above the detection limit of the eROSITA telescope (in $0.5-2$ keV band) assuming an exposure time of $\sim 100$ ks. Therefore, detecting such X-ray bubbles/shells could indicate the possibility of detecting \gray and radio-emitting bubbles too.

\subsection{Summary}
To summarize, I have discussed a variety of observational and theoretical aspects of the Fermi/eROSITA bubbles and related multi-wavelength emissions in our Galaxy. These bubbles are the closest example to us of a large-scale galactic outflow and provide an excellent opportunity to study such outflows in unique multi-wavelength tracers. While there is a plethora of observational features toward the GC, understanding the origin of these bubbles has been a challenging task. It is primarily due to the interpretation of these features since no single theoretical model has been able to reproduce all of them simultaneously. An additional confusion regarding the origin of the North Polar Spur is now solved and is understood to be a part of the Galactic Center activity, thanks to the eROSITA discovery of a southern bubble. A successful theoretical model of the FEBs should reproduce -
\begin{itemize}
    \item The X-ray emitting eROSITA bubbles at $\sim 12$ kpc scale with plasma temperature $\sim 0.3$ keV \citep{Kataoka2013, Predehl2020}.
    \item The \gray emitting Fermi Bubbles at $\sim 8-10$ kpc scale with uniform brightness and sharp edge along with a consistent \gray spectrum \citep{Su2010, Ackerman2014}.
    \item The volume filled (within the FBs) microwave emission at $\sim 20-100$ GHz with CR electron spectral index, $p\approx 2.2$ \citep{Finkbeiner2004, PlanckCollaboration2013}.
    \item A polarized radio plume at $\sim 2$ GHz extending beyond the edge of the FBs and containing synchrotron cooling signatures \citep{Carretti2013}.
    \item Warm ($\sim$ few $\times 10^4$ K) cloud kinematics of $v_{\rm los} \sim \pm 200$ \kmps at $|b|\gtrsim 10\deg$ \citep{Fox2015, Bordoloi2017, Ashley2020}.
    \item The North-South and East-West asymmetry (in terms of brightness and shape) of the NPS and radio plume. 
    \item High velocity atomic and molecular clouds at $\sim$ kpc scale, with $v_{\rm los} \sim \pm 200$ \kmps \citep{DiTeodoro2018}.
    \item The X-shaped X-ray emission at $\sim 0.5$ kpc \citep{Bland-Hawthorn2003}.
    \item The excess ionization of the Magellanic stream \citep{Bland-Hawthorn2019}.
\end{itemize}

At this point, it is still unclear whether the eROSITA bubbles and Fermi Bubbles were generated from a single event or from two separate events. As far as the source of the mechanical energy for powering the bubbles, both SNe and AGN feedback could be the reason for these bubbles. The SNe-driven models require only a factor of a few boost in the SFR (section \ref{chap5:subsec:SFR-models}) compared to the currently estimated value but is consistent with some observations claiming to have detected such an enhanced SFR at the GC (section \ref{chap4:subsec:SFR}). The AGN wind/jet/burst models require an enhancement of the jet activity by a factor of $10^{3-7}$ (section \ref{chap5:subsec:AGN-models}) compared to the current AGN power of the \sgra. There are some indications that suggest that \sgra was indeed $\sim 10^{3-4}$ times brighter $\sim 100$ yr ago but it is found to last only for $\sim 1-10$ yr, i.e. an average enhancement of only $10^{2-3}$ times (section \ref{chap4:subsec:MWBH-acc-rate}). There is also evidence from the excess ionization of the Magellanic steam that \sgra was active $\sim 1-3$ Myr ago (similar to the age of the nuclear young star cluster) and had an accretion rate of $\sim 0.1-1$ \% of the Eddington rate, although the actual duration of this activity and hence the total energy produced is uncertain. 

From energetic considerations alone, one can devise two different types of origin scenarios for the FEBs - first, a continuous wind scenario, be it from the star formation or the AGN activity at \sgra, and second, an AGN burst. In the wind scenario, the X-ray temperature constraints suggest that the wind power should be $L_{\rm wind} \sim 10^{40.5-41}$ \ergps (section \ref{chap2:subsec:NPS} and \ref{chap2:subsec:oviii-ovii}) which is achievable using an enhanced (factor of few) star-formation rate or an enhanced (factor of $\sim 10^{3-4}$) wind-like activity from SMBH. The wind models also reproduce the FBs and eROSITA bubbles in the same model. The wind-like power from the \sgra could be generated either from accretion wind or from lower-power jets that are dissipated inside the ISM, or randomly oriented TDEs. The low power of the wind means that the age of the eROSITA bubbles is $t_{\rm erosita}\sim 10^{56}$erg$/10^{41}$ \ergps $\sim 30$ Myr. 
In the AGN-burst scenario, although the models can reproduce the shape and spectrum of the Fermi Bubbles, the temperature of the NPS/eROSITA bubbles is found to be too high compared to the X-ray observations.  Therefore, it is almost certain that an Eddington limited AGN burst \textit{did not} produce the eROSITA bubbles. As for the FBs, it is quite possible that a separate Eddington-limited AGN event (as suggested from the excess ionization in the Magellanic stream; section \ref{chap4:subsec:MWBH-acc-rate}) could have powered them in the last $\sim 3-6$ Myr. However, given that the $450$ pc radio bubbles are also estimated to have a very similar dynamical age, the FBs ($\sim 8-10$ kpc) must be older than these radio bubbles. Therefore, the issue remains - whether the $\approx 6$ Myr old AGN event powered the large-scale FBs or the $450$ pc radio bubbles. Nonetheless, the age of the FBs, $t_{\rm FBs}$, is almost certainly in the range of $\approx 6-30$ Myr and is limited by the age of the $450$ pc radio bubbles and the NPS/eROSITA bubbles. 

As far as the spectral origin of the \gray is concerned, it is almost certain that the \gray emission is produced by leptonic processes rather than hadronic processes. The main drawback of the hadronic process is that it cannot completely explain the microwave haze emission and one requires an extra electron population to explain such emission. Additional constraints are also obtained from hydrodynamical simulations that limit the contribution of the hadronic processes (section \ref{chap5:subsec:SFR-models}). The leptonic process can simultaneously explain the \gray emission as well as the microwave emission. The main issue of short cooling time for the CR electrons can be bypassed by considering \textit{in-situ} acceleration inside the FBs. 

\subsection{Open questions and the way forward}
Several questions need to be answered before one can paint a complete picture of the FEBs.
\begin{itemize}
    \item Are the NPS/Loop-I situated at a Galactic distance ($\sim 8$ kpc) or are they nearby ($\sim 200$ pc) structures? The detection of the X-ray eROSITA bubbles has solidified the fact that these X-ray bubbles are at a Galactic distance. Questions, however, remain regarding the origin of the stellar polarization signatures that suggest a distance of $\sim 200$ pc. Is there a second component of the NPS/Loop-I that lies at a closer distance but has not been detected due to morphological coincidence with the eROSITA bubbles?
    
    \item Is there a southern \gray and microwave/radio bubble corresponding to the southern eROSITA bubble? \gray emission maps for the soft component (figure \ref{chap1:fig:Fermi-skymap}; \citealt{Ackerman2014}) indicates the presence of two horn-like features exactly opposite to the \gray NPS/Loop-I in the northern hemisphere. The polarized microwave emission at $23$ GHz (figure \ref{chap2:fig:xradio-map}; \citealt{Vidal2015}) also shows that the southern eROSITA bubble traces the South Polar Spur. If the \gray and radio structures indeed show a bubble-like emission in the southern hemisphere, they would provide further support for the Galactic Center origin of the NPS/Loop-I. 
    
    \item What is the temperature of the NPS/eROSITA bubbles? A simple assumption of a collisional equilibrium plasma with Solar abundance produces a temperature of $\sim 0.3$ keV \citep{Kataoka2013}. More recent analyses of the NPS have revealed that the plasma temperature could be $\sim 0.2$ keV due to non-Solar abundances \citep{AGupta2023} or $\sim 0.7$ keV, due to non-equilibrium ionization effects \citep{Yamamoto2022}. Careful analysis of the X-ray spectra in addition to creating an accurate non-equilibrium ionization shock model for the FEBs, is required to settle this issue. High-resolution X-ray spectra from \textit{XRISM}/\textit{Athena}/\textit{LEM} will be helpful in this context. Additional information regarding the hot gas velocity in the eROSITA bubbles will be able to determine the shock velocity more accurately. If the NPS expands at a velocity of $v_{\rm exp} \sim 300$ \kmps ($\mathcal{M} \sim 1.5$), an X-ray emission/absorption line through the eROSITA bubbles has a velocity width of $\sim 600$ \kmps, marginally detectable using high-resolution X-ray spectrometers, such as \textit{XRISM} but possible with the Line Emission Mapper (LEM; \citealt{LEM2022}). An X-ray spectrometer with the capability of resolving $\sim 500$ \kmps velocity width will provide a certain answer regarding the expansion speed of these bubbles. Alternatively, a current non-detection of such velocities can also be used to put an upper limit of the $v_{\rm exp}$.
    
    \item What is the characteristic of the X-shaped X-ray emission (figure \ref{chap2:fig:xradio-map}) at the central $0.5$ kpc? Is it produced by thermal emission, if so, then what is the temperature of these features? One would require high-resolution X-ray spectra to understand these features. An estimation of the temperature would be able to inform us about the power of the central activity that produced the FBs. 
    
    \item What is the 3D velocity structure of the atomic/warm clouds inside the FBs? Are they flowing radially or do they have non-radial motions? Recent findings of \cite{Ashley2020} suggest that some of the warm clouds have non-radial motions. Are such velocities common in all the other warm clouds detected in AGN absorption lines? A deep survey tracing the emission signatures of the high-latitude clouds would be able to give us more information regarding the direction of motion (maybe from the head-tail orientation). 
    
    \item What is the star formation rate of the central molecular zone? While source counting of young stellar objects, \ion{H}{i} regions, masers, and SN remnants in the central region suggests SFR$\approx 0.07-0.09$ \mpy in the past few Myr \citep{Henshaw2022}, IR color-magnitude analysis suggests SFR $\approx 0.2-0.8$ \mpy over the past $\sim 30$ Myr \citep{Nogueras-Lara2019}. The resolution of the SFR discrepancy at the CMZ probably requires accurate source counting using high-sensitivity surveys, such as SOFIA, and accurate modeling of the stellar spectra. 
     
    \item From the theoretical side, a complete numerical simulation explaining all the features of the FEBs (as noted in the previous subsection) is awaited. Several theoretical works have reproduced different parts of observational features and none have successfully modeled all of them simultaneously.  One would need MHD simulations with i) accurate CR propagation and spectral evolution of CR protons (and heavier nuclei) and CR electrons, ii) realistic gas density profiles for the ISM and CGM, including the effects of inhomogeneous ISM and gas cooling, and iii) realistic magnetic field orientation in the ISM and CGM to properly model the origin of the FEBs. 
\end{itemize}
The main unanswered questions, however, are i) What is the age of the Fermi Bubbles? ii) Did the Fermi Bubbles and the eROSITA bubbles originate from the same nuclear event or separate events? and finally, iii) What is the source of energy for these bubbles? Are they powered by SNe or the central SMBH?

%% file: chap_7_acknoledgement.tex
\section*{ACKNOWLEDGEMENTS}
I thank Andrew Fox, Santanu Mondal, Matteo Pais, Peter Predehl, Prateek Sharma, Amiel Sternberg, and Nicholas Stone for providing helpful comments that improved the content of the article. I also thank the referee, Joss Bland-Hawthorn, and the scientific editor, Joel Bregman, for providing critical comments that helped improve the content of this article.
I convey my special thanks to Matteo Pais for providing me with the code to calculate shock propagation using Kompaneet's approximation, Santanu Mondal for providing simulation data from \cite{Mondal2022}, and Roland Crocker for providing data from \cite{Crocker&Aharonian2011}. My research in Israel had been supported by the German Science Foundation via DFG/DIP grant STE/ 1869-2 GE/ 625 17-1.